\documentclass[12pt,a4paper,final]{iopart}
\usepackage{iopams}  
\usepackage{graphicx}
\usepackage{hyperref}
\usepackage{cleveref}
\begin{document}
\title[Einstein Flow]{The linear stability of the n+1 dimensional FLRW spacetimes}
\author{Puskar Mondal$^{*}$}

\begin{abstract}
Here we prove the linear stability of a family of `$n+1$'-dimensional cosmological models of general relativity locally isometric to the  Friedmann Lema\^{i}tre Robertson Walker (FLRW) spacetimes including a positive cosmological constant. We show that the solutions to the linearized Einstein-Euler field equations around a class of FLRW metrics with compact spatial topology (negative Einstein spaces and in particular hyperbolic for $n=3$) arising from regular initial data remain uniformly bounded and decay to a family of metrics with constant negative spatial scalar curvature. To accomplish the result, we express the Einstein-Euler system in constant mean extrinsic curvature spatial harmonic gauge and linearize about the chosen FLRW background. Utilizing a Hodge decomposition of the fluid's $n-$velocity 1-form, the linearized system becomes elliptic-hyperbolic (and non-autonomous) in the CMCSH gauge facilitating an application of an energy type argument. Utilizing the estimates derived from the associated elliptic equations, we first prove the uniform boundedness of a energy functional (controlling an appropriate norm of the data) in the expanding direction. Utilizing the uniform boundedness, we later obtain a sharp decay estimate which suggests accelerated expansion (for $\Lambda>0$) of this particular universe model may be sufficient to control the non-linearities (including possible shock formation) of the Einstein-Euler system in a potential future proof of the fully non-linear stability. In addition, the rotational and harmonic parts of the fluid's $n-$velocity field only couple to the remaining degrees of freedom in higher orders, which once again indicates a straightforward extension of current analysis to the fully non-linear setting in the sufficiently small data limit. In addition, our results require a certain integrability condition on the expansion factor and a suitable range of the adiabatic index $\gamma_{ad}$ ($(1,\frac{n+1}{n})$ i.e., $(1,\frac{4}{3})$ in the physically relevant `$3+1$' universe) if the barotropic equation of state $p=(\gamma_{ad}-1)\rho$ is chosen.  
\end{abstract}

\section{Introduction}
Recent astronomical observations \cite{aghanim2018planck, ryan2018constraints} indicate a possibility that the spatial universe may have slightly negative curvature. This leads to the consideration of the well-known FLRW spacetime with negative spatial curvature as a  possible cosmological model. Of course, the spatial manifold of this model is $\mathbb{H}^{3}$and it satisfies the global homogeneity and isotropy criteria as dictated by the cosmological principle. However, the astronomical observations that motivate the cosmological principle are necessarily limited to a fraction (possibly small) of the entire universe and such observations are compatible with spatial metrics being locally but not globally homogeneous and isotropic. Once the restriction on the topology (a global property) is removed, numerous closed manifolds may be constructed as the quotients of $\mathbf{H}^{3}$ by discrete, proper, and torsion-free subgroups of $SO^{+}(1,3)$, with each satisfying the local homogeneity and isotropy criteria but no longer being globally homogeneous or isotropic. In other words, the indication of the spatial universe to have a constant negative curvature opens up the possibility of quite `exotic' topologies. In fact, there are infinitely many closed (compact without boundary) hyperbolic $3-$manifolds topologically distinct from each other. Each such manifold may serve as a possible candidate model for the spatial universe.
These spatially compact model spacetimes may be written as the following warped product form in their `$n+1$' dimensional generalization 
\begin{eqnarray}
\label{eq:background}
^{n+1}g=-dt\otimes dt+a(t)^{2}\gamma_{ij}dx^{i}\otimes dx^{j},
\end{eqnarray}
where $R[\gamma]_{ij}=-\frac{1}{n}\gamma_{ij}$, $t\in (0,\infty)$, $a(t)$ is the scale factor satisfying $\partial_{t}a>0$ for this expanding universe model, and $R[\gamma]_{ij}$ is the Ricci tensor of the metric $\gamma$. Restricting to $n=3$ yields the compact hyperbolic manifolds by Mostow rigidity \cite{lebrun1994einstein}.
However, an important question remains to be answered. Are the FLRW models (variants of FLRW to be precise since spatial manifolds of these models are not globally isotropic and homogeneous only locally so) predicted by general relativity? To conclude that they are may seem to hinge on proof that the purely theoretical FLRW models with compact hyperbolic spatial parts are dynamically stable. A natural first step towards such proof would be to establish stability at the level of linear perturbation theory.

The theory of cosmological perturbations and linearization stability is not new. A substantial amount of work has been done on the topic by several authors (e.g., \cite{hawking1966perturbations, weinberg1972gravitation, bardeen1980gauge, kodama1984cosmological}) since the first study by \cite{lifshitz1992gravitational}. Despite the tremendous amount of existing work in the literature, these are not the `true' linear stability problem rather `formal' mode stability results where the relevant fields are decomposed into modes via eigenfunction (of Laplacian) decomposition and the problem is subsequently reduced to ordinary differential equations in time. However, \textit{the true problem of linear stability concerns
general solutions to the linearized Einstein-Euler system about (\ref{eq:background}) arising from regular initial data, not simply the fixed modes}. Consider the following two linear stability statements \cite{dafermos2019linear}: (a) whether all solutions of the linearized Einstein-Euler system about [\ref{eq:background}] remain bounded for all times in terms of a suitable norm of their initial data, (b) the asymptotic stability i.e., whether all solutions to the linearized equations asymptotically decay. The mode analysis yields necessary but not sufficient conditions for either (a) or (b) of `true' linear stability. In addition, the mode analysis does not work at the non-linear level due to `all to all' coupling (of modes) and therefore a straightforward extension of the linearized mode analysis to study the fully non-linear problem (irrespective of the size of the data) becomes tremendously difficult if not impossible. In particular, we want to extend our analysis in the future to study the fully non-linear problem in our choice of gauge (CMCSH-to be described later) on spacetimes of topological type $M\times \mathbb{R}$ (where $M$ is compact negative Einstein for $n>3$ and in particular compact hyperbolic for $n=3$).

In the regime of fully non-linear stability of Einstein's equations (vacuum or coupled to suitable matter sources), there has been considerable progress in the last three decades. Global nonlinear stability proof of de-Sitter spacetime by \cite{friedrich1986existence} marked the initiation of this progress. This was followed by the proof of the stability of the Minkowski space by \cite{christodoulou1993global}. Later \cite{lindblad2010global} provided with a simplified proof of the same in spacetime harmonic or wave gauge. Andersson and Moncrief \cite{andersson2004future} proved the global existence of `$3+1$' dimensional vacuum Einstein's equations for sufficiently small however fully nonlinear perturbations of spacetimes of the type $\mathbb{R}\times M_{t}$ in constant mean curvature spatial harmonic (CMCSH) gauge, where $M_{t}$ is a compact hyperbolic manifold. Following these fundamental studies, numerous studies have been performed regarding the small data global existence issues associated with Einstein's equations including various sources \cite{ringstrom2008future, oliynyk2016future, fajman2021attractors, fajman2017nonvacuum, rodnianski2018regime, lefloch2021nonlinear}. Two of the studies that will be most relevant to the current study are that by Rodnianski and Speck \cite{rodnianski2009stability} and Speck \cite{speck2012nonlinear}. Rodnianski and Speck \cite{rodnianski2009stability} studied the small data perturbations of the $\mathbb{R}\times \mathbb{T}^{3}$ type FLRW model with a positive cosmological constant in spacetime harmonic gauge, where an irrotational perfect fluid model was assumed. Later Speck \cite{speck2012nonlinear} extended this analysis on the same topology (and same gauge) to include non-zero vorticity utilizing suitably defined energy currents. Commonly, the perfect fluid matter model is considered to be `bad' because of its finite time shock singularity formation property without even coupling to gravity (e.g., \cite{brauer1994cosmic, christodoulou2007formation}). One would surely not hope that while coupled to gravity, this property may be 'suppressed' by gravity since gravity itself may yield finite-time singularity. However, \cite{speck2013stabilizing, rodnianski2009stability, speck2012nonlinear} have utilized the property that the presence of a positive cosmological constant induces an accelerated expansion of the physical universe, which avoids shocks in the regime of a small data limit (rapid expansion obstructs energy concentration by non-linearities). With an accelerated expansion, there are universally decaying `good' terms in the evolution equations that dominate the nonlinearities and drive the universe towards a stable configuration. Recently, \cite{fajman2021slowly} proved a nonlinear stability result for Milne universe including dust fluid source. This model is devoid of a positive cosmological constant and therefore slowly expanding.     

Our analysis indicates that if the expansion factor $a(t)$ obeys a suitable integrability condition and the adiabatic index $\gamma_{ad}$ lies in the range $(1,\frac{n+1}{n})$ ($(1,4/3)$ in the physically relevant `3+1' case; $\gamma_{ad}=1$ and $\gamma_{ad}=4/3$ correspond to a pressure-less fluid or `dust' and a `radiation` fluid, respectively), then the Lyapanuv functional remains uniformly bounded by its initial value and moreover the perturbations decay in the expanding direction. This is in a sense equivalent to the fact that if the perturbations to the matter part are restricted within a smaller cone contained in the `sound' cone (in the tangent space; to be defined later), that is, they do not propagate at speed higher than $\sqrt{\frac{1}{n}}$ (in the unit of light speed), then the asymptotic stability holds. On the other hand, there exists a seemingly contradictory result in the purely non-relativistic case of a self-gravitating fluid system. For a self gravitating non-relativistic 3-dimensional fluid body (Euler-Poisson system), Chandrasekhar \cite{chandrasekhar2013hydrodynamic} used a virial identity argument to show that the static isolated compact solution of the Euler-Poisson system is stable for $\gamma_{ad}>\frac{4}{3}$. However, if one observes carefully, our result and Chandrasekhar's result are not contradictory since the notions of stability are different in the two contexts. In the context of Chandrasekhar's argument, the stability is defined by the negativity of total energy i.e., the gravitational energy dominates (stable $\iff$ gravitationally bound system). Now of course, a system with strong gravity is stable in the sense of Chandrasekhar (since total energy is dominated by gravitational energy and therefore negative) which is nonsensical in our fully relativistic context (ultra high gravity could focus and blow up forming a singularity, indicating instability in our context).

The stability criteria is satisfied by the universe model in which one includes a positive cosmological constant $\Lambda$ and in such case, one obtains a uniform decay at the linear level. However, turning off the cosmological constant results in losing the uniform boundedness property simply because one does not have the integrability property for the scale factor. This may be attributed to the fact that a positive cosmological constant induced accelerated expansion wins over the gravitational effect of the fluid source at the level of linear theory. However, we do not claim that turning off the cosmological constant leads to instability but simply we are unable to reach a definite conclusion with the currently available method. In addition, we also note that in the borderline case $\gamma_{ad}=\frac{n+1}{n}$, we can only prove uniform boundedness of the energy functional (no decay). We will investigate the borderline cases of $\gamma_{ad}=0$ and $\gamma_{ad}=\frac{n+1}{n}$ using special techniques in the future. Considering this stability criterion holds, we hope to extend our analysis in a potential future proof of non-linear stability of (\ref{eq:background}) in the CMCSH gauge using the energy current method developed by Christodoulou \cite{christodoulou2016action}.

\subsection{\textbf{Notations and facts}}
The `$n+1$' dimensional spacetime manifold is denoted by $\tilde{M}$. We are interested in spacetime manifolds $\tilde{M}$ with the product topology $\mathbb{R}\times M$, where $M$ denotes the $n$-dimensional spatial manifold diffeomorphic to a Cauchy hypersurface (i.e., every inextendible causal curve intersects $M$ exactly once). We denote the space of the Riemannian metrics on $M$ by $\mathcal{M}$. A subspace $\mathcal{M}_{-1}$ of $\mathcal{M}$ is defined as follows 
\begin{eqnarray}
\mathcal{M}_{-1}=\{\gamma\in \mathcal{M}| R(\gamma)=-1\},\nonumber
\end{eqnarray}
with $R(\gamma)$ is the scalar curvature associated with $\gamma\in \mathcal{M}$. We explicitly work in the $L^{2}$ (with respect to a background metric $\gamma$) Sobolev space $H^{s}$ for $s>\frac{n}{2}+1$. The $L^{2}$ inner product on the fibres of the bundle of symmetric covariant 2-tensors on $(M,\gamma)$ with co-variant derivative $\nabla[\gamma]$ is defined as 
\begin{eqnarray}
\langle u|v\rangle_{L^{2}}:=\int_{M}u_{ij}v_{kl}\gamma^{ik}\gamma^{jl}\mu_{\gamma}.
\end{eqnarray}
The standard norms are defined naturally as follows 
\begin{eqnarray}
||u||_{L^{2}}:=\left(\int_{M}u_{ij}u_{kl}\gamma^{ik}\gamma^{jl}\mu_{\gamma}\right)^{\frac{1}{2}}\\
||u||_{L^{\infty}}:=\sup_{M}(u_{ij}u_{kl}\gamma^{ik}\gamma^{jl})^{1/2}
\end{eqnarray}
and so on. The inner product on the derivatives is defined 
\begin{eqnarray}
\langle\nabla[\gamma] u|\nabla[\gamma]v\rangle_{L^{2}}&=&\int_{M}\nabla[\gamma]_{m}u_{ij}\nabla[\gamma]_{n}v_{kl}\gamma^{mn}\gamma^{ik}\gamma^{jl}\mu_{\gamma}\nonumber,
\end{eqnarray} 
where $\mu_{\gamma}$ is the volume form associated with $\gamma\in\mathcal{M}$
\begin{eqnarray}
\mu_{\gamma}=\sqrt{\det(\gamma_{ij})}dx^{1}\wedge dx^{2}\wedge dx^{3}\wedge.........\wedge dx^{n}.\nonumber
\end{eqnarray}
we will use $\mu_{\gamma}$ to denote both the volume form as well as $\sqrt{\det(\gamma_{ij})}$ (slight abuse of notation). We define the rough Laplacian $\Delta_{\gamma}$ on $(M,\gamma)$ for any $\gamma\in \mathcal{M}$ in the following way so that it has a non-negative spectrum i.e., \begin{eqnarray}
\Delta_{\gamma}\equiv-\gamma^{ij}\nabla_{i}\nabla_{j}.
\end{eqnarray}
For $a,b \in R_{>0}$, $a\lesssim b$ is defined to be $a\leq Cb$ for some constant $0<C<\infty$. This essentially means that $a$ is controlled by $b$ (in this context $a$ and $b$ could be norms of two different entities). For example $||A||_{L^{2}}\lesssim ||B||_{L^{2}}$ means 
$||A||_{L^{2}}\leq C||B||_{L^{2}}$, where $0<C<\infty$.  
The spaces of symmetric covariant 2-tensors and vector fields on $M$ are denoted by $S^{0}_{2}(M)$ and $\mathfrak{X}(M)$, respectively. The adiabatic index in denoted by $\gamma_{ad}$.

\subsection{\textbf{Overview of the paper}}
Section 2 provides the necessary detail about the formulation of the gauge fixed Einstein-Euler field equations including a positive cosmological constant $\Lambda$, the motivation of the study, and the main result. Note that we do not however discuss the technicalities associated with the main result in this section rather a few physical consequences. 

Next in section $3$, we introduce the re-scaled system of Einstein-Euler field equations with a positive cosmological constant and state the key properties of the re-scaled background solutions describing a class of fixed points. In addition, we describe the center manifold of the re-scaled dynamics when the Einstein moduli space is non-trivial. Lastly, we describe the kinematics of the perturbations under the assumption of a shadow gauge condition (introduced by \cite{andersson2011einstein}) when the center manifold is finite-dimensional in addition to computing a crucial term that appears in one of the elliptic equations obtained through imposing the gauge condition. 

In section $4$ we linearize the re-scaled field equations about the background solutions described in sections $2$ and $3$. Let us for the moment denote the field equations by $\mathcal{O}[\mathcal{A}]=0$ where $\mathcal{A}$ denotes the associated fields with their background values being $\mathcal{A}_{B.G}$. The linearized equations about the background $\mathcal{A}_{B.G}$ are obtained by setting the first variation to zero i.e., $D\mathcal{O}|_{\mathcal{A}_{B.G}}\cdot \delta \mathcal{A}=\frac{d}{dt}\mathcal{O}(A_{B.G}+t\delta \mathcal{A})|_{t=0}=0$, where the first variations to the fields are denoted by $\delta \mathcal{A}$. The linearized equations are studied with the assumption that the suitable function space norms of the higher-order terms are significantly smaller compared to the linear terms. Note that the relativistic Euler's equations are essentially a hyperbolic system of partial differential equations that are derivable from a Lagrangian (see \cite{moncrief1977hamiltonian, demaret1980hamiltonian, bao1985hamiltonian} for a Hamiltonian structure of a relativistic perfect fluid). At the linearized level, this may be shown explicitly by employing Hodge decomposition of the $n-$velocity vector field  (projection of the $n+1-$velocity field onto the spatial hypersurface $M$). Doing so we reduce the relativistic Euler's equations to a wave equation for the associated scalar potential of the irrotational part (with respect to the associated `sound' metric) while the rotational part and the harmonic part (a topological contribution) of the velocity field become decoupled and satisfy ordinary differential equations in time. Therefore the linearized coupled Einstein-Euler field equations become an `elliptic- hyperbolic' system in our choice of gauge. Lastly, we close this section by stating a local well-posedness theorem for this coupled system and necessary elliptic estimates. 

Section $5$ is the most important chapter in terms of the proof of the main theorem of the paper. To address stability issues of a dynamical system, the use of energy functional is indispensable. On this particular occasion, we model the energy functional after the wave equation type energy introduced by \cite{andersson2011einstein}. We construct a similar wave equation type energy and its higher-order analog. First, we show the uniform boundedness of this energy functional in the expanding direction, and later utilizing this uniform boundedness we prove the decay property of the desired fields. Doing so we derive the required conditions for the stability results to hold. Finally, we close the section with a sketch of the proof of the geodesic completeness of the perturbed spacetimes at the level of linear perturbation theory.

\section{Complete Einstein-Euler system and gauge fixing} 

The ADM formalism splits the spacetime described by an `$n+1$' dimensional Lorentzian manifold $\tilde{M}$ into $\mathbb{R}\times M$. Here each level set $\{t\}\times M$ of the time function $t$ is an orientable n-manifold diffeomorphic to a Cauchy hypersurface (assuming the spacetime admits a Cauchy hypersurface) and equipped with a Riemannian metric. Such a split may be executed by introducing a lapse function $N$ and shift vector field $X$ belonging to suitable function spaces on $\tilde{M}$ and defined such that
\begin{eqnarray}
\partial_{t}&=&N\mathbf{n}+X,
\end{eqnarray}
where $t$ and $\mathbf{n}$ are time and a hypersurface orthogonal future directed timelike unit vector i.e., $\hat{g}(\mathbf{n},\mathbf{n})=-1$, respectively. The above splitting writes the spacetime metric $\hat{g}$ in local coordinates $\{x^{\alpha}\}_{\alpha=0}^{n}=\{t,x^{1},x^{2},....,x^{n}\}$ as 
\begin{eqnarray}
\hat{g}&=&-N^{2}dt\otimes dt+g_{ij}(dx^{i}+X^{i}dt)\otimes(dx^{j}+X^{j}dt)
\end{eqnarray} 
and the stress-energy tensor as
\begin{eqnarray}
\mathbf{T}&=&E\mathbf{n}\otimes \mathbf{n}+2\mathbf{J}\odot \mathbf{n}+\mathbf{S},
\end{eqnarray}
where $\mathbf{J}\in \mathfrak{X}(M)$, $\mathbf{S}\in S^{2}_{0}(M)$, and $A\odot B=\frac{1}{2}(A\otimes B+B\otimes A$). Here, $\mathfrak{X}(M)$ and $S^{2}_{0}(M)$ are the space of vector fields and the space of symmetric covariant 2-tensors, respectively. Here $E:=\mathbf{T}(\mathbf{n},\mathbf{n})$ is the energy density observed by a time-like observer with $n+1$-velocity $\mathbf{n}$, $\mathbf{J}_{i}=-\mathbf{T}(\partial_{i},\mathbf{n})$ is the momentum density, $\mathbf{J}_{i}=-\mathbf{T}(\mathbf{n},\partial_{i})=$ is the energy flux density, and $\mathbf{S}_{ij}=\mathbf{T}(\partial_{i},\partial_{j})$ is the momentum flux density (with respect to the chosen constant $t$ hypersurface $M$). The choice of a spatial slice in the spacetime leads to consideration of the second fundamental form $k_{ij}$ which describes how the slice is curved in the spacetime. The trace of the second fundamental form ($\tau:=\tr_{g}k$) is the mean extrinsic curvature of the slice, which will play an important role in the analysis. Under such decomposition, the Einstein equations 
\begin{eqnarray}
R_{\mu\nu}-\frac{1}{2}R\hat{g}_{\mu\nu}+\Lambda \hat{g}_{\mu\nu}&=&T_{\mu\nu}
\end{eqnarray}
take the form ($8\pi G=c=1$)
\begin{eqnarray}
\partial_{t}g_{ij}&=&-2Nk_{ij}+L_{X}g_{ij},\\\nonumber
\partial_{t}k_{ij}&=&-\nabla_{i}\nabla_{j}N+N\{R_{ij}+\tau k_{ij}-2k_{ik}k^{k}_{j}\\\nonumber
&&-\frac{1}{n-1}(2\Lambda-S+E)g_{ij}-\mathbf{S}_{ij}\}+L_{X}k_{ij}
\end{eqnarray}
along with the constraints (Gauss and Codazzi equations)
\begin{eqnarray}
\label{eq:HC}
R(g)-|k|^{2}+\tau^{2}&=&2\Lambda+2E,\\
\label{eq:MC}
\nabla_{j}k^{j}_{i}-\nabla_{i}\tau&=&-\mathbf{J}_{i},
\end{eqnarray} 
where $S=g^{ij}\mathbf{S}_{ij}$. The vanishing of the covariant divergence of the stress energy tensor i.e., $\nabla_{\nu}T^{\mu\nu}=0$ is equivalent to the continuity equation and equations of motion of the matter
\begin{eqnarray}
\frac{\partial E}{\partial t}&=&L_{X}E+NE\tau-N\nabla_{i}\mathbf{J}^{i}-2\mathbf{J}^{i}\nabla_{i}N+N\mathbf{S}^{ij}k_{ij},\\
\frac{\partial \mathbf{J}^{i}}{\partial t}&=&L_{X}\mathbf{J}^{i}+N\tau \mathbf{J}^{i}-\nabla_{j}(N\mathbf{S}^{ij})+2Nk^{i}_{j}\mathbf{J}^{j}-E\nabla^{i}N.
\end{eqnarray}
We want to study the Einstein-Euler system. Let the `$n+1$' velocity field of a perfect fluid be denoted by $\mathbf{u}$ which satisfies the normalization condition $\hat{g}(\mathbf{u},\mathbf{u})=-1$. One may for convenience decompose the $n+1$ velocity $\mathbf{u}$ into its component parallel and perpendicular to constant $t$ hypersurface $M$ as follows 
\begin{eqnarray}
\mathbf{u}=v-\hat{g}(\mathbf{u},\mathbf{n})\mathbf{n}
\end{eqnarray}
where $v$ is a vector field parallel to the spatial manifold $M$ i.e., $v\in \mathfrak{X}(M)$. Importantly note that $\mathbf{u}_{i}=v_{i}$ but $\mathbf{u}^{i}\neq v^{i}$ unless the shift vector field $X$ vanishes. 
The stress energy tensor for a perfect fluid with $n+1$ velocity field $\mathbf{u}$ reads
\begin{eqnarray}
\mathbf{T}=(P+\rho)\mathbf{u}\otimes\mathbf{u}+P\hat{g},
\end{eqnarray}
where $P$ and $\rho$ are the pressure and the mass energy density, respectively and $\hat{g}$ is the spacetime metric. After projecting onto the spatial manifold $M$, several components of the stress energy tensor may be computed as follows
\begin{eqnarray}
E=(P+\rho)(N\mathbf{u}^{0})^{2}-P, 
\mathbf{J}^{i}=(P+\rho)\sqrt{1+g(v,v)}v^{i},\\\nonumber 
\mathbf{S}=(P+\rho)v\otimes v+P g. 
\end{eqnarray}
However, notice that the field equations do not close with the information of the stress-energy tensor alone, and therefore one needs an equation of state relating pressure $P$ and mass-energy density $\rho$. The choice of the equation of state is a non-trivial fact and there is numerous study about the equation of state alone in cosmology literature (e.g., \cite{babichev2004dark, chavanis2018simple, nakamura1999determining}). In the standard model of cosmology, one frequently uses a barotropic equation of state of the type $P=(\gamma_{ad}-1)\rho$, where $\gamma_{ad}$ is the adiabatic index. The speed of sound $C_{s}$ is defined as $C^{2}_{s}:=\frac{\partial P}{\partial\rho}$, where the derivative is computed at constant entropy. However, for the barotropic equation of state (for which pressure is only a function of mass-energy density) chosen here, the entropy equation decouples from the field equations and therefore plays no role in our analysis (e.g., see \cite{christodoulou2016action}). The speed of sound is then a constant $\sqrt{\gamma_{ad}-1}$. Due to causality, one must have $0\leq C^{2}_{s}\leq 1$ yielding $\gamma_{ad}\in [1,2]$. This is equivalent to the fact that the sound cone is contained within the light cone in the tangent space. Here $\gamma_{ad}=1$ corresponds to pressure-less fluid or `dust' and $\gamma_{ad}=2$ corresponds to a ultra-relativistic stiff fluid. We note an important fact that choosing this equation of state, we ignore the rest energy of the fluid which may be relevant in the late epoch of the universe evolution. In other words, the equation of state $P=(\gamma_{ad}-1)(\rho-ne)$, $n$ being the baryon number density and $e$ the fluid rest energy per particle, may be more appropriate \cite{eosnew}. Here we will stick with the equation of state $P=(\gamma_{ad}-1)\rho$ here for simplicity and in such case, the baryon number conservation equation $\nabla_{\mu}(nu^{\mu})=0$ is a simple consequence of the Euler's equations i.e., $\nabla_{\nu}T^{\mu\nu}=0$ \cite{christodoulou2016action}. With the barotropic equation of state, the components of the stress energy tensor is computed to be     
\begin{eqnarray}
E=(P+\rho)(1+g(v,v))-P=\rho+\gamma_{ad}\rho g(v,v)\\\nonumber
\mathbf{J}^{i}=\gamma_{ad}\rho\sqrt{1+g(v,v)}v^{i},
\mathbf{S}=\gamma_{ad}\rho v\otimes v+(\gamma_{ad}-1)\rho g.
\end{eqnarray}
The complete system of evolution and constraint equations is expressible as follows
\begin{eqnarray}
 \partial_{t}\rho+\frac{\gamma\rho N}{[1+g(v,v)]^{1/2}}\nabla_{i}v^{i}=L_{X}\rho\nonumber-\frac{\gamma_{ad}\rho g(\partial_{t}v,v)}{1+g(v,v)}+\frac{\gamma_{ad}\rho Nk(v,v)}{1+g(v,v)}-\frac{2\gamma_{ad}\rho v_{i}L_{X}v^{i}}{1+g(v,v)}\\\nonumber 
-\frac{\gamma_{ad}N\rho L_{v}N}{[1+g(v,v)]^{1/2}}+Nk^{i}_{i}\rho-\frac{NL_{v}\rho}{[1+g(v,v)]^{1/2}}.
 \end{eqnarray}
and
\begin{eqnarray}
\gamma_{ad}\rho\left[u^{0}\partial_{t}v^{i}-2Nu^{0}k^{i}_{j}v^{j}-u^{0}(L_{X}v)^{i}+v^{j}\nabla_{j}v^{i}+\frac{1+g(v,v)}{N}\nabla^{i}N\right]\\\nonumber 
+(\gamma_{ad}-1)\left[\nabla^{i}\rho+v^{i}L_{v}\rho+u^{0}v^{i}(\partial_{t}\rho-L_{X}\rho )\right]=0
\end{eqnarray}
\begin{eqnarray}
\label{eq:evol1}
\partial_{t}g_{ij}&=&-2Nk_{ij}+L_{X}g_{ij},\\
\label{eq:evol2}
\partial_{t}k_{ij}&=&-\nabla_{i}\nabla_{j}N+N\left\{R_{ij}+\tau k_{ij}-2k_{ik}k^{k}_{j}-\frac{2\Lambda}{n-1} g_{ij}\right.\\\nonumber
&&\left.-\gamma_{ad}\rho v_{i}v_{j}+\frac{\gamma_{ad}-2}{n-1}\rho g_{ij}\right\}
+L_{X}k_{ij},
\end{eqnarray}
\begin{eqnarray}
\label{eq:HC}
R(g)-|k|^{2}+\tau^{2}&=&2\Lambda+2\rho\left\{1+\gamma_{ad} g(v,v)\right\},\\
\label{eq:MC}
\nabla_{i}k^{ij}-\nabla^{j}\tau&=&-\gamma_{ad}\rho\sqrt{1+g(v,v)}v^{i}.
\end{eqnarray} 

In order to obtain equations satisfied by the lapse function and the shift vector field, we need to fix a choice of gauge. Setting the mean extrinsic curvature ($\tau$) of the hypersurface $M$ to a constant allows it to be a suitable time function. This choice of temporal gauge yields an elliptic equation for the lapse function $N$. However, not all spacetimes admit a constant mean extrinsic curvature (CMC) hypersurface. Existence of a CMC slice is far from obvious and is a part of active mathematical research (for detail see \cite{bartnik1988remarks, galloway2018existence, rendall1996constant, andersson1999existence}). Luckily the background spacetimes (\ref{eq:background}) that we are interested in, do admit a CMC slice which is verified through explicit calculations. Therefore simply setting the mean extrinsic curvature of the perturbed spacetimes to be constant on each spatial slice seems to be a reasonable choice
\begin{eqnarray}
\label{eq:uniformmean}
\partial_{i}\tau&=&0.
\end{eqnarray}
Such an assumption can be made without loss of generality for solutions that are close to the background solution \cite{fajman2016local, rodnianski2018stable}. 
This choice allows $\tau$ to play the role of time i.e.,
\begin{eqnarray}
t=monotonic~function~of~\tau.
\end{eqnarray}
To utilize the property $\partial_{i}\tau=0$, one may simply compute the entity $\frac{\partial\tau}{\partial t}$ to yield
\begin{eqnarray}
\frac{\partial \tau}{\partial t}=\Delta_{\gamma}N+\left\{|k|^{2}_{g}+[\frac{(n\gamma-2)}{n-1}+\gamma_{ad} g(v,v)]\rho-\frac{2\Lambda}{n-1}\right\}N+L_{X}\tau.
\end{eqnarray}
Using equation (\ref{eq:uniformmean}), we obtain the desired elliptic equation for the lapse function
\begin{eqnarray}
\Delta_{g}N+\left\{|k|^{2}_{g}+[\frac{(n\gamma-2)}{n-1}+\gamma_{ad} g(v,v)]\rho-\frac{2\Lambda}{n-1}\right\}N=&\frac{\partial \tau}{\partial t}.
\end{eqnarray}
Once we have fixed the temporal gauge, we need to fix the spatial gauge which would yield an equation for the shift vector field. We follow the work of \cite{andersson2004future, andersson2011einstein} regarding spatial gauge fixing. Let $\zeta: (M,g)\to (M,\gamma)$ be a harmonic map with the Dirichlet energy $\frac{1}{2}\int_{M}g^{ij}\frac{\partial \zeta^{k}}{\partial x^{i}}\frac{\partial \zeta^{l}}{\partial x^{j}}\gamma_{kl}\mu_{g}$. Since the harmonic maps are the critical points of the Dirichlet energy functional, $\zeta$ satisfies the following formal Euler-Lagrange equation 
\begin{eqnarray}
g^{ij}\left(\partial_{i}\partial_{j}\zeta^{k}-\Gamma[g]_{ij}^{l}\partial_{l} \zeta^{k}+\Gamma[\gamma]_{\alpha\beta}^{k}\partial_{i} \zeta^{\alpha}\partial_{j} \zeta^{\beta}\right)=0.
\end{eqnarray}
Now, we fix the gauge by imposing the condition that $\zeta=id$, which leads to the following equation 
\begin{eqnarray}
\label{eq:sh}
-g^{ij}(\Gamma[g]_{ij}^{k}-\Gamma[\gamma]_{ij}^{k})&=&0.
\end{eqnarray}
where $\hat{\Gamma}[\gamma]_{ij}^{k}$ is the connection with respect to some arbitrary background Riemannian metric $\gamma$. Choice of this spatial harmonic gauge yields the following elliptic equation for the shift vector field $X$ after time differentiating equation (\ref{eq:sh})  
\begin{eqnarray}
\Delta_{g}X^{i}-R^{i}_{j}X^{j}&=&-2k^{ij}\nabla_{j}N-2N\nabla_{j}k^{ij}+\tau\nabla^{i}N\\\nonumber
&&+(2Nk^{jk}-2\nabla^{j}X^{k})(\Gamma[g]^{i}_{jk}-\Gamma[\gamma]^{i}_{jk})-g^{ij}\partial_{t}\Gamma[\gamma]_{ij}^{k}.
\end{eqnarray}
However, one natural question arises while imposing the spatial harmonic gauge. Does there exist a harmonic map between the manifolds $(M,g)$ and $(M,\gamma)$? This is an extremely important and nontrivial issue. There are explicit results in geometry about the existence of harmonic maps between Riemannian manifolds. The one that is relevant to the current study is that of Eells and Sampson \cite{eells1964harmonic} who proved if $M$ is a compact manifold with non-positive Riemann curvature, then any continuous map from a compact manifold into $M$ is homotopic to a harmonic map. Andersson and Moncrief \cite{andersson2003elliptic} applied this gauge to prove a local existence theorem for the vacuum Einstein's equations in arbitrary dimensions. Later they have successfully implemented this gauge to prove a small data nonlinear stability result of the `Milne' model and its higher-dimensional generalization (so-called `Lorentz cone spacetimes') \cite{andersson2004future, andersson2011einstein}. We designate our gauge fixed system as \textbf{CMCSH Einstein-Euler-$\Lambda$} system.



\subsection{\textbf{Motivation of the study and main result}}
In this section, we describe the motivation for the current study. Firstly, for a universe model to be a viable candidate for a physical cosmological model, it must be stable against perturbations in a suitable sense. Therefore, one should ideally prove the fully non-linear stability assuming certain smallness condition on the data. A natural first step towards such stability would be to establish the stability at the level of linear perturbation theory. In addition to this obvious motivation, based on a few preliminary studies \cite{fischer2002hamiltonian,  moncrief2019could, mondal2019asymptotic} we are led to believe that these expanding (accelerated expansion since $\Lambda>0$) spacetimes may be stable under small perturbations. From a physical viewpoint, the accelerated expansion tends to kill off the perturbations. Let us describe this point in a slightly more formal way.  First, we decompose the second fundamental form $k$ as follows
\begin{eqnarray}
k_{ij}=k^{TT}_{ij}+\frac{tr_{g}k}{n}g_{ij}+(L_{Y}g-\frac{2}{n}(\nabla\cdot Y)g)_{ij},
\end{eqnarray}
where $k^{TT}$ is a transverse-traceless (with respect to the metric $g$) symmetric $2-$tensor, $tr_{g}k$ is the trace of the second fundamental form with respect to the metric $g$, and $Y$ is a vector field tangent to $M$. 
Now consider the Hamiltonian constraint (\ref{eq:HC}) 
\begin{eqnarray}
\label{eq:HC1}
\frac{n-1}{n}(\tau^{2}-\frac{2n\Lambda}{n-1})=|k^{TT}|^{2}_{g}+|L_{Y}g-\frac{2}{n}(\nabla\cdot Y)g|^{2}_{g}\\\nonumber 
+2\rho(1+\gamma_{ad} g(v,v))-R(g)
\end{eqnarray}
and construct the following integral 
\begin{eqnarray}
\frac{n-1}{n}\int_{M}(\tau^{2}-\frac{2n\Lambda}{n-1})\mu_{g}=\int_{M}(|k^{TT}|^{2}_{g}+|L_{Y}g-\frac{2}{n}(\nabla\cdot Y)g|^{2}_{g}\\\nonumber 
+2\rho(1+\gamma_{ad} g(v,v))-R(g))\mu_{g}>0,
\end{eqnarray}
then we readily observe that the integral contains the $L^{2}$ norm of $k^{tr}=k^{TT}_{ij}+(L_{Y}g-\frac{2}{n}(\nabla\cdot Y)g)_{ij}$ in addition to the physical energy density and is strictly positive since the re-scaled volume is positive (note that $\tau^{2}>\frac{2n\Lambda}{n-1}$ for spacetimes of interest; spatial manifold is of negative Yamabe type; see \cite{moncrief2019could} for detail about Yamabe classification; this inequality can be showed as follows: The left hand side of (\ref{eq:HC1}) contains non-negative term except $-R(g)$. But since $M$ is assumed to be negative Yamabe, $R(g)<0$ must hold at some point on $M$. But $\tau^{2}-\frac{2n\Lambda}{n-1}$ is constant on $M$ due to constant mean extrinsic curvature assumption on each spatial slice. Therefore $\tau^{2}>\frac{2n\Lambda}{n-1}$). Motivated by such property, \cite{fischer2002hamiltonian} studied the monotonic decay of the entity $\int_{M}\tau^{2}\mu_{g}$ for pure vacuum case and \cite{moncrief2019could} studied the same for $\int_{M}(\tau^{2}-\frac{2n\Lambda}{n-1})\mu_{g}$ in the vacuum case with $\Lambda>0$ in the CMC gauge (also see \cite{fajmaneinstein1, fajmaneinstein2}). In each of these cases, this re-scaled volume acted as a Hamiltonian while expressed in terms of suitable ``reduced" phase space variables.
Subsequently, \cite{mondal2019asymptotic} studied the monotonic decay (in the expanding direction) of the entity $\int_{M}(\tau^{2}-\frac{2n\Lambda}{n-1})\mu_{g}$ in the presence of matter sources satisfying suitable energy conditions in particular perfect fluids, where this entity played the role of a Lyapunov functional.

Naively, monotonic decay of this Lyapunov functional indicates the stability of a class of expanding solutions (on which the Lyapunov functional attains its infimum) against arbitrary large data perturbations. Motivated by this result, we are interested in pursuing the stability property of a perfect fluid-filled spatially compact universe model. Before proceeding to the fully nonlinear study of the stability problem, it is important to prove the linear stability in a rigorous way (not simply the mode stability) and obtain decay estimates if possible. In a sense, the linear stability and decay should these hold provide motivation towards studying the fully non-linear stability and may also become useful in handling the higher-order couplings and defining appropriate energies. However, this does not imply stability since the Lyapunov functional does not control the norm that is required by the local existence theorem (for vacuum and $\Lambda$-vacuum the function space required for $(g,k)$ by the local existence is  $H^{s}\times H^{s-1}, s>\frac{n}{2}+1$). It only controls the minimum regularity ($H^{1}\times L^{2}$ of $(g,k)$) and therefore is not sufficient to state a global existence for arbitrarily large data (one may therefore call this Lyapunov functional as a `weak' Lyapunov functional). Even though the re-scaled volume functional controlling the minimum norm decays monotonically in the expanding direction and attains its infimum for a particular class of spacetimes (see the relevant sections of \cite{fischer2002hamiltonian, moncrief2019could, mondal2019asymptotic} for details), the classical solution may form a finite time singularity via curvature focusing. In other words, a suitable norm of arbitrarily large data (physically interesting solutions) may be obstructed to decay monotonically to its infimum since pure gravity could blow up before the spatial volume of the universe tends to infinity. The addition of perfect fluids adds additional problems since they are known to form shock singularities in the fully non-linear setting. Christodoulou proved that perfect fluids form shock on Minkowski background for arbitrary small data \cite{christodoulou2007formation}. Of course, Minkowski space does not have the expansion property (accelerated expansion including $\Lambda>0$) like the spacetimes (\ref{eq:background}) of the current investigation do. We denote a ball centered at $(x,y,z,..)$ with radius $\delta$ by $B_{\delta}(x,y,z,..)$. The Einstein structure of the manifold $M$ (or the deformation space of an arbitrary Einstein metric $\gamma$ on $M$) is denoted by $\mathcal{N}$. This will be defined later in a rigorous way. Now we summarize the result in terms of the following `rough' version which is equivalent to the main theorem stated in the later part of the paper. 

\textbf{A rough version of the main Theorem:} \textit{Let $(g_{0},k^{tr}_{0},\rho_{0},v_{0})\in H^{s}\times H^{s-1}\times H^{s-1}\times H^{s-1},~s>\frac{n}{2}+2$ be the initial data for the re-scaled Einstein-Euler-$\Lambda$ system with $\Lambda>0$. Let us also consider that the Cauchy data for the re-scaled background solutions are $(g_{B.G},k^{tr}_{B.G},\rho_{B.G},v_{B.G})$ which satisfy $R[g_{B.G}]_{ij}=-\frac{1}{n}(g_{B.G})_{ij}, k^{tr}_{B.G}=0, \rho_{B.G}=C_{\rho}, v^{i}_{B.G}=0$, $C_{\rho}$ is a constant. Now assume that the following smallness condition holds $||g_{0}-g_{B.G}||_{H^{s}}+||k^{tr}_{0}-k^{tr}_{B.G}||_{H^{s-1}}+||\rho_{0}-\rho_{B.G}||_{H^{s-1}}+||v_{0}-v_{B.G}||_{H^{s-1}}<\delta$, $\delta$ is chosen sufficiently small. If $t \mapsto (g(t), k^{tr}(t), \rho(t), v(t))$ is the maximal development of the Cauchy problem for the linearized re-scaled Einstein-Euler-$\Lambda$ system about the background solutions $(g_{B.G},k^{tr}_{B.G},\rho_{B.G},v_{B.G})$ in constant mean extrinsic curvature spatial harmonic gauge (CMCSH) (\ref{eq:metric1}-\ref{eq:MCNW}) with initial data $(g_{0},k^{tr}_{0},\rho_{0},v_{0})$, then the following holds in the limit of infinite time
\begin{eqnarray}
\lim_{t\to\infty}(g(t),k^{tr}(t),\rho(t),v(t))=(\gamma^{\dag},0,\rho^{'},0),
\end{eqnarray}
where $\gamma^{\dag}$ satisfies $R[\gamma^{\dag}]=-1$ and $\rho^{'}$ depends on the initial re-scaled density $\rho_{0}$ and $\partial_{i}\rho^{'}\neq 0$ in general.
}

Let us now explain the main theorem putting aside the technical details. The statement essentially deals with the \textit{asymptotic} stability of the solutions that are characterized by constant negative scalar curvature ($-1$ to be precise in this occasion) and sufficiently close to the FLRW solutions In other words this corresponds to a \textit{Lyapunov} stability of the FLRW solutions. In other words, the solutions of linearized Einstein-Euler-$\Lambda$ equations about the background FLRW solutions (\ref{eq:background}) decay to nearby solutions which have constant negative scalar curvature. While the background solutions the FLRW spacetimes (\ref{eq:background}) are described by the property that the spatial metric $g_{B.G}$ is negative Einstein i.e., $R[g_{B.G}]_{ij}=-\frac{1}{n}(g_{B.G})_{ij}$, the perturbed solutions decay in infinite time to the space which does not share the same property. More specifically, the spatial metric $\gamma^{\dag}$ of the asymptotic solution satisfies $R[\gamma^{\dag}]=-1$. Now, the space of negative Einstein metrics is a subset of the space of metrics with constant negative scalar curvature. Simple use of triangle inequality shows that the asymptotic state should be the space of constant negative scalar curvature sufficiently close to and containing the background FLRW solutions. Since the latter is a subspace of the former, one may fine-tune the initial conditions to obtain the asymptotic solution to be exactly of FLRW type (i.e., the solution has a spatial metric that is negative Einstein). But such a procedure proves to be too restrictive and physically undesirable. 

If for the moment we focus on the physical $3+1$ case, then ideally we want the asymptotic solution to have constant negative sectional curvature (i.e., hyperbolic or negative Einstein by Mostow rigidity) not just constant negative scalar curvature and a spatially uniform background energy density. In other words, if one perturbs an FLRW solution, it does not generically come back to an FLRW solution in infinite time. Now notice that the local spatial homogeneity and isotropy criteria required constancy of the sectional curvature and as such in 3 spatial dimensions, a negative Einstein metric automatically has constant sectional curvature. However, space of metrics with constant negative scalar curvature is a much larger space and does not necessarily exhibit the local spatial homogeneity and isotropy criteria (only a subspace that is described by negative Einstein metrics or hyperbolic metrics do; these correspond to compact hyperbolic manifolds that are locally homogeneous and isotropic by construction). This result is not completely satisfactory but on the other hand, maybe physically expected as these inhomogeneous and anisotropic characteristics may be attributed to the structure formation (see  \cite{lifshitz1992gravitational, peebles2020large} about the role of cosmological perturbation theory in structure formation). This is a remarkable fact that indicates there is a dynamical mechanism at work within the Einstein-Euler-$\Lambda$ flow that naturally drives the physical universe to an anisotropic and inhomogeneous state leading to cosmological structure formation. We do not of course claim such a strong result only based on our linear stability analysis.   

We note that \cite{roy2011global} reported this behavior of the FLRW spacetimes as well considering `dust' as the matter source. They constructed a finite-dimensional dynamical system by spatially averaging out the inhomogeneities and showed that \textit{FLRW cosmologies are unstable in some relevant cases: averaged models are driven away from them through structure formation and accelerated expansion}. However, presently we do not claim that our theorem indicates such a strong result since we need to execute the full nonlinear analysis to incorporate the possibility of shock formation. Nevertheless, the phenomenon of accelerated expansion driving the solutions away from the background is previously noted in \cite{mondal2019attractors} which dealt with a particular case of vacuum gravity (background solutions were described by the negative Einstein spaces) including a positive cosmological constant. The solutions decay asymptotically to the nearby ones described by constant negative scalar curvature. On the other hand, in the same problem if the cosmological constant is turned off, then the perturbed solutions do come back to the background solution asymptotically \cite{andersson2004future, andersson2011einstein}. Nevertheless, the presence of a positive cosmological constant yields universal decay rates that do not depend on the background geometry unlike the zero cosmological constant case, where the decay rate explicitly depended on the spectra of a Lichnerowicz type Laplacian operator defined on the background geometry (see \cite{andersson2011einstein} for detail).

\section{Re-scaled field equations and background solutions}
The background solution we are interested in is the constant negative sectional spatial curvature FLRW model in `$3+1$' dimensions. In higher dimensions ($n>3$), however, the spatial metrics are negative Einstein (not necessarily hyperbolic). The solution is expressible in the following warped product form   
 \begin{eqnarray}
 \label{eq:fixed_space}
 ^{n+1}g&=&-dt\otimes dt+a(t)^{2}\gamma_{ij}dx^{i}\otimes dx^{j},
 \end{eqnarray}
 where $R_{ij}(\gamma)=-\frac{1}{n}\gamma_{ij}$ and $t\in (0,\infty)$. Clearly we have $N=1,~X^{i}=0,~g_{ij}=a(t)^{2}\gamma_{ij}$. These spacetimes are globally foliated by constant mean curvature slices which is evident from the following simple calculations 
 \begin{eqnarray}
 k_{ij}&=&-\frac{1}{2N}\partial_{t}g_{ij}=-a\dot{a}\gamma_{ij},\\\nonumber
 \tau(t)&=&k_{ij}g^{ij}=-a\dot{a}\gamma_{ij}\frac{1}{a(t)^{2}}\gamma^{ij}=-\frac{n\dot{a}}{a}.
 \end{eqnarray}
Note that in order to study the linearized perturbations, we need to re-scale the field equations. We want to re-scale in such a way as to make the background time independent and then analyze the associated re-scaled dynamical system. For this particular purpose, we will need the background density and $n-$velocity. Let us now explicitly calculate the scale factor $a(t)$, the background density, and the velocity. Note that we are only interested in the asymptotic behaviour of the scale factor at $t\to\infty$ for our stability analysis instead of its precise expression. The following lemma states the asymptotic behaviour of the scale factor.\\
 \textbf{Lemma 1:}~\textit{Let $\gamma_{ad}$ be the adiabatic index. The background energy density, $n-$velocity vector field, and the scale factor satisfy the following 
 \begin{eqnarray}
 \rho(t)a(t)^{n\gamma_{ad}}=\nonumber C_{\rho},\\\nonumber
 v^{i}=0,\\\nonumber
 a\sim 
 e^{\alpha t}~~for~~ \Lambda>0,\\\nonumber 
 a\sim t~~as~~t\to \infty~~for~~\Lambda=0,
 \end{eqnarray}
 where $\alpha:=\sqrt{\frac{2\Lambda}{n(n-1)}}$, $a(0):=a(t=0)$, and $\infty>C_{\rho}>0$ is a constant.}
 \\
 \textbf{Proof:} The energy equation yields 
 \begin{eqnarray}
 \frac{\partial\rho}{\partial t}&=&-n\gamma_{ad}\frac{\dot{a}}{a}\rho,
 \end{eqnarray} 
 integration of which leads to
 \begin{eqnarray}
\rho a(t)^{n\gamma_{ad}}=C_{\rho},
\end{eqnarray}
for some finite positive constant $C_{\rho}$. The momentum constraint (\ref{eq:MC}) equation yields 
\begin{eqnarray}
v^{i}=0
\end{eqnarray}
since $k^{tr}=0$.
Utilizing the Hamiltonian constraint, we obtain the following ODE for the scale factor
\begin{eqnarray}
\label{eq:backint}
\frac{1}{a}\frac{da}{dt}=\sqrt{\frac{1}{n(n-1)}\left(2\Lambda+\frac{1}{a^{2}}+\frac{2C_{\rho}}{a^{n\gamma_{ad}}}\right)},
\end{eqnarray}
integration of which yields at $t\to \infty$
\begin{eqnarray}
a\sim e^{\alpha t}
\end{eqnarray}
where $\alpha:=\sqrt{\frac{2\Lambda}{n(n-1)}}$. Taking $\Lambda= 0$, integration of (\ref{eq:backint}) 
yields
\begin{eqnarray}
\label{eq:asymp}
a\sim a(0)+\frac{t}{\sqrt{n(n-1)}}
\end{eqnarray}
as $t\to\infty$, which completes the proof of the lemma.

Now let us focus on obtaining a re-scaled system of evolution and constraint equations. Recall that the metric in local coordinate $(t,x)$ may be written as 
 \begin{eqnarray}
 \hat{g}&=&-N^{2}dt\otimes dt+g_{ij}(dx^{i}+X^{i}dt)\otimes(dx^{j}+X^{j}dt)
 \end{eqnarray}
 and the background solutions are written as 
 \begin{eqnarray}
 ^{n+1}g_{background}=-dt\otimes dt+a^{2}\gamma_{ij}dx^{i}\otimes dx^{j},
 \end{eqnarray}
where $R[\gamma]_{ij}=-\frac{1}{n}\gamma_{ij}$ and $t\in (0,\infty)$. If we assign $a$ the dimension of length (natural) (similar to the re-scaling used by \cite{andersson2011einstein}), the dimensions of the rest of the entities follow 
$t\sim length,~x^{i}\sim (length)^{0},~g_{ij}\sim (length)^{2},~N\sim (length)^{0},~X^{i}\sim (length)^{-1}, ~\tau \sim(length)^{-1}$,~$\kappa^{tr}_{ij}\sim length$. Therefore, the following scaling follows naturally 
\begin{eqnarray}
\tilde{g}_{ij}=a^{2} g_{ij},
\tilde{N}=N,
\tilde{X}^{i}=\frac{1}{a}X^{i},
\tilde{k^{tr}}_{ij}=ak^{tr}_{ij}.
\end{eqnarray}
Here, we denote the dimension-full entities with a tilde sign and dimensionless entities are left alone for simplicity. The Hamiltonian constraint (\ref{eq:HC}) and the momentum constraint (\ref{eq:MC}) take the following forms 
 \begin{eqnarray}
 R(g)-|k^{tr}|^{2}+1+2C_{\rho}a^{2-n\gamma}&=&2a(t)^{2}\tilde{\rho}\left(1+\gamma_{ad} \tilde{g}(\tilde{v},\tilde{v})\right),\\\nonumber
 \nabla_{j}k^{trij}&=&a(t)^{3}\gamma_{ad}\tilde{\rho}\sqrt{1+\tilde{g}(\tilde{v},\tilde{v})}\tilde{v}^{i}.
 \end{eqnarray}
 Notice that we are yet to re-scale the matter fields i.e., $(\rho,v)$. Now, of course, in lemma 1, we have already seen that the background energy density satisfies $\rho(t)=\frac{C_{\rho}}{a^{n\gamma_{ad}}}$ for some constant $C_{\rho}>0$ depending on the initial density. Therefore, we naturally scale the energy density by $\frac{1}{a^{n\gamma_{ad}}}$ i.e.,
 \begin{eqnarray}
 \tilde{\rho}&=&\frac{1}{a(t)^{n\gamma_{ad}}}\rho
 \end{eqnarray}
 so that the background density becomes time independent. Notice that $C_{\rho}$ is a constant with dimension of $(length)^{n\gamma_{ad}-2}$.
now, since, the background $n-$velocity vanishes for the background solution, we do not have a straightforward scaling. However, we have the normalization condition for the $n+1-$velocity field  $\textbf{u}=-^{n+1}g(u,n)n+v$
 \begin{eqnarray}
 -(\tilde{N}\tilde{\textbf{u}}^{0})^{2}+\tilde{g}_{ij}\tilde{v}^{i}\tilde{v}^{j}&=&-1,
 \end{eqnarray}
 and therefore scaling of the metric $\tilde{g}_{ij}=a^{2}g_{ij}$ yields the following natural scaling of $v^{i}$
 \begin{eqnarray}
 \tilde{v}^{i}&=&\frac{1}{a}v^{i},
 \end{eqnarray}
 since the right hand side is simply a constant. 
 The complete evolution and constraints of the gravity coupled fluid system may be expressed in the following re-scaled form 
 \begin{eqnarray}
 \label{eq:fixed_new}
 \frac{\partial g_{ij}}{\partial t}&=&2\frac{\dot{a}}{a}\left(N-1\right)g_{ij}-\frac{2}{a}Nk^{tr}_{ij}+\frac{1}{a}(L_{X}g)_{ij},
 \end{eqnarray}
 \begin{eqnarray}
\frac{\partial k^{tr}_{ij}}{\partial t}&=&-\underbrace{\frac{\dot{a}}{a}\left((N-1)(n-2)+n-1\right)}_{should~be~<0~to~generate~decay}k^{tr}_{ij}-\frac{2}{a}Nk^{tr}_{ik}k^{trk}_{j} +\underbrace{\frac{1}{a}NR_{ij}}_{gauge~fixing~is ~required}\\\nonumber 
 &&-\left(\frac{a}{n^{2}}(\tau^{2}-\frac{2n\Lambda}{n-1})+\frac{n\gamma_{ad}-2}{n(n-1)}a^{1-n\gamma_{ad}}C_{\rho}-\frac{N(1+2C_{\rho}a^{2-n\gamma})}{a(n-1)}\right.\\\nonumber
 &&\left.+\frac{\gamma_{ad}-2}{n-1}a^{1-n\gamma_{ad}}N\rho g_{ij}\right)g_{ij}-\nonumber \frac{1}{a}\nabla_{i}\nabla_{j}N-a^{1-n\gamma_{ad}}N\gamma\rho v_{i}v_{j} +\frac{1}{a}(L_{X}k^{tr})_{ij},
 \end{eqnarray}

\begin{eqnarray}
\partial_{t}\rho+\frac{\gamma_{ad}\rho N\nabla_{i}v^{i}}{a[1+g(v,v)]^{1/2}}+\underbrace{\frac{(N-1)n\gamma_{ad}\rho\dot{a}}{a}}_{should~contribute~to~higher~order~since ~N-1\leq 0}\\\nonumber=\frac{1}{a}L_{X}\rho-\frac{\gamma_{ad}\rho g(\partial_{t}v,v)}{1+g(v,v)}-\frac{NL_{v}\rho}{a[1+g(v,v)]^{1/2}}+\frac{\gamma_{ad}\rho N}{1+g(v,v)}\left(\frac{1}{a}k^{tr}(v,v)-\frac{\dot{a}}{a}g(v,v)\right)\\\nonumber+\frac{\gamma_{ad}\rho v_{i}L_{X}v^{i}}{a[1+g(v,v)]}-\frac{\gamma_{ad}N\rho L_{v}N}{a[1+g(v,v)]^{1/2}},
 \end{eqnarray}
 \begin{eqnarray}
 \label{eq:eomf}
 \gamma_{ad}\rho u^{0}\partial_{t}v^{i}+\underbrace{\gamma_{ad}\rho u^{0}[2N-1-n(\gamma_{ad}-1)]\frac{\dot{a}}{a}}_{should~be~>0~to~generate~decay}v^{i}-\frac{2\gamma_{ad}\rho N u^{0}k^{tri}_{j}v^{j}}{a}\nonumber-\frac{\gamma_{ad}\rho u^{0}}{a}L_{X}v^{i}\\\nonumber
+\frac{\gamma_{ad}\rho}{a}v^{j}\nabla_{j}v^{i}+\frac{\gamma_{ad}\rho[1+g(v,v)]}{aN}\nabla^{i}N+\frac{(\gamma_{ad}-1)}{a}\left(\nabla^{i}\rho+v^{i}L_{v}\rho+au^{0}v^{i}\partial_{t}\rho\right.\\\nonumber
\left.-u^{0}v^{i}L_{X}\rho\right)=0,
 \end{eqnarray}
 \begin{eqnarray}
 \label{eq:hamilton}
 R(g)-|k^{tr}|^{2}-2a^{2-n\gamma_{ad}}\rho\{1+\gamma_{ad} g(v,v)\}+1+2C_{\rho}a^{2-n\gamma}=0,\\
 \label{eq:mc}
\nabla_{j}k^{trij}=-\gamma_{ad} a^{2-n\gamma_{ad}}\rho\sqrt{1+g(v,v)} v^{i},
\end{eqnarray}
 \begin{eqnarray}
 \label{eq:lapse1}
 \Delta_{g}N+\left(\underbrace{|k^{tr}|^{2}+a(t)^{2}(\frac{\tau^{2}}{n}-\frac{2\Lambda}{n-1})+a^{2-n\gamma_{ad}}[\frac{(n\gamma_{ad}-2)}{n-1}+\gamma_{ad}\nonumber g(v,v)]\rho}_{should~be~>0~for~ elliptic~regularity}\right)N\\
 =a^{2}\frac{\partial\tau}{\partial t},
 \end{eqnarray}
 \begin{eqnarray}
 \label{eq:shiftin}
\Delta_{g}X^{i}-R^{i}_{j}X^{j}=(1-\frac{2}{n})\tau \nabla^{i}N-2\nabla^{j}Nk^{tri}_{j}-2N\nabla_{j}k^{trij}\\\nonumber
+(2Nk^{trjk}-2\nabla^{j}X^{k})\left(\Gamma[g]^{i}_{jk}-\Gamma[\gamma]^{i}_{jk}\right)-ag^{ij}\partial_{t}\Gamma[\gamma]_{ij}^{k}.
\end{eqnarray}
Note that an straightforward maximum principle argument applied to the lapse equation yields $N\leq 1$.
In CMCSH gauge, the time function is obtained by setting $\tau=$monotonic function of $t$ alone. Now, we have $\tau$ evaluated for the background solution which is a monotonic function of $t$ alone since utilizing the Hamiltonian constraint and the lapse equation at the background solution one obtains 
\begin{eqnarray}
\frac{\partial \tau}{\partial t}&=&\frac{1}{n}(\tau^{2}-\frac{2n\Lambda}{n-1})+\frac{n\gamma_{ad}-2}{n-1}a^{-n\gamma_{ad}}C_{\rho}\\\nonumber 
&=&\frac{1}{a(t)^{2}(n-1)}\left(1+n\gamma_{ad} C_{\rho}a^{2-n\gamma_{ad}}\right)>0.
\end{eqnarray}
We set the time function to be the solution of this equation
\begin{eqnarray}
\frac{\partial \tau}{\partial t}=\frac{1}{a(t)^{2}(n-1)}\left(1+n\gamma_{ad} C_{\rho}a(t)^{2-n\gamma_{ad}}\right),
\end{eqnarray}
 i.e., $t=t(\tau)$ which settles the business of choosing a time coordinate. Let us analyze the background solutions in a bit more detail.\\
 \textbf{Lemma 2(a):}~\textit{The background re-scaled solution of the Einstein-Euler system verifies the following equations
 \begin{eqnarray}
 N_{B.G}=1,X^{i}_{B.G}=0,R[g_{B.G}]_{ij}=-\frac{1}{n}g_{B.Gij}, k^{tr}_{B.G}=0, v^{i}_{B.G}=0, \partial_{i}\rho_{B.G}=0,
 \end{eqnarray}
 and the spacetime is foliated by CMC slices.
 }\\
 \textbf{Proof:} The background spacetime  
 \begin{eqnarray}
 ^{n+1}g&=&-dt\otimes dt+a(t)^{2}\gamma_{ij}dx^{i}\otimes dx^{j}
 \end{eqnarray}
 yields $\tilde{N}_{B.G}=1,~\tilde{X}^{i}_{B.G}=0,~\tilde{g}_{B.Gij}=a^{2}\gamma_{ij}$ and therefore yields $N_{B.G}=1,~X^{i}_{B.G}=0,~g_{B.Gij}=\gamma_{ij}$. A simple calculation yields 
 \begin{eqnarray}
 \tilde{k}_{B.Gij}=-a\dot{a}\gamma_{ij}
 \end{eqnarray}
 and therefore 
 \begin{eqnarray}
 k^{tr}_{B.G}=0,~\tau=-\frac{n\dot{a}}{a}.
 \end{eqnarray}
The momentum constraint (\ref{eq:mc}) yields
 \begin{eqnarray}
 v^{i}_{B.G}=0
 \end{eqnarray}
 which together with the equations of motion (\ref{eq:eomf}) for the fluid leads to 
 \begin{eqnarray}
 \partial_{i}\rho_{B.G}=0.
 \end{eqnarray}
 From the previous calculation $\tau=-\frac{\dot{a}}{a}$, it is obvious that the spacetime is foliated by constant mean curvature spatial slices.\\
 \textbf{Lemma 2(b):}~\textit{The transverse-traceless part of the traceless second fundamental form vanishes for the background solutions i.e., $k^{tr}_{B.G}=k^{TT}+L_{Y}g_{B.G}-\frac{2}{n}(\nabla[g_{B.G}]_{m}Y^{m})g_{B.G}$ reduces to $k^{tr}=k^{TT}=0$ and moreover $Y\equiv0$ for background solutions.}\\
 \textbf{Proof:}
The vanishing of the re-scaled trace-less second fundamental form $k^{tr}=0$ for the background solutions yields the following through the decomposition of the symmetric trace-less $2-$ tensors
 \begin{eqnarray}
 k^{TT}+L_{Y}\gamma-\frac{2}{n}(\nabla[\gamma]_{m}Y^{m})\gamma=0,
 \end{eqnarray}
 which may be converted to 
 \begin{eqnarray}
 \int_{M}(k^{TT}+L_{Y}\gamma-\frac{2}{n}\nabla[\gamma]_{m}Y^{m}\gamma)_{ij}(k^{TT}+L_{Y}\gamma-\frac{2}{n}\nabla[\gamma]_{m}Y^{m}\gamma)^{ij}\mu_{\gamma}&=&0\\\nonumber
 \int_{M}|k^{TT}|^{2}_{\gamma}\mu_{\gamma}+\int_{M}|L_{Y}\gamma-\frac{2}{n}\nabla[\gamma]_{m}Y^{m}\gamma|^{2}_{\gamma}\mu_{\gamma}&=&0
 \end{eqnarray}
 since $k^{TT}$ and $L_{Y}\gamma-\frac{2}{n}\nabla[\gamma]_{m}Y^{m}\gamma$ are $L^{2}$ orthogonal with respect to the metric $\gamma$.
 Therefore we readily obtain
 \begin{eqnarray}
 k^{TT}\equiv0,\\
 \label{eq:killing}
 L_{Y}\gamma-\frac{2}{n}\nabla[\gamma]_{m}Y^{m}\gamma\equiv0,
 \end{eqnarray}
 where we have used the fact that $k^{TT}$ and $L_{Y}\gamma-\frac{2}{n}\nabla[\gamma]_{m}Y^{m}$ are $L^{2}$ orthogonal (with respect to the background metric $\gamma$). Now the covariant divergence of the conformal Killing equation (\ref{eq:killing}) yields 
 \begin{eqnarray}
 \nabla[\gamma]^{i}\left(\nabla[\gamma]_{i}Y_{j}+\nabla[\gamma]_{j}Y_{i}-\frac{2}{n}(\nabla[\gamma]_{m}Y^{m})\gamma_{ij}\right)&=&0\\\nonumber
 \int_{M}\left(Y_{j}\nabla[\gamma]^{i}\nabla[\gamma]_{i}Y^{j}+R^{j}_{i}Y^{i}Y_{j}+(1-\frac{2}{n})Y_{j}\nabla[\gamma]^{j}(\nabla[\gamma]_{i}Y^{i})\right)\mu_{\gamma}&=&0\\\nonumber 
 -\int_{M}\left(\nabla[\gamma]_{i}Y_{j}\nabla[\gamma]^{i}Y^{j}+\frac{1}{n}\gamma_{ij}Y^{i}Y^{j}+(1-\frac{2}{n})|\nabla[\gamma]_{i}Y^{i}|^{2}\right)\mu_{\gamma}&=&0\\\nonumber 
 \Rightarrow Y\equiv 0
 \end{eqnarray}
 throughout $M$ since $n\geq 3$. Here we have used  $R(\gamma)_{ij}=-\frac{1}{n}\gamma_{ij}$ together with integration by parts and the Stokes' theorem for a closed (compact with $\partial M=\{0\}$) manifold. This is equivalent to the fact that the isometry group of $(M,\gamma)$ is discrete. The spacetimes (\ref{eq:background}) admit a timelike conformal Killing field $K:=K^{\mu}\frac{\partial}{\partial x^{\mu}}=a(t)\frac{\partial}{\partial t}$ i.e.,
 \begin{eqnarray}
 L_{a(t)\frac{\partial}{\partial t}}~^{n+1}g=2\dot{a}~^{n+1}g.
 \end{eqnarray}
 The results obtained so far yields the following theorem characterizing a class of \textit{fixed points} of the \textit{re-scaled} Einstein-Euler-$\Lambda$ system.
 
 \textbf{Theorem:} \textit{Let $M$ be a closed (compact without boundary) connected orientable manifold that admits a negative Einstein metric. Then the fixed point solutions of the re-scaled Einstein-Euler-$\Lambda$ flow (\ref{eq:fixed_new}-\ref{eq:shiftin}) in constant mean extrinsic curvature spatial harmonic gauge (CMCSH) on $(t_{-},t_{+})\times M,~0\leq t_{-}<t_{+}< \infty$ have the Cauchy data $(g_{B.G},k^{tr}_{B.G},N_{B.G},X_{B.G},\rho_{B.G},v_{B.G})$ that satisfy the following equations
 \begin{eqnarray}
 R_{ij}[g_{B.G}]=-\frac{1}{n}(g_{B.G})_{ij},~k^{tr}_{B.G}=0,~N_{B.G}=1,~X_{B.G}\nonumber=0,~\rho_{B.G}=C_{\rho},\\ 
 \label{eq:backgroundsolution}
 ~v^{i}_{B.G}=0.
 \end{eqnarray}
 The physical spacetimes (\ref{eq:fixed_space}) constructed by this Cauchy data admit $K:=a(t)\partial_{t}$ as a globally defined time-like conformal Killing vector field. Here $a(t)$ is the scale factor. 
} 
 
\subsection{\textbf{Center Manifold of the dynamics}}
The following mathematics concerning the center manifold dynamics in the setting of Einstein flow was first established by the work of Andersson-Moncrief \cite{andersson2004future}. We now present some of this work and adapt it to the Einstein-Euler-Lambda setting. A center manifold of a dynamical system is associated with fixed points of the dynamics. Essentially a center manifold of a fixed point corresponds to the nearby solutions (in phase space) that do not exhibit exponential growth or decay. This may be zero-dimensional or a subspace of the phase space and admits a manifold structure (so the name center `manifold'). For a more precise and mathematically rigorous definition, we refer the reader to \cite{marsden2012hopf, lanford1973bifurcation}. There is a crucial theorem namely the `center manifold theorem' which plays a significant role in the analysis of dynamical systems. In general, center manifolds do not have the uniqueness property, unlike stable and unstable manifolds which do \cite{kelley1967stable, takens1974singularities, osipenko2011center}. In a finite-dimensional setting, this roughly corresponds to the case when the linearization of the flow vector field has purely imaginary spectra. In an infinite-dimensional setting, moduli spaces naturally play the role of center manifold (in an appropriate sense of course). As we shall see, the center manifold in this particular occasion is played by the Einstein moduli space, which consists of non-isolated fixed points of the Einstein-Euler-$\Lambda$ flow. 

Unlike the $n=3$ dimensional case where the negative Einstein spaces are hyperbolic (and the center manifold of the gravitational dynamics consists of a point), the higher dimensional case is more interesting. The matter degrees of freedom corresponding to the background solution (in its re-scaled version) in CMCSH gauge are essentially described by $v^{i}=0, \rho=\rho_{B.G}$. It becomes more interesting when we consider the gravitational degrees of freedom. In the case of $n=3$, a negative Einstein structure implies hyperbolic structure through Mostow rigidity theorem and therefore the manifold describing the fixed point of the dynamics is zero-dimensional. In other words, the particular solution $(\{\gamma|Ricci[\gamma]=-\frac{1}{3}\gamma, k^{tr}=0, \rho=C_{\rho}, v^{i}=0)$  serves as an isolated fixed point of the Einstein-Euler dynamics. However, in the higher dimensional cases ($n>3$), a new possibility arises of having non-hyperbolic negative Einstein spaces. When we linearize about any member of such a non-isolated family of negative Einstein metrics, the linearized equations will always admit a finite-dimensional space of neutral modes. Naturally, this represents (modulo the matter degrees of freedom) the tangent space to the background spacetimes (\ref{eq:background}). These smooth families of background spacetimes determined by the corresponding families of negative Einstein metrics and zero-dimensional matter degrees of freedom ($v^{i}=0, \rho=C_{\rho}$) form the `center manifold' for the dynamical system defined by the re-scaled Einstein-Euler-$\Lambda$ equations. A family of background solutions of the Einstein-Euler-$\Lambda$ system in CMCSH gauge is the spacetimes as described in the previous section. The spatial metric component of these spacetimes is a negative Einstein metric i.e., the spatial metric satisfies 
\begin{eqnarray}
\label{eq:ein}
R_{ij}(\gamma)=-\frac{1}{n}\gamma_{ij}.
\end{eqnarray}
Let us denote the space of metrics satisfying equation (\ref{eq:ein}) by $\mathcal{E}in_{-\frac{1}{n}}$

Let $\gamma^{*}\in \mathcal{E}in_{-\frac{1}{n}}$ and $\mathcal{V}$ be its connected component. Also consider $\mathcal{S}_{\gamma}$ to be the harmonic slice of the identity diffeomorphism i.e., the set of $\gamma\in\mathcal{E}in_{-\frac{1}{n}}$ for which the identity map $id: (M,\gamma)\to (M,\gamma^{*})$ is harmonic (since any metric $\gamma\in \mathcal{E}in_{-\frac{1}{n}}$ verifies the fixed point criteria $R[\gamma]_{ij}=-\frac{1}{n}\gamma_{ij}$, it should also satisfy the CMCSH gauge condition with respect to a background metric and in this case the background metric is simply chosen to be $\gamma^{*}$, another element of $\mathcal{E}in_{-\frac{1}{n}}$). This condition is equivalent to the vanishing of the tension field $-V^{k}$ that is
\begin{eqnarray}
\label{eq:harmonic}
-V^{k}=-\gamma^{ij}(\Gamma[\gamma]^{k}_{ij}-\Gamma[\gamma^{*}]^{k}_{ij})=0.
\end{eqnarray} 
For $\gamma\in \mathcal{E}in_{-\frac{1}{n}}$, $\mathcal{S}_{\gamma}$ is a submanifold of $\mathcal{M}$ for $\gamma$ sufficiently close to $\gamma^{*}$ (easily proven using standard procedure and therefore we omit the proof, see \cite{andersson2011einstein} for detail). The deformation space $\mathcal{N}$ of $\gamma^{*}\in \mathcal{E}in_{-\frac{1}{n}}$ is defined as the intersection of the $\gamma^{*}-$connected component $ \mathcal{V}\subset \mathcal{E}in_{-\frac{1}{n}}$ and the harmonic slice $\mathcal{S}_{\gamma}$ i.e.,
\begin{eqnarray}
\label{eq:ds}
\mathcal{N}:= \mathcal{V}\cap \mathcal{S}_{\gamma}.
\end{eqnarray}
$\mathcal{N}$ is assumed to be smooth (i.e., equipped with $C^{\infty}$ topology). 
In the case of $n=3$, following the Mostow rigidity theorem, the negative Einstein structure is rigid \cite{lebrun1994einstein}. This rigid Einstein structure in $n=3$ corresponds to the hyperbolic structure up to isometry. For higher genus Riemann surfaces $\Sigma_{genus}$ ($genus>1$), space of distinct hyperbolic structures is the classical Teichm\"uller space diffeomorphic to $\mathbf{R}^{6genus-6}$. Tangent space to any point in the Teichm\"uller space corresponds to the `neutral modes' of $2+1$ gravity defined by transverse-traceless tensors. However, $2+1$ gravity is fundamentally different from the higher dimensional cases since the former is devoid of gravitational waves degrees of freedom. For $n>3$, the centre manifold $\mathcal{N}$ is a finite dimensional submanifold of $\mathcal{M}$ (and in particular of $\mathcal{M}_{-1}$). 
Following the analysis of \cite{andersson2011einstein}, the formal tangent space $T_{\gamma}\mathcal{N}$ in local coordinates is expressible as 
\begin{eqnarray}
\label{eq:tangentspace}
\frac{\partial \gamma}{\partial q^{a}}&=&l^{TT||}_{a}+L_{W^{||}_{a}}\gamma,
\end{eqnarray}
where $l^{TT||}_{a}\in \ker(\mathcal{L})=C^{TT||}(S^{2}M)\subset C^{TT}(S^{2}M)$, $W^{||}\in \mathfrak{X}(M)$ satisfies
through the time derivative of the CMCSH gauge condition $-\gamma^{ij}(\Gamma[\gamma]^{k}_{ij}-\Gamma[\gamma^{*}]^{k}_{ij})=0$
\begin{eqnarray}
-[\nabla[\gamma]^{m}\nabla[\gamma]_{m}W^{||i}+R[\gamma]^{i}_{m}W^{||m}]+(l^{TT||}\nonumber+L_{W^{||}}\gamma)^{mn}(\Gamma[\gamma]^{k}_{mn}-\Gamma[\gamma^{*}]^{k}_{mn})\\\nonumber
=0,\nonumber 
\end{eqnarray} 
and, $\{q^{a}\}_{a=1}^{dim(\mathcal{N})}$ is a local chart on $\mathcal{N}$, $\mathfrak{X}(M)$ is the space of vector fields on $M$ (in a suitable function space setting), $C^{TT}(S^{2}M)$ is the space of $\gamma-$fransverse-traceless $2-$tensors on $M$. An important thing to note is that all known examples of closed negative Einstein spaces have integrable deformation spaces and that the deformation spaces are stable i.e., $Spec\{\mathcal{L}_{\gamma,\gamma}\}\geq 0$. The spectrum of the operator $\mathcal{L}_{\gamma,\gamma}$ plays an important role determining the decay rates in the pure vacuum case considered by \cite{andersson2004future, andersson2011einstein}. 
\begin{center}
\begin{figure}
\begin{center}
\includegraphics[width=13cm,height=60cm,keepaspectratio,keepaspectratio]{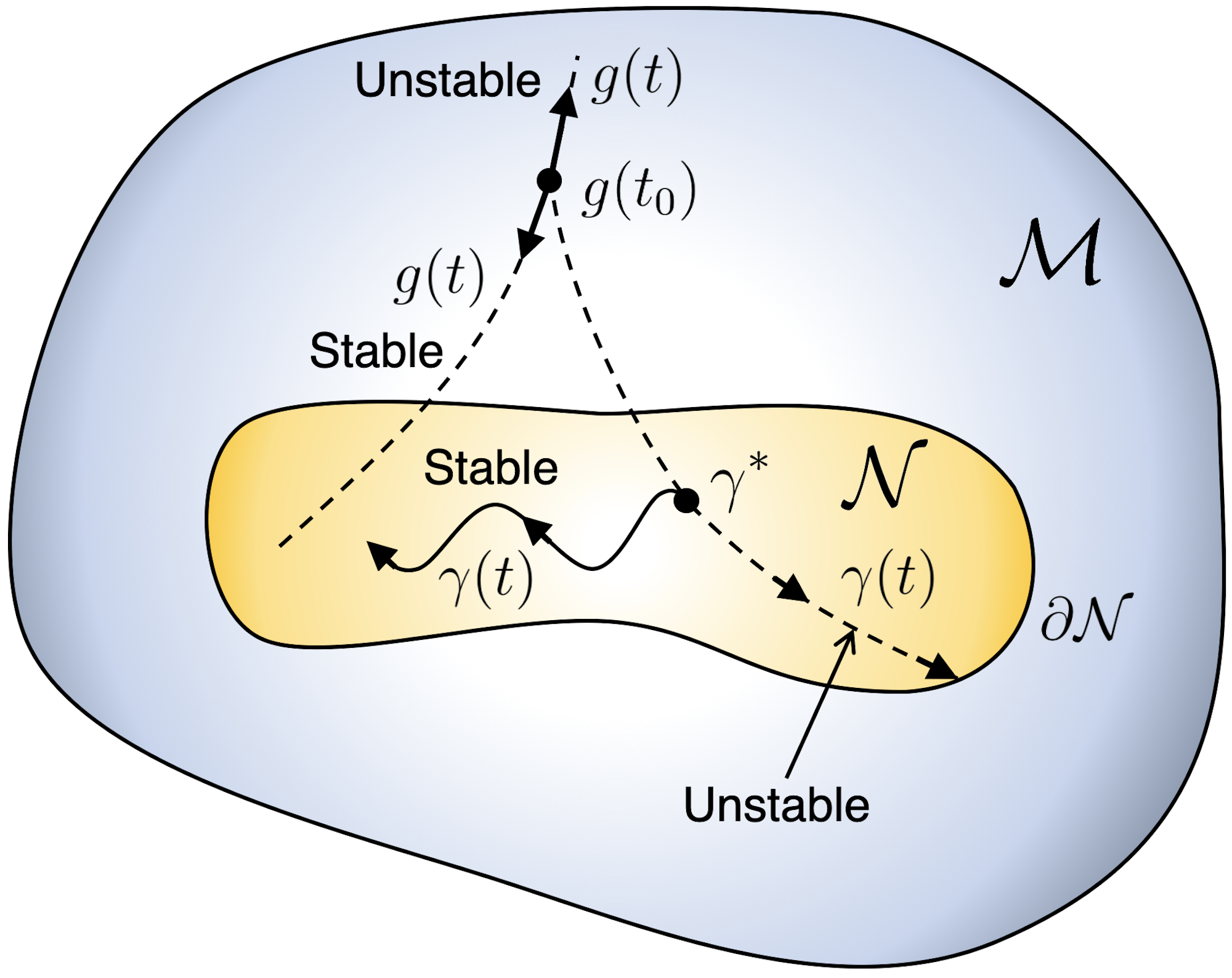}
\end{center}
\begin{center}
\caption{Kinematics of the perturbations for the Einstein-Euler system. Instability materializes in $\mathcal{N}$ iff $\gamma(t)$ leaves every compact set of $\mathcal{N}$ i.e., approaches $\partial\mathcal{N}$. Stability occurs when $\gamma(t)$ evolves from one point of $\mathcal{N}$ to simply another point of $\mathcal{N}$. The instability of $g(t)$ is naturally defined by the unbounded growth of $g(t)-\gamma(t)$ in a suitable function space.}
\label{fig:pdf}
\end{center}
\end{figure}
\end{center}

\subsection{\textbf{Kinematics of the perturbations at the linear level}}
As discussed in the previous section, the Einstein moduli space is finite-dimensional for $n>3$ and serves as the centre manifold of the dynamics (the deformation space $\mathcal{N}$ to be precise). On the other hand, for the $n=3$ case, following Mostow rigidity, negative Einstein structure implies hyperbolicity and therefore the moduli space reduces to a point. We will consider the general case when the moduli space is finite-dimensional since the zero-dimensional moduli space case is trivial. Let $\mathcal{N}$ be the deformation space of $\gamma^{*}$. Now suppose, we perturb the metric about $\gamma^{*}$. The perturbation is defined to be $h:=g-\gamma^{*}$ such that $||h||_{H^{s}}<\delta$ for a suitable $\delta>0$. Due to the finite dimensionality of $\mathcal{N}$, $g$ in fact could be an element of $\mathcal{N}$ lying within a $\delta-$ball of $\gamma^{*}$. But if $g$ lies on $\mathcal{N}$, then corresponding solution satisfies the fixed point condition $Ric(g)=-\frac{1}{n}g$. Therefore these `trivial' perturbations do not see the dynamics. The question is how to treat such perturbations which are tangent to $\mathcal{N}$. This was precisely handled by invoking a shadow gauge condition in the study of a small data fully nonlinear stability problem of the vacuum Einstein' equations by \cite{andersson2011einstein} and later used in \cite{mondal2019attractors} to handle the case with a positive cosmological constant. The shadow gauge reads    
\begin{eqnarray}
(g-\gamma)\perp \mathcal{N}
\end{eqnarray}
for $\gamma\in\mathcal{N}$, where the orthogonality is in the $L^{2}$ sense with respect to the metric $\gamma$. This precisely takes care of the motion of the orthogonal perturbations by studying the time evolution of $h:=g-\gamma$, while the motion of the tangential perturbations is taken care of by evolving $\gamma\in\mathcal{N}$. This does not really affect the characteristics of the background spacetimes as $\gamma$ does not leave the moduli space i.e., it is still a negative Einstein metric with Einstein constant $-\frac{1}{n}$. However, if we write the perturbations as $h_{ij}=g_{ij}-\gamma_{ij}$, then $\frac{\partial g_{ij}}{\partial t}=\frac{\partial h_{ij}}{\partial t}+\frac{\partial \gamma_{ij}}{\partial t}$ and therefore, $\frac{\partial \gamma_{ij}}{\partial t}$ enters into the evolution equation for the metric. But an important question is how exactly to control this term since it will enter into the time derivative of the energy as well. We will obtain the necessary estimates in a later section when we study the energy inequality. But how does the instability (if it occurs) materialize in $\mathcal{N}$? The instability is precisely described by the fact that the curve $\gamma(t)$ runs off the edge of $\mathcal{N}$ (leaves every compact set of $\mathcal{N}$) similar to the case of $2+1$ gravity where the solution curve runs off the edge of the Teichm\"uller space (played by $\mathcal{N}$ in the current context) at the limit of the big-bang singularity \cite{moncrief2007relativistic, mondal2020thurston}. See figure (1) which depicts the kinematics of the perturbations (notice that this notion of stability/ instability is general; here we specialize to the linearized problem). This notion of stability/instability is true in a general non-linear setting as well. Linearization is a special case which we will consider here.

One subtle point left to discuss is how to control the time derivative term `$g^{ij}\partial_{t}\Gamma[\gamma]_{ij}^{k}$' (or `$\gamma^{ij}\partial_{t}\Gamma[\gamma]_{ij}^{k}$' at the linear level) in the equation for the shift vector field (\ref{eq:shiftin}). By a simple calculation, we will show that this contribution vanishes at the linear level. $\frac{\partial \gamma_{ij}}{\partial t}$ may be written as (equation (\ref{eq:tangentspace})) 
\begin{eqnarray}
\frac{\partial \gamma_{ij}}{\partial t}=l^{TT||}_{ij}+(L_{Z^{||}}\gamma)_{ij},
\end{eqnarray}
where $Z^{||}$ satisfies through the CMCSH gauge condition (\ref{eq:sh}), 
\begin{eqnarray}
-[\nabla[\gamma]^{m}\nabla[\gamma]_{m}Z^{||i}+R[\gamma]^{i}_{m}Z^{||m}]+(l^{TT||}\nonumber+L_{Z^{||}}\gamma)^{mn}(\Gamma[\gamma]^{k}_{mn}-\Gamma[\gamma^{*}]^{k}_{mn})=0.
\end{eqnarray}
At the linear level, however, the term `$(l^{TT||}+L_{Z^{||}}\gamma)^{mn}(\Gamma[\gamma]^{k}_{mn}-\Gamma[\gamma^{*}]^{k}_{mn})$' vanishes since it is of second order. This yields 
\begin{eqnarray}
\nabla[\gamma]^{m}\nabla[\gamma]_{m}Z^{||i}+R[\gamma]^{i}_{m}Z^{||m}=0
\end{eqnarray}
leading to 
\begin{eqnarray}
-\int_{M}(\nabla[\gamma]_{i}Z^{||}_{j}\nabla[\gamma]^{i}Z^{||j}+\frac{1}{n}\gamma_{ij}Z^{||i}Z^{||j})\mu_{\gamma}=0\Rightarrow Z^{||}=0
\end{eqnarray}
on M.
Therefore, the pure gauge part of the $\mathcal{N}-$tangential velocity $\frac{\partial\gamma_{ij}}{\partial t}$ vanishes yielding
\begin{eqnarray}
\frac{\partial \gamma_{ij}}{\partial t}=l^{TT||}_{ij}.
\end{eqnarray} 
A direct calculation yields the following 
\begin{eqnarray}
\gamma^{ij}\partial_{t}\Gamma[\gamma]^{k}_{ij}&=&\gamma^{ij}D\Gamma[\gamma]^{k}_{ij}\cdot \frac{\partial\gamma}{\partial t}\\\nonumber 
&=&\gamma^{ki}\nabla[\gamma]^{j}(\partial_{t}\gamma_{ij})-\frac{1}{2}\nabla[\gamma]^{k}tr_{\gamma}(\partial_{t}\gamma)\\\nonumber 
&=&\gamma^{ki}\nabla[\gamma]^{j}l^{TT||}_{ij}-\frac{1}{2}\nabla[\gamma]^{k}tr_{\gamma}(l^{TT||})\\\nonumber 
&=&0,
\end{eqnarray}
since $l^{TT||}$ is transeverse-traceless with respect to $\gamma$. We are left with one task of estimating the tangential velocity $l^{TT||}$, which will be done in a later chapter.

\section{Linearized Einstein-Euler system}
In this section we linearize the complete Einstein-Euler system and state a local existence theorem for such system. The background (re-scaled) solution obtained previously reads 
$(g_{B.G},k^{tr}_{B.G},N_{B.G}, X_{B.G}, \rho_{B.G},v_{B.G})=(\gamma,0,1,0,C_{\rho},0)$, where $R_{ij}(\gamma)=-\frac{1}{n}$ and $C_{\rho}$ is a positive constant. Of course, in the previous section, we have recalled that the Einstein structure is not rigid for the $n>3$ cases. The perturbations to the background solutions may be described as follows $h:=g-\gamma,~ \delta k^{tr}=k^{tr}-k^{tr}_{B.G}=k^{tr}-0=k^{tr},~\delta N=N-N_{B.G}=N-1, \delta X:=X-X_{B.G}=X-0=X,~ \delta\rho:=\rho-\rho_{B.G}=\rho-C_{\rho},~\delta v:=v-v_{B.G}=v-0=v$. Now we linearize the Einstein-Euler system about this background solution. The linearized Euler equations read 
\begin{eqnarray}
 \frac{\partial\delta\rho}{\partial t}=-\frac{C_{\rho}n\gamma_{ad}\dot{a}}{a}\delta N-\frac{1}{a}\gamma_{ad}C_{\rho}\nabla[\gamma]_{i}v^{i},\\
 \label{eq:velocityeqn}
 \frac{\partial v^{i}}{\partial t}=\frac{n\dot{a}}{a}(\gamma_{ad}-\frac{n+1}{n})v^{i}-\frac{1}{a}\nabla[\gamma]^{i}\delta N-\frac{\gamma_{ad}-1}{a\gamma_{ad} C_{\rho}}\nabla[\gamma]^{i}\delta\rho.
\end{eqnarray}
Now, an obvious problem is that the $n-$velocity vector field $v^{i}$ consists of irrotational, rotational and harmonic parts. Utilizing a Hodge decomposition \cite{warner2013foundations} of the $n-$velocity $1-$form on the spatial manifold $M$, one may simply write the velocity vector field as follows 
 \begin{eqnarray}
 v=(d\phi)^{\sharp}+(\delta\psi)^{\sharp}+(\mathcal{H})^{\sharp},
 \end{eqnarray}
 where $\phi\in \Omega^{0}(M), \psi\in \Omega^{2}(M), \mathcal{H}\in\Omega^{1}(M)$, and $\Delta\mathcal{H}=(d\delta+\delta d)\mathcal{H}=0$ (in the weak sense). Existence of a metric ensures the required isomorphism between the space of vector fields and the space of 1-forms and such isomorphism is denoted by $\sharp$.  
 At the linear level, the covariant divergence of the evolution equation for the velocity (\ref{eq:velocityeqn}) yields 
 \begin{eqnarray}
 \Delta_{\gamma}\left(\frac{\partial\phi}{\partial t}-\frac{n\dot{a}}{a}(\gamma_{ad}-\frac{n+1}{n})\phi+\frac{1}{a}\delta N+\frac{\gamma-1}{a\gamma_{ad} C_{\rho}}\delta\rho\right)=0.
 \end{eqnarray}
 Now clearly, the kernel of the rough Laplacian $\Delta_{\gamma}$ on a closed connected $M$ consists of constant functions only and therefore 
 \begin{eqnarray}
 \frac{\partial\phi}{\partial t}-\frac{n\dot{a}}{a}(\gamma_{ad}-\frac{n+1}{n})\phi+\frac{1}{a}\delta N+\frac{\gamma-1}{a\gamma_{ad} C_{\rho}}\delta\rho=C(t).
 \end{eqnarray}
One may absorb $C(t)$ in the scalar field $\phi$ since $\phi$ is unique up to functions constant on $M_{t}$. Let us make the transformation $\phi\mapsto \phi+F(t)$. The evolution equation for $\phi$ yields 
 \begin{eqnarray}
 \frac{\partial\phi}{\partial t}-\frac{n\dot{a}}{a}(\gamma_{ad}-\frac{n+1}{n})\phi+\frac{1}{a}\delta N+\frac{\gamma-1}{a\gamma_{ad} C_{\rho}}\delta\rho\nonumber+\frac{d F(t)}{dt}-\frac{n\dot{a}}{a}(\gamma_{ad}-\frac{n+1}{n})F(t)\\\nonumber 
 =C(t).
 \end{eqnarray}
 Now setting $\frac{dF}{dt}-\frac{n\dot{a}}{a}(\gamma_{ad}-\frac{n+1}{n})F(t) 
 =C(t)$ (which has a unique solution $F(t)=F(t_{0})(\frac{a(t)}{a(t_{0})})^{n(\gamma_{ad}-\frac{n-1}{n})}+a(t)^{n(\gamma_{ad}-\frac{n-1}{n})}\int_{t_{0}}^{t}\frac{C(t^{'})}{a(t^{'})^{n(\gamma_{ad}-\frac{n-1}{n})}}dt^{'}$) , the fluid equations reduce to the following wave equation for the scalar potential $\phi$ (propagating with the speed of sound $c_{s}=\sqrt{\gamma_{ad}-1}$)
 \begin{eqnarray}
 \frac{\partial\phi}{\partial t}=\frac{n\dot{a}}{a}(\gamma_{ad}-\frac{n+1}{n})\phi-\frac{1}{a}\delta N-\frac{\gamma_{ad}-1}{a\gamma_{ad} C_{\rho}}\delta\rho,\\\nonumber \frac{\partial\delta\rho}{\partial t}=-\frac{C_{\rho}n\gamma_{ad}\dot{a}}{a}\delta N+\frac{1}{a}\gamma_{ad} C_{\rho}\Delta_{\gamma}\phi.
 \end{eqnarray}
 Notice that this is precisely the wave equation with respect to the `sound' metric $S_{\mu\nu}:=\hat{g}_{\mu\nu}+(1-c^{2}_{s})\mathbf{u}_{\mu}\mathbf{u}_{\nu}$ (this sound metric was first derived by \cite{moncrief1980} in the context of the stability of a black hole accretion problem).
Therefore, we observe that the irrotational part of the $n-$velocity field is decoupled from the rotational and the harmonic parts. The question remains how do we extract the evolution equations for the remaining parts of the velocity vector field. Let us now focus on the harmonic contribution. For closed (closed negative Einstein in this case to be precise) manifolds, the space of harmonic forms (1-forms in this particular case) is a finite-dimensional vector space. Therefore, the harmonic form contribution of the $n-$velocity 1-form field may be written as 
 \begin{eqnarray}
 \mathcal{H}=\sum_{k=1}^{L}m^{k}\mathcal{A}_{k},
 \end{eqnarray}
 where $\{\mathcal{A}\}_{k=1}^{l}$ form a basis of the space of harmonic forms. Here $L$ is the first Betti number or the dimension of the first de-Rham cohomology group (or the space of the first singular cohomology group $H^{1}(M)$ with real coefficients). Now we may utilize the $L^{2}$ orthogonality property of the three constituents in the Hodge decomposition of the velocity field with respect to the natural Riemannian metric $\gamma$ i.e., 
 \begin{eqnarray}
 \int_{M}\langle d\phi,\mathcal{H}\rangle_{\gamma}= \int_{M}\langle\delta\psi,\mathcal{H}\rangle_{\gamma}=0
 \end{eqnarray}
 since $\mathcal{H}\in \ker(d\delta+\delta d)$.
 Therefore, the harmonic contribution decouples to yield the following set of ordinary differential equations for each harmonic part 
 \begin{eqnarray}
 \frac{dm^{k}}{dt}=\frac{n\dot{a}}{a}(\gamma_{ad}-\frac{n+1}{n})m^{k}.
 \end{eqnarray}
 Integration of this equation yields 
 \begin{eqnarray}
 m^{k}(t)=m^{k}(t_{0})\left(\frac{a(t)}{a(t_{0})}\right)^{n(\gamma_{ad}-\frac{n+1}{n})}.
 \end{eqnarray}
Once we have extracted the time evolution of the harmonic contribution, the pure rotational part may be extracted by the application of a double `curl' operator $\delta\circ d$. Notice that $(\delta \psi)^{\sharp}$, the pure rotational contribution and $\mathcal{H}^{\sharp}$, the harmonic contribution to the velocity field, both have zero covariant divergence (see the appendix). Let us denote the vector field counterparts $(\delta\psi)^{\sharp}$ of $\delta\psi$ and $\mathcal{H}^{\sharp}$ of $\mathcal{H}$ by $\xi$ and $\chi$, respectively i.e., 
 \begin{eqnarray}
 v^{i}=\nabla[\gamma]^{i}\phi+\xi^{i}+\chi^{i}.
 \end{eqnarray}
 Here $\xi^{i}$ and $\chi^{i}$ satisfy (in the weak sense of course since they belong to $H^{s-1}, s>\frac{n}{2}+1$)
 \begin{eqnarray}
 \nabla[\gamma]_{i}\xi^{i}=0,\\
 \nabla[\gamma]_{i}\chi^{i}=0.
 \end{eqnarray}
 (i.e., $\exists \Theta\in C^{\infty}(M)$ such that $\int_{M}(\nabla[\gamma]_{i}\Theta)\xi^{i}=0=\int_{M}(\nabla[\gamma]_{i}\Theta)\chi^{i}$)
 The action of double curl $\delta\circ d$ on the evolution equation for the velocity (\ref{eq:velocityeqn}) yields 
 \begin{eqnarray}
 \frac{\partial [(\delta\circ d v^{\flat})^{\sharp }]^{i}}{\partial t}=\frac{n\dot{a}}{a}(\gamma_{ad}-\frac{n+1}{n})[(\delta\circ d v^{\flat})^{\sharp}]^{i},
 \end{eqnarray}
 since the action of curl annihilates the gradient by definition (exterior derivative to be precise). Noting $(\delta A)_{i}=\nabla[\gamma]^{j}A_{ij}~\forall A\in S^{0}_{2}(M)$, $d^{2}\phi=0=d\chi^{\flat}$, One may explicitly evaluate $[(\delta\circ dv^{\flat})^{\sharp}]^{i}$ to yield 
 \begin{eqnarray}
  [(\delta\circ d v^{\flat})^{\sharp }]^{i}&=&\nabla[\gamma]_{j}(\nabla[\gamma]^{i}\xi^{j}-\nabla[\gamma]^{j}\xi^{i}),\\\nonumber
  &=&-\nabla[\gamma]_{j}\nabla[\gamma]^{j}\xi^{i}+\nabla[\gamma]^{i}\nabla[\gamma]_{j}\xi^{j}+R^{i}_{j}\xi^{j}\\\nonumber 
  &=&(\delta^{i}_{j}\Delta_{\gamma}+R[\gamma]^{i}_{j})\xi^{j},
 \end{eqnarray}
where we have used the divergence free property of $\xi$. Therefore, one obtains the following evolution equation for $\xi$
  \begin{eqnarray}
 (\delta^{i}_{j}\Delta_{\gamma}+R[\gamma]^{i}_{j})\left(\frac{d\xi^{j}}{dt}-\frac{n\dot{a}}{a}(\gamma_{ad}-\frac{n+1}{n})\xi^{j}\right)=0.
  \end{eqnarray}
 Now, the kernel of the Hodge Laplacian $(d\delta+\delta d)$ is precisely the space of harmonic forms. Therefore, 
 \begin{eqnarray}
\frac{\partial\xi^{i}}{\partial t}-\frac{n\dot{a}}{a}(\gamma_{ad}-\frac{n+1}{n})\xi^{i}=\zeta^{i}(t,x)
 \end{eqnarray}
 for $\zeta^{\flat}\in \ker(d\delta+\delta d)$. On the other hand, $\xi^{\flat}$ is always unique up to a harmonic form $\kappa^{\flat}$ since $\delta \xi^{\flat}=\delta (\xi^{\flat}+\kappa^{\flat})$. Therefore, if we make a transformation $\xi(t,x)\mapsto \xi(t,x)+\kappa(t,x)$, the evolution equation becomes 
 \begin{eqnarray}
\frac{\partial\xi^{i}}{\partial t}-\frac{n\dot{a}}{a}(\gamma_{ad}-\frac{n+1}{n})\xi^{i}+\left(\frac{\partial \kappa^{i}}{\partial t}-\frac{n\dot{a}}{a}(\gamma_{ad}-\frac{n+1}{n})\kappa^{i}\right)=\zeta^{i}(t,x).
\end{eqnarray}
 Now setting $\frac{\partial\kappa^{i}}{\partial t}-\frac{n\dot{a}}{a}(\gamma_{ad}-\frac{n+1}{n})\kappa^{i}=\zeta^{i}(t,x)$ (which has a unique solution $\kappa^{i}(t,x)=\kappa^{i}(t_{0},x)(\frac{a(t)}{a(t_{0})})^{n(\gamma_{ad}-\frac{n-1}{n})}+a(t)^{n(\gamma_{ad}-\frac{n-1}{n})}\int_{t_{0}}^{t}\frac{\zeta^{i}(t^{'},x)}{a(t^{'})^{n(\gamma_{ad}-\frac{n-1}{n})}}dt^{'}$),
the evolution equation for the rotational part $\xi$ reduces to
\begin{eqnarray}
\label{eq:curleqn}
\frac{d\xi^{i}}{dt}-\frac{n\dot{a}}{a}(\gamma_{ad}-\frac{n+1}{n})\xi^{i}=0
\end{eqnarray}
from which one obtains the following stability criteria
 \begin{eqnarray}
 \label{eq:adiabat}
 \gamma_{ad}< \frac{n+1}{n}.
 \end{eqnarray}
 Direct integration of the evolution equation for the curl part (\ref{eq:curleqn}) yields 
 \begin{eqnarray}
 \xi^{i}(t)=\xi^{i}(t_{0})\left(\frac{a(t)}{a(t_{0})}\right)^{n(\gamma_{ad}-\frac{n+1}{n})}.
 \end{eqnarray}
 Clearly, the curl part satisfies an ordinary differential equation in time and therefore we may control its Sobolev norm of arbitrary order. Provided that the adiabatic index $\gamma_{ad}$ lies in the suitable range (\ref{eq:adiabat}), the curl part of the $n-$velocity field decays. 
 
 Notice another remarkable fact that the three parts of the velocity field are decoupled at the linear level. In other words, the coupling is necessarily of higher order. Now we turn to the linearized Einstein's equations imposing constant mean extrinsic curvature spatially harmonic (CMCSH) gauge. Note that the following lemma holds\\ 
 \textbf{Lemma 3:} \textit{Let $h_{ij}$ be the perturbation to the metric $\gamma_{ij}$. The perturbation to the Ricci tensor in CMCSH gauge satisfies 
 \begin{eqnarray}
 \delta R_{ij}+\frac{1}{n}h_{ij}=DR[\gamma]_{ij}\cdot h+\frac{1}{n}h_{ij}=\frac{1}{2}\mathcal{L}_{\gamma,\gamma}h_{ij},
 \end{eqnarray}
 where $\mathcal{L}_{\gamma,\gamma}h_{ij}$ is defined to be 
 \begin{eqnarray}
 \mathcal{L}_{\gamma,\gamma}h_{ij}:=\Delta_{\gamma}h_{ij}+(R[\gamma]^{k}~_{ij}~^{m}+R[\gamma]^{k}~_{ji}~^{m})h_{km}.
 \end{eqnarray}
 }\\ 
 \textbf{Proof:} Using the Frechet derivative formula for the connection coefficents $D\Gamma[\gamma]^{i}_{jk}\cdot h=\frac{1}{2}\gamma^{il}(\nabla[\gamma]_{j}h_{lk}+\nabla[\gamma]_{k}h_{jl}-\nabla[\gamma]_{l}h_{ij}$), one may readily obtain the first variation of the Ricci tensor in the direction of $h$
 \begin{eqnarray}
 \delta R_{ij}=DR[\gamma]_{ij}\cdot h=\frac{1}{2}\{\Delta_{\gamma}h_{ij}+(R[\gamma]^{k}~_{ij}~^{m}+R[\gamma]^{k}~_{ji}~^{m})h_{km}\\\nonumber +\nabla[\gamma]_{j}\nabla[\gamma]_{m}h^{m}_{i} +\nabla[\gamma]_{i}\nabla[\gamma]_{m}h^{m}_{j}+R[\gamma]_{kj}h^{k}_{i}+R[\gamma]_{ki}h^{k}_{j}-\nabla[\gamma]_{i}\nabla[\gamma]_{j}tr_{\gamma}h\}.
 \end{eqnarray}
The linearized version of the spatial harmonic gauge `$id: (M,\gamma+h)\to (M,\gamma)$ is harmonic' yields 
\begin{eqnarray}
\gamma^{ij}D\Gamma[\gamma]^{k}_{ij}\cdot h=0\\\nonumber 
2\nabla[\gamma]_{j}h^{ij}-\nabla[\gamma]^{i}tr_{\gamma}h=0.
\end{eqnarray}
Therefore, the potentially problematic term $\nabla[\gamma]_{j}\nabla[\gamma]_{m}h^{m}_{i}+\nabla[\gamma]_{i}\nabla[\gamma]_{m}h^{m}_{j}$ obstructing the ellipticity of $h_{ij}\mapsto DR[\gamma]_{ij}\cdot h$ becomes 
\begin{eqnarray}
\nabla[\gamma]_{j}\nabla[\gamma]_{m}h^{m}_{i}+\nabla_{i}\nabla_{m}h^{m}_{j}&=&\frac{1}{2}(\nabla[\gamma]_{i}\nabla[\gamma]_{j}+\nabla[\gamma]_{i}\nabla[\gamma]_{j})tr_{\gamma}h\\\nonumber 
&=&\nabla[\gamma]_{i}\nabla[\gamma]_{j}tr_{\gamma}h,
\end{eqnarray}
which gets cancelled with $-\nabla[\gamma]_{i}\nabla[\gamma]_{j}tr_{\gamma}h$
and therefore 
\begin{eqnarray}
\delta R_{ij}=DR[\gamma]_{ij}\cdot h&=&\frac{1}{2}\{\Delta_{\gamma}h_{ij}+(R[\gamma]^{k}~_{ij}~^{m}+R[\gamma]^{k}~_{ji}~^{m})h_{km}\nonumber +R[\gamma]_{kj}h^{k}_{i}\\\nonumber
&&+R[\gamma]_{ki}h^{k}_{j}\}\\\nonumber 
&=&\frac{1}{2}\{\Delta_{\gamma}h_{ij}+(R[\gamma]^{k}~_{ij}~^{m}+R[\gamma]^{k}~_{ji}~^{m})h_{km}\nonumber\}\\\nonumber
&&-\frac{1}{n}h_{ij},
\end{eqnarray}
i.e.,
\begin{eqnarray}
\delta R_{ij}+\frac{1}{n}h_{ij}=DR[\gamma]_{ij}\cdot h+\frac{1}{n}h_{ij}=\mathcal{L}_{\gamma,\gamma}h_{ij},
\end{eqnarray}
where we have used the fact that the background metric is negative Einstein ($R[\gamma]_{ij}=-\frac{1}{n}\gamma_{ij}$). This concludes the proof of the lemma.

Therefore, we establish that $h\mapsto DR[\gamma]_{ij}\cdot h$ is linear elliptic, which will help us to cast the Einstein evolution equations as a linear hyperbolic system. The linearized Einstein-Euler evolution and constraint equations read 
\begin{eqnarray}
\label{eq:metric1}
\frac{\partial h_{ij}}{\partial t}
 &=&-\frac{2}{a}k^{tr}_{ij}+\frac{2\dot{a}}{a}\delta N \gamma_{ij}+\underbrace{l^{TT||}_{ij}}_{I}+\frac{1}{a}(L_{X}\gamma)_{ij},\\ 
 \frac{\partial k^{tr}_{ij}}{\partial t}
 &=&-(n-1)\frac{\dot{a}}{a}k^{tr}_{ij}\nonumber-\frac{1}{a}\nabla[\gamma]_{i}\nabla[\gamma]_{j}\delta N+
+\frac{\gamma-2}{n-1}a^{1-n\gamma_{ad}}\delta\rho \gamma_{ij}+\frac{1}{2a     }\mathcal{L}_{\gamma,\gamma}h_{ij}\\
\label{eq:SF}
&&+\frac{1}{a}\left\{\frac{1}{n-1}(1+\gamma_{ad} a^{2-n\gamma_{ad}}C_{\rho})\gamma_{ij}-\frac{1}{n}\gamma_{ij}\right\}\delta N,\\
\frac{\partial\phi}{\partial t}&=&\underbrace{\frac{n\dot{a}}{a}(\gamma_{ad}-\frac{n+1}{n})}_{<0}\phi-\frac{1}{a}\delta N-\frac{\gamma_{ad}-1}{C_{\rho}\gamma_{ad} a}\delta\rho,\\
 \frac{\partial\delta\rho}{\partial t}&=&-\frac{C_{\rho}n\gamma_{ad}\dot{a}}{a}\delta N+\frac{C_{\rho}\gamma_{ad}}{a} \Delta_{\gamma}\phi,\\
 \frac{d\xi^{i}}{dt}&=&\underbrace{\frac{n\dot{a}}{a}(\gamma_{ad}-\frac{n+1}{n})}_{<0}\xi^{i},\\ \frac{d\chi^{i}}{dt}&=&\underbrace{\frac{n\dot{a}}{a}(\gamma_{ad}-\frac{n+1}{n})}_{<0}\chi^{i},
 \end{eqnarray}
 \begin{eqnarray}
\Delta_{\gamma}tr_{\gamma}h+\nabla[\gamma]^{i}\nabla[\gamma]^{j}h_{ij}-R[\gamma]_{ij}h^{ij}&=&2a^{2-n\gamma_{ad}}\delta\rho,\\
\label{eq:MCNW}
 \nabla[\gamma]_{j}k^{trij}&=&-\gamma_{ad} a^{2-n\gamma_{ad}}C_{\rho}v^{i}.
 \end{eqnarray}
 along with the elliptic equations for the lapse function (perturbation) and the shift vector field arising as a consequence of the gauge fixing 
 \begin{eqnarray}
 \label{eq:lapsenew}
 \Delta_{\gamma}\delta N+\frac{1}{n-1}\left(1+C_{\rho}n\gamma_{ad} a^{2-n\gamma_{ad}}\right)\delta N=-\frac{n\gamma_{ad}-2}{n-1}a^{2-n\gamma_{ad}}\delta\rho,\\
 \label{eq:shiftnew}
 \Delta_{\gamma} X^{i}-R[\gamma]^{i}_{j} X^{j}=-(n-2)\frac{\dot{a}}{a}\nabla[\gamma]^{i}\delta N-2\nabla[\gamma]_{j}k^{trij}.
\end{eqnarray}
Notice that the negativity of the underlined terms in the linearized evolution equations is crucial in our analysis since these will generate uniform decay of the matter fields.   
Here we have used the fact that the term `$g^{ij}\partial_{t}\Gamma[\gamma]^{k}_{ij}$'in the shift equation (\ref{eq:shiftin}) vanishes at the linear level (as shown in section 3.2).
Notice that a problematic term $\nabla[\gamma]_{j}k^{trij}$ arises in the right hand side of the shift equation, which requires a higher regularity of the trace-less second fundamental form $k^{tr}$. However, we may immediately utilize the momentum constraint (\ref{eq:MC}) to get rid of this term and obtain
\begin{eqnarray}
\label{eq:shift1}
\Delta_{\gamma} X^{i}-R[\gamma]^{i}_{j} X^{j}=-(n-2)\frac{\dot{a}}{a}\nabla[\gamma]^{i}\delta N+2\gamma_{ad} C_{\rho} a^{2-n\gamma_{ad}}v^{i}.
\end{eqnarray}
Here $v=(d\phi)^{\sharp}+\xi+\chi$. Observe that the $\mathcal{N}-$tangential velocity $l^{TT||}$ entered into the linearized evolution equation for the metric perturbations $h_{ij}$ and we need to estimate it (term $I$ in the equation). Choosing a basis $\frac{\partial\gamma}{\partial q^{a}}=m^{TT||}_{a}+L_{W^{||}_{a}}\gamma$ for the formal tangent space of $\mathcal{N}$ at $\gamma$, the time derivative of the shadow gauge condition yields at the linear level 
\begin{eqnarray}
\langle\partial_{t}h,m^{TT||}_{a}+L_{W^{||}_{a}}\gamma\rangle_{\gamma}=0\\\nonumber 
\langle-\frac{2}{a}k^{tr}+\frac{2\dot{a}}{a}\delta N \gamma+l^{TT||}+\frac{1}{a}L_{X}\gamma,m^{TT||}_{a}+L_{W^{||}_{a}}\gamma\rangle_{\gamma}=0\\\nonumber 
\langle-\frac{2}{a}k^{tr}+l^{TT||},m^{TT||}_{a}+L_{W^{||}_{a}}\gamma\rangle_{\gamma}=0\\\nonumber
\langle-\frac{2}{a}(k^{TT}+L_{Y}\gamma-\frac{2}{n}\nabla[\gamma]_{m}Y^{m}\gamma)+l^{TT||},m^{TT||}_{a}+L_{W^{||}_{a}}\gamma\rangle_{\gamma}=0\\\nonumber 
\langle-\frac{2}{a}k^{TT}+l^{TT||},m^{TT||}_{a}\rangle_{\gamma}=0\Rightarrow l^{TT||}=\frac{2}{a}k^{TT||}.
\end{eqnarray}
Here we have utilized the $L^{2}$ orthogonality between $A^{TT}$ and $L_{B}\gamma$ for $B\in \mathfrak{X}(M)$ and $A^{TT}$ is a TT-tensor with respect to $\gamma$. $k^{TT||}$ denotes the projetion of the transverse-traceless part of $k^{tr}$ onto the kernel of $\mathcal{L}_{\gamma,\gamma}$ (as we have discussed in section 3.1). In the energy inequality, we will observe that the term $l^{TT||}$  is rather innocuous at the linear level since action of $\mathcal{L}_{\gamma,\gamma}$ annihilates it (and only contributes at higher order).   

\subsection{\textbf{Local Well-posedness}}
\cite{andersson2003elliptic} proved a well-posedness theorem for the Cauchy problem for a family of elliptic-hyperbolic systems that included the `$n+1$' dimensional vacuum Einstein equations in CMCSH gauge. \cite{moncrief2019could} sketched how to apply the theorem of \cite{andersson2003elliptic} to a gauge fixed system of `Einstein-$\Lambda$' field equations. However, a local existence theorem for the fully non-linear Einstein-Euler system in CMCSH gauge remains open till today. \cite{rodnianski2018stable} utilized a CMC transported spatial coordinate condition (zero shift) to establish a local and subsequently a global existence result for the sufficiently small however fully non-linear irrotational Einstein-Euler system on $(0,\infty)\times \mathbb{T}^{3}$. In the case of an irrotational fluid, one readily has a nonlinear wave equation (in a `sound metric') for the scalar potential. One potential complication that arises in the case of a general Einstein-Euler system is that the rotational contribution and the harmonic contribution of the $n-$velocity vector field are coupled non-linearly to the scalar field. Such questions are not however relevant at the linear level and left for future study. At this linear level, the system (\ref{eq:metric1})-(\ref{eq:shift1}) forms a coupled linear elliptic hyperbolic system. A straightforward energy argument together with a contraction mapping on the Banach space $C^{0}([0,T];H^{s}\times H^{s-1}\times H^{s-1}\times H^{s}\times H^{s-1}\times H^{s-1})$ (i.e., $(h(t),k^{tr}(t),\delta \rho(t),\phi(t),\xi(t),\chi(t))\in H^{s}\times H^{s-1}\times H^{s-1}\times H^{s}\times H^{s-1}\times H^{s-1}$) with $s>\frac{n}{2}+1$ would yield an existence result. However, a subtlety is that in the energy expressions, we will have the shift vector field $X$ and the perturbation to the lapse function $\delta N$. Therefore, we need to ensure that they are uniquely determined or in other words that the following elliptic equations have unique solutions 
\begin{eqnarray}
\Delta_{\gamma}\delta N+\frac{1}{n-1}\left(1+C_{\rho}n\gamma_{ad} a^{2-n\gamma_{ad}}\right)\delta N=-\frac{n\gamma_{ad}-2}{n-1}a^{2-n\gamma_{ad}}\delta\rho,\\
\Delta_{\gamma} X^{i}-R[\gamma]^{i}_{j} X^{j}=-(n-2)\frac{\dot{a}}{a}\nabla[\gamma]^{i}\delta N+2\gamma_{ad} C_{\rho} a^{2-n\gamma_{ad}}v^{i}
\end{eqnarray}
in $H^{s+1}\times H^{s+1}$. But, this is equivalent to showing that the operators $\Delta_{\gamma}+\frac{1}{n-1}\left(1+C_{\rho}n\gamma_{ad} a^{2-n\gamma_{ad}}\right)id$ and $\delta^{i}_{j}\Delta_{\gamma}-R[\gamma]^{i}_{j}$ have trivial kernels on $H^{s+1}\times H^{s+1}$. However, this follows from the fact that $1+C_{\rho}n\gamma_{ad}a^{2-n\gamma_{ad}}>0$ and $R[\gamma]_{ij}=-\frac{1}{n}\gamma_{ij}$  \cite{andersson2003elliptic, andersson2011einstein}. Note that once the perturbation to the lapse is uniquely determined in terms of $\delta \rho$, the shift vector field is then uniquely determined in terms of $v$ $(\phi,\xi,\chi)$ and $\delta\rho$. The detailed elliptic estimates are stated in the next section. Uniqueness and the continuation criteria is straightforward for this linear elliptic-hyperbolic system. The only difference between our equations here and the vacuum work studied by \cite{andersson2003elliptic} is the
appearance of matter terms. However these appear with good signs and so the analysis proceeds in the same way. For completeness we state the following theorem concerning the local well-posedness.

\textbf{Local well-posedness theorem:} \textit{Let $s>\frac{n}{2}+1$. The CMCSH Cauchy problem of the linearized Einstein-Euler$-\Lambda$ system with initial data $(h_{0},k^{tr}_{0},\delta\rho_{0},\phi_{0},\xi_{0},\chi_{0})\in H^{s}\times H^{s-1}\times H^{s-1}\times H^{s}\times H^{s-1}\times H^{s-1}$ and $\Lambda\geq 0$ is strongly well posed in $C^{0}([0,T^{\dag}]; H^{s}\times H^{s-1}\times H^{s-1}\times H^{s}\times H^{s-1}\times H^{s-1})$. In particular, there exists a time $T^{\dag}>0$ depending on $||h_{0}||_{H^{s}}, ||k^{tr}_{0}||_{H^{s-1}}, ||\delta\rho_{0}||_{H^{s-1}}, ||\phi_{0}||_{H^{s}}, ||\xi_{0}||_{H^{s-1}}, ||\chi_{0}||_{H^{s-1}}$ such that the solution map $(h_{0},k^{tr}_{0},\delta\rho_{0},\phi_{0},\xi_{0},\chi_{0})\mapsto (h(t),k^{tr}(t),\delta\rho(t),\phi(t),\xi(t),\chi(t),N(t),X(t))$ is continuous as a map 
\begin{eqnarray}
H^{s}\times H^{s-1}\times H^{s-1}\times H^{s}\times H^{s-1}\times H^{s-1}\to \\\nonumber H^{s}\times H^{s-1}\times H^{s-1}\times H^{s}\times H^{s-1}\times H^{s-1}\times H^{s+1}\times H^{s+1}.
\end{eqnarray}
Let $T^{\dag}$ be the maximal time of existence of the solution to the CMCSH Cauchy problem with data $(h_{0},k^{tr}_{0},\delta\rho_{0},\phi_{0},\xi_{0},\chi_{0})$. Then either $T^{\dag}=+\infty$ or 
\begin{eqnarray}
\lim_{t\to T^{\dag}} \sup \max(||h(t)||_{H^{s}}, ||k^{tr}(t)||_{H^{s-1}}, ||\delta\rho(t)||_{H^{s-1}},\\\nonumber  ||\phi(t)||_{H^{s}}, ||\xi(t)||_{H^{s-1}}, ||\chi(t)||_{H^{s-1}})=+\infty. 
\end{eqnarray}
}
In addition to the local well-posedness, one needs to ensure the conservation of gauges and constraints, that is, the following entities must vanish identically along the solution curve
\begin{eqnarray}
A^{i}:=\nabla[\gamma]^{i}\tau,\\
B^{k}:=\gamma^{ij}(\Gamma[\gamma+h]^{k}_{ij}-\Gamma[\gamma]^{k}_{ij})=\gamma^{ij}D\Gamma^{k}[\gamma]_{ij}\cdot h,\\
D:=\Delta_{\gamma}tr_{\gamma}h+\nabla^{i}\nabla^{j}h_{ij}-R[\gamma]_{ij}h^{ij}-2a^{2-n\gamma}\delta\rho,\\
E^{i}:=\nabla[\gamma]_{j}k^{trij}+\gamma a^{2-n\gamma}Cv^{i}.
\end{eqnarray}
The vanishing of the entity $A^{i}$ precisely states the imposition of the constant mean extrinsic curvature gauge i.e., $\tau$ is a function of time $t$ alone. Vanishing of $D$ is essentially the conservation of the linearized version of the Hamiltonian constraint. One might naturally expect the conservation of the gauges and constraints due to the Bianchi identities. However, utilizing the evolution equations of the relevant fields, one may also directly obtain a system of coupled PDEs after a lengthy but straightforward calculation. Utilizing an energy argument similar to \cite{andersson2003elliptic}, one may show that if $(A,B,D,E)=0$ for the initial data $(h_{0}, k^{tr}_{0}, \delta\rho_{0}, v_{0})$, then $(A,B,D,E)\equiv 0$ along the solution curve $t\mapsto (h(t),k^{tr}(t), \delta\rho(t), v(t), \delta N(t), X(t))$. At the linear level, one sees trivially that $A^{i}\equiv0$. Here, of course, by $v$ we mean the triple $(\phi,\xi,\chi)$.
\subsection{\textbf{Elliptic Estimates}} 
In this section, we estimate $\delta N$ and $X^{i}$ from their respective elliptic equations. These estimates are necessary to derive the desired energy inequality. At each step, we carefully keep track of the time dependent expansion factor.\\
\textbf{Lemma 4:} \textit{Let $s>\frac{n}{2}+1$ and assume ($\delta N, X^{i}, h_{ij}, k^{tr}_{ij}, \delta \rho, v^{i}$) solves the linearized Einstein-Euler system expressed in CMCSH gauge. The following estimate holds for the linearized perturbation to the lapse function $\delta N$
\begin{eqnarray}
 ||\delta N||_{H^{s+1}}\leq Ca^{2-n\gamma_{ad}}||\delta\rho||_{H^{s-1}}, 
\end{eqnarray}
where $a$ is the scale factor for the background solution and $C>0$ is a suitable constant dependent on the background solution (\ref{eq:background}).}\\
\textbf{Proof:} The elliptic equation for the perturbation to the lapse function reads 
\begin{eqnarray}
\Delta_{\gamma}\delta N+\frac{1}{n-1}\left(1+C_{\rho}n\gamma_{ad} a^{2-n\gamma_{ad}}\right)\delta N=-\frac{n\gamma_{ad}-2}{n-1}a^{2-n\gamma_{ad}}\delta\rho.
\end{eqnarray}
Now, we have shown in the previous section that the operator $\Delta_{\gamma}+\frac{1}{n-1}\left(1+C_{\rho}n\gamma_{ad} a^{2-n\gamma_{ad}}\right)id: H^{s+1}\to H^{s-1}$ is injective and following its self adjointness, also surjective (it has closed range). Therefore, following this isomorphism, we immediately obtain
\begin{eqnarray}
||\delta N||_{H^{s+1}}\leq Ca^{2-n\gamma_{ad}}||\delta\rho||_{H^{s-1}}.
\end{eqnarray}
A short proof may be as follows. Let us denote the operator $\Delta_{\gamma}+\frac{1}{n-1}\left(1+C_{\rho}n\gamma_{ad} a^{2-n\gamma_{ad}}\right)id$ by $A$. Assume that an estimate of the type $||u||_{H^{s+1}}\lesssim ||Au||_{H^{s-1}}$ does not hold. Then there exists a sequence $\{u_{i}\}_{i=1}^{\infty}$ with $||u_{i}||_{H^{s+1}}=1$ and $||Au_{i}||_{H^{s-1}}\to 0$ as $i\to \infty$. Now $M$ is compact and therefore $H^{s+1}$ is compactly embedded into $H^{s-1}$ (notice $s>\frac{n}{2}+1$). This yields a sub-sequence $\{u_{i_{j}}\}$ converging to $u^{*}\in H^{s+1}$ strongly in $H^{s-1}$, which by construction satisfies $Au^{*}=0$. This contradicts the fact that the operator $A$ is injective. Therefore an estimate of the type $||u||_{H^{s+1}}\lesssim ||Au||_{H^{s-1}}$ holds $\forall s\geq 1$ (which is satisfied here since $s>\frac{n}{2}+1$). 
Here, the constant $C$ depends on the background entities as evident from the elliptic equation.\\ 
\textbf{Lemma 5:} \textit{Let $s>\frac{n}{2}+1$ and assume ($\delta N, X^{i}, h_{ij}, k^{tr}_{ij}, \delta\rho, v^{i}$) solves the linearized Einstein-Euler system expressed in CMCSH gauge. The following estimate holds for the linearized perturbation to the shift vector field 
\begin{eqnarray}
 ||X||_{H^{s+1}}\leq C(\dot{a}a^{1-n\gamma_{ad}}||\delta\rho||_{H^{s-1}}+C_{\rho}a^{2-n\gamma_{ad}}||v||_{H^{s-1}}), 
\end{eqnarray}
where $a$ is the scale factor for the background solution and $C>0$ is a suitable constant dependent on the background solution (\ref{eq:background}).}\\
\textbf{Proof:} The elliptic equation for the shift vector field reads
\begin{eqnarray}
\Delta_{\gamma} X^{i}-R[\gamma]^{i}_{j}X^{j}=-(n-2)\frac{\dot{a}}{a}\nabla[\gamma]^{i}\delta N+2\gamma_{ad} C_{\rho} a^{2-n\gamma_{ad}}v^{i}.
\end{eqnarray}
Now given that $\gamma$ is negative Einstein i.e., $R[\gamma]_{ij}=-\frac{1}{n}\gamma_{ij}$, the operator
$P:=\delta^{i}_{j}\Delta_{\gamma}-R[\gamma]^{i}_{j}: H^{s+1}\to H^{s-1}$ has trivial kernel. We may use a similar argument as that of the previous case to obtain the following estimate
\begin{eqnarray}
||X||_{H^{s+1}}&\leq& C\left(||\frac{-(n-2)\dot{a}}{a}\nabla[\gamma]^{i}\delta N+2\gamma_{ad} C_{\rho}a^{2-n\gamma_{ad}}v^{i}||_{H^{s-1}}\right)\\\nonumber 
&\leq& C\left(\frac{\dot{a}}{a}||\nabla[\gamma]^{i}\delta N||_{H^{s-1}}+C_{\rho}a^{2-n\gamma_{ad}}||v^{i}||_{H^{s-1}}\right)\\\nonumber 
&\leq& C\left(\frac{\dot{a}}{a}||\delta N||_{H^{s}}+C_{\rho}a^{2-n\gamma_{ad}}||v^{i}||_{H^{s-1}}\right)\\\nonumber 
&\leq& C\left(\dot{a}a^{1-n\gamma_{ad}}||\delta\rho||_{H^{s-1}}+C_{\rho}a^{2-n\gamma_{ad}}||v||_{H^{s-1}}\right).
\end{eqnarray}
Here we have used the fact that $||A||_{H^{s_{1}}}\lesssim ||A||_{H^{s_{2}}}$ for $s_{1}<s_{2}$ and the estimate for the perturbation to the lapse function from the previous lemma (4). This concludes the proof of the lemma.

\section{Energy functional}
In the study of hyperbolic PDE, the use of an energy functional is indispensable. Since the linearized Einstein-Euler system is a linear coupled hyperbolic system (except that the rotational and harmonic parts of the velocity field are decoupled), one may naturally define a wave equation type of energy and its higher-order extension which controls the desired norm of the data $(h,k^{tr},\delta\rho, \phi,\xi,\chi)$. Note that at the linear level, the rotational ($\xi$) and the harmonic ($\chi$) parts of the velocity field simply satisfy ordinary differential equations in time and therefore, we may control their Sobolev norm of any order. Let us first define the lowest order energy $\mathcal{E}_{1}$
\begin{eqnarray}
\mathcal{E}_{1}&:=&\frac{(\gamma_{ad}-1)}{2\gamma_{ad}^{2}C_{\rho}^{2}}\langle\delta\rho,\delta\rho\rangle_{L^{2}}+\frac{1}{2}\langle\phi,\Delta_{\gamma}\phi\rangle_{L^{2}}+\frac{1}{2}\langle\xi,\xi\rangle_{L^{2}}\\\nonumber 
&&+\frac{1}{2}\langle\mathcal{H},\mathcal{H}\rangle_{L^{2}}+\frac{1}{2}\langle k^{tr},k^{tr}\rangle_{L^{2}}+\frac{1}{8}\langle h,\mathcal{L}_{\gamma,\gamma}h\rangle_{L^{2}}.
 \end{eqnarray}
 Here $\Delta_{\gamma},\Delta$, and $\mathcal{L}_{\gamma,\gamma}$ are defined as follows 
 \begin{eqnarray}
 \Delta_{\gamma}:=-\gamma^{ij}\nabla[\gamma]_{i}\nabla[\gamma]_{j}, \Delta\xi^{i}:=\Delta_{\gamma}\xi^{i}+R[\gamma]^{i}_{j}\xi^{j}, \mathcal{L}_{\gamma,\gamma}h_{ij}:=\Delta_{\gamma}h_{ij}\\\nonumber+(R[\gamma]^{k}~_{ij}~^{m}+R[\gamma]^{k}~_{ji}~^{m})h_{km}.
 \end{eqnarray}
 Clearly, $\Delta$ and $\mathcal{L}_{\gamma,\gamma}$ are the Hodge Laplacian and a Lichnerowicz type Laplacian, respectively. They both have non-negative spectrum on $M$. The lowest order energy may explicitly be written as follows 
 \begin{eqnarray}
 \mathcal{E}_{1}=\frac{(\gamma_{ad}-1)}{2\gamma_{ad}^{2}C_{\rho}^{2}}\int_{M}(\delta\rho)^{2}\mu_{\gamma}+\frac{1}{2}\int_{M}\nabla[\gamma]_{i}\phi\nabla[\gamma]_{j}\phi\gamma^{ij}\mu_{\gamma}\nonumber+\frac{1}{2}\int_{M}\xi^{i} \xi^{j}\gamma_{ij}\mu_{\gamma}\\\nonumber+\frac{1}{2}\int_{M}\mathcal{H}_{i}\mathcal{H}_{j}\gamma^{ij}\mu_{\gamma}+\frac{1}{2}\int_{M}k^{tr}_{ij}k^{tr}_{kl}\gamma^{ik}\gamma^{jl}\mu_{\gamma}\\\nonumber 
+\frac{1}{8}\int_{M}(\nabla[\gamma]_{k}h_{ij}\nabla[\gamma]_{l}h_{mn}\gamma^{kl}\gamma^{im}\gamma^{jn}+(R[\gamma]^{k}~_{ij}~^{m}+R[\gamma]^{k}~_{ji}~^{m})h_{km}h_{ln}\gamma^{il}\gamma^{jn})\mu_{\gamma}.
 \end{eqnarray}
 Using the structure of the lowest order energy, the higher order energies may be defined as follows 
 \begin{eqnarray}
\mathcal{E}_{i}&:=&\frac{(\gamma_{ad}-1)}{2\gamma_{ad}^{2}C_{\rho}^{2}}\langle\delta\rho,\Delta^{i-1}_{\gamma}\delta\rho\rangle_{L^{2}}+\frac{1}{2}\langle\phi,\Delta^{i}_{\gamma}\phi\rangle_{L^{2}}+\nonumber\frac{1}{2}\langle\xi,\Delta^{i-1} \xi\rangle_{L^{2}}\\\nonumber 
&&+\frac{1}{2}\langle\mathcal{H},\Delta^{i-1}_{\gamma}\mathcal{H}\rangle_{L^{2}}+\frac{1}{2}\langle k^{tr},\mathcal{L}^{i-1}_{\gamma,\gamma}k^{tr}\rangle_{L^{2}}+\frac{1}{8}\langle h,\mathcal{L}^{i}_{\gamma,\gamma}h\rangle_{L^{2}},
 \end{eqnarray}
 for $1\leq i\leq s$.
The total energy is naturally defined as 
\begin{eqnarray}
\mathcal{E}:=\sum_{i=1}^{s}\mathcal{E}_{i}.
\end{eqnarray}
The energy is positive semi-definite and it only vanishes precisely when $(h,k^{tr},\delta\rho,\phi,\xi,\chi)\equiv 0$. The first variation of the energy about $0$ in the direction of $\mathcal{X}:=(u,v,w,x,y,z)$ vanishes 
 \begin{eqnarray}
 D\mathcal{E}(0)\cdot \mathcal{X}=0,
 \end{eqnarray}
 that is, $0$ is a critical point of $\mathcal{E}$ in the space $\mathcal{Q}:=H^{s}\times H^{s-1}\times H^{s-1}\times H^{s}\times H^{s-1}\times H^{s-1}$. The Hessian of $\mathcal{E}$ at $0$ reads 
 \begin{eqnarray}
 D^{2}\mathcal{E}(0)\cdot (\mathcal{X},\mathcal{X})=\frac{(\gamma_{ad}-1)}{\gamma_{ad}^{2}C_{\rho}^{2}}\sum_{i=1}^{s}\langle w,\Delta^{i-1}_{\gamma}w\rangle_{L^{2}}+\sum_{i=1}^{s}\langle x,\Delta^{i}_{\gamma}x\rangle_{L^{2}}\\\nonumber+\nonumber\sum_{i=1}^{s}\langle y,\Delta^{i-1}y\rangle_{L^{2}} 
+\sum_{i=1}^{s}\langle z,\Delta^{i-1}_{\gamma}z\rangle_{L^{2}}+\sum_{i=1}^{s}\langle v,\mathcal{L}^{i-1}_{\gamma,\gamma}v\rangle_{L^{2}}\\\nonumber +\frac{1}{4}\sum_{i=1}^{s}\langle u,\mathcal{L}^{i}_{\gamma,\gamma}u\rangle_{L^{2}}\geq 0.
 \end{eqnarray}
 The equality hold iff $u=u^{TT||}, v=x=y=z=0$ (at the linear level $u^{TT||}=0$). Therefore, the positive semi-definiteness of the Lyapunov function on $B_{\delta}(0)-\{0\}\subset \mathcal{Q}$ is heavily dependent on the non-negativity of the spectrum of the associated rough Laplacian $\Delta_{\gamma}$, Hodge Laplacian $\Delta$, and the Lichnerowicz type Laplacian $\mathcal{L}_{\gamma,\gamma}$. Now, since the hessian is positive definite on $B_{\delta}(0)-\{0\}$, the map $D^{2}\mathcal{E}(0):B_{\delta}(0)-\{0\}\to Image (D^{2}\mathcal{E}(0))$ is an isomorphism yielding 
\begin{eqnarray}
||h||^{2}_{H^{s}}+||k^{tr}||^{2}_{H^{s-1}}+||\delta\rho||^{2}_{H^{s-1}}+||\phi||^{2}_{H^{s}}+||\xi||^{2}_{H^{s-1}}+||\chi||^{2}_{H^{s-1}}\\\nonumber
\lesssim D^{2}\mathcal{E}(0)\cdot (\mathcal{X},\mathcal{X}).
\end{eqnarray}
Now, using a version of the Morse lemma (Hilbert space version) on the non-degenerate critical point $(0,0)$, we obtain that there exists a $\delta>0$ such that for variations lying within $B_{\delta}(0)$, the following holds up to a possibly non-linear diffeomorphism $\mathcal{E}=\mathcal{E}(0)+D^{2}\mathcal{E}(0)\cdot (\mathcal{X},\mathcal{X})=D^{2}\mathcal{E}(0)\cdot (\mathcal{X},\mathcal{X})$ (notice that $\mathcal{E}(0)=0$). Therefore, we prove that the energy controls the desired norm of the perturbations i.e.,
\begin{eqnarray}
||h||^{2}_{H^{s}}+||k^{tr}||^{2}_{H^{s-1}}+||\delta\rho||^{2}_{H^{s-1}}+||\phi||^{2}_{H^{s}}+||\xi||^{2}_{H^{s-1}}+||\chi||^{2}_{H^{s-1}} \lesssim \mathcal{E}.
\end{eqnarray}  
\subsection{\textbf{Uniform boundedness of the energy functional}}
In this section, we prove the uniform boundedness of the energy functional and it's possible in time decay. In the process, we obtain the necessary and sufficient conditions for such estimates to hold. To obtain the uniform boundedness property of the energy, the necessary ingredients such as elliptic estimates are at hand. We therefore explicitly compute the time derivative for the lowest order energy defined in the previous section. An analogous calculation holds for the time derivative of the higher-order energies as well. We will use the Sobolev embedding on a compact domain $||A||_{L^{\infty}}\lesssim ||A||_{H^{a}}$ for $a>\frac{n}{2}$ and the following product estimates
\begin{eqnarray}
\label{eq:inequality1}
||AB||_{H^{s}}\lesssim (||A||_{L^{\infty}}||B||_{H^{s}}+||A||_{H^{s}}||B||_{L^{\infty}}),s>0,\\
||AB||_{H^{s}}\lesssim ||A||_{H^{s}}||B||_{H^{s}}, s>\frac{n}{2},\\
\label{eq:inequality3}
||[P,A]B||_{H^{a}}\lesssim (||\nabla A||_{L^{\infty}}||B||_{H^{s+a-1}}+||A||_{H^{s+a}}||B||_{L^{\infty}}),\\
P\in \mathcal{OP}^{s}, s>0, a\geq 0,\nonumber
\end{eqnarray}
where $\mathcal{OP}^{s}$ denotes the pseudo-differential operators with symbol in the Hormander class $S^{s}_{1,0}$ (see \cite{Taylor} for details). The first and second inequalities essentially emphasize the algebra property of $H^{s}\cap L^{\infty}$ for $s>0$ and of $H^{s}$ for $s>\frac{n}{2}$, respectively. In addition, we of course use integration by parts, Holder's and Minkowski's inequalities whenever necessary.\\
 \textbf{Lemma 6:}~\textit{Let $s>\frac{n}{2}+1$, $\gamma\in\mathcal{E}in_{-\frac{1}{n}}$ be the shadow of $g\in\mathcal{M}$ and assume $(h=g-\gamma,k^{tr},\delta\rho, \phi,\xi,\chi)$ satisfy the linearized Einstein-Euler system. Also assume there exists a $\delta>0$ such that $(h,k^{tr},\delta\rho, \phi, \xi,\chi)\in \mathcal{B}_{\delta}(0)\subset H^{s}\times H^{s-1}\times H^{s-1}\times H^{s}\times H^{s-1}\times H^{s-1}$. The time derivative of the lowest order energy reads 
 \begin{eqnarray}
 \frac{\partial \mathcal{E}_{1}}{\partial t}&=&-(n-1)\frac{\dot{a}}{a}\langle k^{tr},k^{tr}\rangle_{L^{2}}+\frac{n\dot{a}}{a}(\gamma_{ad}-\frac{n+1}{n})\langle \phi,\Delta_{\gamma}\phi\rangle_{L^{2}}\\\nonumber 
 &&+\frac{n\dot{a}}{a}(\gamma_{ad}-\frac{n+1}{n})\langle\xi,\Delta\xi\rangle_{L^{2}}+\frac{n\dot{a}}{a}(\gamma_{ad}-\frac{n+1}{n})\langle\mathcal{H},\Delta_{\gamma}\mathcal{H}\rangle_{L^{2}}\\\nonumber 
 &&+\mathcal{B},
\end{eqnarray}
where $\mathcal{B}$ satisfies 
\begin{eqnarray}
|\mathcal{B}|\lesssim (\gamma_{ad}-1)\frac{\dot{a}}{a}||\delta\rho||_{H^{s-1}}||\delta N||_{H^{s+1}}+\frac{1}{a}||\phi||_{H^{s}}||\delta N||_{H^{s+1}}+\frac{1}{a}\nonumber||\delta N||_{H^{s+1}}||k^{tr}||_{H^{s-1}}\\\nonumber
+\frac{\dot{a}}{a}||\delta N||_{H^{s+1}}||h||_{H^{s}}+\frac{1}{a}||X||_{H^{s+1}}||h||_{H^{s}}+\frac{1}{a}(||k^{tr}||_{H^{s-1}}+C_{\rho}a^{2-n\gamma_{ad}}||v||_{H^{s-1}})\mathcal{E}_{1}.
\end{eqnarray}
 }
 \textbf{Proof:} A direct calculation using the evolution equations (\ref{eq:metric1})-(\ref{eq:MCNW}) yields 
 \begin{eqnarray}
 \frac{\partial \mathcal{E}_{1}}{\partial t}=\frac{\gamma_{ad}-1}{\gamma_{ad}^{2}C^{2}_{\rho}}\langle\partial_{t}\delta\rho,\delta\rho\rangle_{L^{2}}+\langle\partial_{t}\phi,\Delta_{\gamma}\phi\rangle_{L^{2}}+\langle\partial_{t}\xi,\xi\rangle_{L^{2}}\\\nonumber 
 +\langle\partial_{t}\chi,\chi\rangle_{L^{2}}+\langle\partial_{t}k^{tr},k^{tr}\rangle_{L^{2}}+\frac{1}{4}\langle\partial_{t}h,\mathcal{L}_{\gamma,\gamma}h\rangle_{L^{2}}+\mathcal{ER}\\\nonumber
 =\frac{\gamma_{ad}-1}{\gamma_{ad}^{2}C^{2}_{\rho}}\langle-\frac{C_{\rho}n\gamma_{ad}\dot{a}}{a}\delta N+\frac{1}{a}C_{\rho}\gamma_{ad} \Delta_{\gamma}\phi,\delta\rho\rangle_{L^{2}}\\\nonumber 
 +\langle\frac{n\dot{a}}{a}(\gamma_{ad}-\frac{n+1}{n})\phi-\frac{1}{a}\delta N-\frac{\gamma_{ad}-1}{aC_{\rho}\gamma_{ad}}\delta\rho,\Delta_{\gamma}\phi\rangle_{L^{2}}\\\nonumber 
 +\langle\frac{n\dot{a}}{a}(\gamma_{ad}-\frac{n+1}{n})\xi,\xi\rangle_{L^{2}}+\langle\frac{n\dot{a}}{a}(\gamma_{ad}-\frac{n+1}{n})\chi,\chi\rangle_{L^{2}}\\\nonumber 
 +\langle-(n-1)\frac{\dot{a}}{a}k^{tr}-\frac{1}{a}\nabla[\gamma]\otimes\nabla[\gamma]\delta N+\frac{1}{a}\left\{\frac{1}{n-1}(1+\nonumber\gamma_{ad} a^{2-n\gamma_{ad}}C_{\rho})\gamma-\frac{1}{n}\gamma\right\}\delta N\\\nonumber
 +\frac{\gamma_{ad}-2}{n-1}a^{1-n\gamma_{ad}}\delta\rho \gamma+\frac{1}{2a}\mathcal{L}_{\gamma,\gamma}h_{ij},k^{tr}\rangle_{L^{2}}+\frac{1}{4}<-\frac{2}{a}k^{tr}+\frac{2\dot{a}}{a}\delta N \gamma+l^{TT||}\\\nonumber
 +\frac{1}{a}L_{X}\gamma,\mathcal{L}_{\gamma,\gamma}h\rangle_{L^{2}}+\mathcal{ER}\\\nonumber 
 =\frac{n\dot{a}}{a}(\gamma_{ad}-\frac{n+1}{n})\langle\phi,\Delta_{\gamma}\phi\rangle_{L^{2}}+\frac{n\dot{a}}{a}(\gamma_{ad}-\frac{n+1}{n})\langle\xi,\xi\rangle_{L^{2}}\\\nonumber 
 +\frac{n\dot{a}}{a}(\gamma_{ad}-\frac{n+1}{n})\langle\chi,\chi\rangle_{L^{2}}-(n-1)\frac{\dot{a}}{a}\langle k^{tr},k^{tr}\rangle_{L^{2}}\\\nonumber 
 -\frac{(\gamma_{ad}-1)n}{\gamma_{ad} C_{\rho}}\frac{\dot{a}}{a}\langle\delta N,\delta\rho\rangle_{L^{2}}-\frac{1}{a}\langle\delta N,\Delta_{\gamma}\phi\rangle_{L^{2}}-\frac{1}{a}\langle\nabla[\gamma]\otimes \nabla[\gamma]\delta N,k^{tr}\rangle_{L^{2}}\\\nonumber 
 +\frac{1}{2}\frac{\dot{a}}{a}\langle\delta N\gamma,\mathcal{L}_{\gamma,\gamma}h\rangle_{L^{2}}+\frac{1}{4a}\langle L_{X}\gamma,\mathcal{L}_{\gamma,\gamma}h\rangle_{L^{2}}+\mathcal{ER}.
 \end{eqnarray}
Here $(\nabla[\gamma]\otimes \nabla[\gamma]\delta N)_{ij}:=\nabla[\gamma]_{i} \nabla[\gamma]_{j}\delta N$. Now the rationale behind choosing the coefficients of different terms in the energy expression (such as $\frac{\gamma_{ad}-1}{\gamma_{ad}^{2}C^{2}_{\rho}}$ term with $\langle\delta\rho,\delta\rho\rangle_{L^{2}}$) may be seen as follows. Note that in the time evolution of the lowest order energy, we have terms like $\langle\delta\rho,\Delta_{\gamma}\phi\rangle_{L^{2}}$ and $\langle k^{tr},\mathcal{L}_{\gamma,\gamma}h\rangle_{L^{2}}$. These terms are dangerous because, when we reach the term with highest Sobolev regularity, the higher order versions of these mixed terms $\langle\delta\rho,\Delta^{s}_{\gamma}\phi\rangle_{L^{2}}$, $\langle k^{tr},\mathcal{L}^{s}_{\gamma,\gamma}h\rangle_{L^{2}}$ require regularity more than the maximum available regularity. Therefore, we want these terms to get cancelled with their negative counterparts. Note that with this particular choice of the positive coefficients of different terms in the energy expression, these dangerous terms are cancelled point-wise. Therefore, we are able to control the time derivative of the energy functional in terms of the maximum available Sobolev norms of the perturbations.

There are additional terms that appear when we apply $\partial_{t}$ on the volume form $\mu_{\gamma}$ since $\partial_{t}\gamma=l^{TT||}=\frac{1}{a}k^{TT||}$. However, since $\partial_{t}\mu_{\gamma}=\frac{1}{2}\mu_{\gamma}\tr_{\gamma}(\partial_{t}\gamma)$, this term does not contribute as a result of $\gamma-$trace-less property of $k^{TT||}$. In addition there are terms that arise due to appearance of the metric $\gamma$ in the definition of the inner product. We denote these additional terms by $\mathcal{ER}$.  Making use of the $\gamma-$traceless property of $k^{tr}$ and $k^{TT||}$ i.e., $\gamma^{ij}k^{tr}_{ij}=0=\gamma^{ij}k^{TT||}_{ij}$ and $\mathcal{L}_{\gamma,\gamma}k^{TT||}=0$, the error term $\mathcal{ER}$ may be written as follows (using Sobolev embedding $||k^{TT||}||_{L^{\infty}}\lesssim ||k^{TT||}||_{H^{s-1}}$ for $s>\frac{n}{2}+1$)  
\begin{eqnarray}
|\mathcal{ER}|\lesssim \frac{1}{a}||k^{TT||}||_{H^{s-1}}\mathcal{E}_{1}.
\end{eqnarray}
Now using the momentum constraint (\ref{eq:MCNW}), $k^{tr}_{ij}=k^{TT}_{ij}+(L_{Y}\gamma-\frac{2}{n}\nabla_{m}Y^{m}\gamma)_{ij}$, $(L_{Y}\gamma)_{ij}:=\gamma_{ik}\nabla[\gamma]_{j}Y^{k}+\gamma_{ki}\nabla[\gamma]_{i}Y^{k}$, and $2\nabla_{m}Y^{m}=\tr_{\gamma}(L_{Y}\gamma)$, we obtain 
\begin{eqnarray}
|\mathcal{ER}|\lesssim \frac{1}{a}\left(||k^{tr}||_{H^{s-1}}+||Y||_{H^{s}}\right)\mathcal{E}_{1}
\end{eqnarray}
where $Y$ satisfies the following elliptic equation
\begin{eqnarray}
\nabla[\gamma]^{i}\nabla[\gamma]_{i}Y^{j}+R[\gamma]^{j}_{i}Y^{i}+(1-\frac{2}{n})\nabla[\gamma]^{j}(\nabla[\gamma]_{i}Y^{i})=a^{2-n\gamma_{ad}}C_{\rho} v^{i}.
\end{eqnarray}
Elliptic regularity yields the following estimate for $Y$
\begin{eqnarray}
||Y||_{H^{s+1}}\lesssim C_{\rho}a^{2-n\gamma_{ad}}||v||_{H^{s-1}}.
\end{eqnarray}
Therefore, $\mathcal{ER}$ satisfies $|\mathcal{ER}|\lesssim \frac{1}{a}(||k^{tr}||_{H^{s-1}}+C_{\rho}a^{2-n\gamma_{ad}}||v||_{H^{s-1}})\mathcal{E}_{1}$. Putting everything together, we obtain
\begin{eqnarray}
\frac{\partial\mathcal{E}_{1}}{\partial t}&=&-(n-1)\frac{\dot{a}}{a}<k^{tr},k^{tr}>_{L^{2}}+\frac{n\dot{a}}{a}(\gamma_{ad}-\frac{n+1}{n})<\phi,\Delta_{\gamma}\phi>_{L^{2}}\\\nonumber 
 &&+\frac{n\dot{a}}{a}(\gamma_{ad}-\frac{n+1}{n})<\xi,\xi>_{L^{2}}+\frac{n\dot{a}}{a}(\gamma_{ad}-\frac{n+1}{n})<\mathcal{H},\mathcal{H}>_{L^{2}}\\\nonumber 
 &&+\mathcal{B}.
\end{eqnarray}
Trivial power counting of $a$ and $\dot{a}$ yields the desired estimate of $\mathcal{B}$
\begin{eqnarray}
|\mathcal{B}|\lesssim (\gamma_{ad}-1)\frac{\dot{a}}{a}||\delta\rho||_{H^{s-1}}||\delta N||_{H^{s+1}}+\frac{1}{a}||\phi||_{H^{s}}||\delta N||_{H^{s+1}}+\frac{1}{a}\nonumber||\delta N||_{H^{s+1}}||k^{tr}||_{H^{s-1}}\\\nonumber
+\frac{\dot{a}}{a}||\delta N||_{H^{s+1}}||h||_{H^{s}}+\frac{1}{a}||X||_{H^{s+1}}||h||_{H^{s}}+\frac{1}{a}(||k^{tr}||_{H^{s-1}}+C_{\rho}a^{2-n\gamma_{ad}}||v||_{H^{s-1}})\mathcal{E}_{1}.
\end{eqnarray}
Similarly, the time derivatives of the higher order energies may be computed. The following lemma states the time derivatives of the higher order energies.\\
\textbf{Lemma 7:}~\textit{Let $s>\frac{n}{2}+1$, $\gamma\in\mathcal{E}in_{-\frac{1}{n}}$ be the shadow of $g\in\mathcal{M}$ and assume $(h=g-\gamma,k^{tr},\delta\rho, \phi,\xi,\chi)$ satisfy the linearized Einstein-Euler system.  Also assume there exists a $\delta>0$ such that $(h,k^{tr},\delta\rho, \phi, \xi,\chi)\in \mathcal{B}_{\delta}(0)\subset H^{s}\times H^{s-1}\times H^{s-1}\times H^{s}\times H^{s-1}\times H^{s-1}$. The time derivative of the higher order energy reads
\begin{eqnarray}
 \frac{\partial \mathcal{E}_{i}}{\partial t}&=&-(n-1)\frac{\dot{a}}{a}\langle k^{tr},\mathcal{L}^{i-1}_{\gamma,\gamma}k^{tr}\rangle_{L^{2}}\nonumber+\frac{n\dot{a}}{a}(\gamma_{ad}-\frac{n+1}{n})\langle\phi,\Delta^{i}_{\gamma}\phi\rangle_{L^{2}}\\\nonumber 
 &&+\frac{n\dot{a}}{a}(\gamma_{ad}-\frac{n+1}{n})\langle\xi,\Delta^{i-1}\xi\rangle_{L^{2}}+\frac{n\dot{a}}{a}(\gamma_{ad}-\frac{n+1}{n})\langle\mathcal{H},\Delta^{i-1}_{\gamma}\mathcal{H}\rangle_{L^{2}}\\\nonumber 
 &&+\mathcal{B}_{i},
\end{eqnarray}
where $\mathcal{B}$ satisfies 
\begin{eqnarray}
|\mathcal{B}_{i}|\lesssim (\gamma_{ad}-1)\frac{\dot{a}}{a}||\delta\rho||_{H^{s-1}}||\delta N||_{H^{s+1}}+\frac{1}{a}||\phi||_{H^{s}}||\delta N||_{H^{s+1}}+\frac{1}{a}\nonumber||\delta N||_{H^{s+1}}||k^{tr}||_{H^{s-1}}\\\nonumber
+\frac{\dot{a}}{a}||\delta N||_{H^{s+1}}||h||_{H^{s}}+\frac{1}{a}||X||_{H^{s+1}}||h||_{H^{s}}+\frac{1}{a}(||k^{tr}||_{H^{s-1}}+C_{\rho}a^{2-n\gamma_{ad}}||v||_{H^{s-1}})\mathcal{E}_{i}
\end{eqnarray}
for $1\leq i\leq s$.
}\\
\textbf{Proof:} With a similar type of calculation using the evolution equations of the Einstein-Euler system together with the inequalities (\ref{eq:inequality1}-\ref{eq:inequality3}) whenever necessary, we obtain the time derivatives of the higher order energies. The estimate of the residue terms $\mathcal{B}_{i}$ may be obtained by carefully keeping track of the explicitly time dependent terms $a$ and $\dot{a}$.\\
Utilizing the previous two lemmas, we may now obtain the time derivative of the total energy. The total energy satisfies the following differential equation
\begin{eqnarray}
\frac{\partial \mathcal{E}}{\partial t}\\
=-(n-1)\frac{\dot{a}}{a}\sum_{i=1}^{s}\langle k^{tr},\mathcal{L}^{i-1}_{\gamma,\gamma}k^{tr}\rangle_{L^{2}}\nonumber+\frac{n\dot{a}}{a}(\gamma_{ad}-\frac{n+1}{n})\sum_{i=1}^{s}\langle\phi,\Delta^{i}_{\gamma}\phi\rangle_{L^{2}}\\\nonumber 
 +\frac{n\dot{a}}{a}(\gamma_{ad}-\frac{n+1}{n})\sum_{i=1}^{s}\langle\xi,\Delta^{i-1}\xi\rangle_{L^{2}}+\frac{n\dot{a}}{a}(\gamma_{ad}-\frac{n+1}{n})\sum_{i=1}^{s}\langle\mathcal{H},\Delta^{i-1}_{\gamma}\mathcal{H}\rangle_{L^{2}}\\\nonumber
 +\mathcal{G},
\end{eqnarray}
where $\mathcal{G}$ satisfies 
\begin{eqnarray}
|\mathcal{G}|\lesssim (\gamma_{ad}-1)\frac{\dot{a}}{a}||\delta\rho||_{H^{s-1}}||\delta N||_{H^{s+1}}+\frac{1}{a}||\phi||_{H^{s}}||\delta N||_{H^{s+1}}+\frac{1}{a}\nonumber||\delta N||_{H^{s+1}}||k^{tr}||_{H^{s-1}}\\\nonumber
+\frac{\dot{a}}{a}||\delta N||_{H^{s+1}}||h||_{H^{s}}+\frac{1}{a}||X||_{H^{s+1}}||h||_{H^{s}}+\frac{1}{a}(||k^{tr}||_{H^{s-1}}+C_{\rho}a^{2-n\gamma_{ad}}||v||_{H^{s-1}})\mathcal{E}.
\end{eqnarray}
Now we make use of the elliptic estimates (lemma 4-lemma 5). An estimate of the shift vector field reads 
\begin{eqnarray}
||X||_{H^{s+1}}\leq C(\dot{a}a^{1-n\gamma_{ad}}||\delta\rho||_{H^{s-1}}+C_{\rho}a^{2-n\gamma_{ad}}||v||_{H^{s-1}}). 
\end{eqnarray}
Now the $n-$velocity field $v$ may be estimated in terms of its irrotational, rotational, and harmonic parts as follows 
\begin{eqnarray}
||v||_{H^{s-1}}&\leq& ||(d\phi)^{\sharp}||_{H^{s-1}}+||\xi||_{H^{s-1}}+||\mathcal{H}||_{H^{s-1}}\\\nonumber 
&\leq& ||\phi||_{H^{s}}+||\xi||_{H^{s-1}}+||\mathcal{H}||_{H^{s-1}}.
\end{eqnarray}
Therefore, the shift vector field may be estimated as follows 
\begin{eqnarray}
||X||_{H^{s+1}}\lesssim \dot{a}a^{1-n\gamma_{ad}}||\delta\rho||_{H^{s-1}}+a^{2-n\gamma_{ad}}(||\phi||_{H^{s}}+||\xi||_{H^{s-1}}+||\mathcal{H}||_{H^{s-1}})
\end{eqnarray}
which together with the estimate of the lapse perturbation (lemma 4)
\begin{eqnarray}
||\delta N||_{H^{s+1}}\lesssim a^{2-n\gamma_{ad}}||\delta\rho||_{H^{s-1}}, 
\end{eqnarray}
yields
\begin{eqnarray}
\mathcal{G}&\lesssim& \dot{a}a^{1-n\gamma_{ad}}||\delta\rho||^{2}_{H^{s-1}}+a^{1-n\gamma_{ad}}(||\phi||^{2}_{H^{s}}+||\delta\rho||^{2}_{H^{s-1}})+a^{1-n\gamma_{ad}}(||\delta\rho||^{2}_{H^{s-1}}\\\nonumber &&+||k^{tr}||^{2}_{H^{s-1}})+\dot{a}a^{1-n\gamma_{ad}}(||\delta\rho||^{2}_{H^{s-1}}+||h||^{2}_{H^{s}})+a^{1-n\gamma_{ad}}(||\phi||^{2}_{H^{s}}+||\xi||^{2}_{H^{s-1}}\\\nonumber
&&+||\chi||^{2}_{H^{s-1}}+||h||^{2}_{H^{s}})+\frac{1}{a}(||k^{tr}||_{H^{s-1}}+C_{\rho}a^{2-n\gamma_{ad}}||v||_{H^{s-1}})\mathcal{E}\\\nonumber 
&\lesssim&(\dot{a}a^{1-n\gamma_{ad}}+a^{1-n\gamma_{ad}})||\delta\rho||^{2}_{H^{s-1}}+a^{1-n\gamma_{ad}}||\phi||^{2}_{H^{s}}+(\dot{a}a^{1-n\gamma_{ad}}+a^{1-n\gamma_{ad}})||h||^{2}_{H^{s}}\\\nonumber &&+a^{1-n\gamma_{ad}}||k^{tr}||^{2}_{H^{s-1}}+\frac{1}{a}(||k^{tr}||_{H^{s-1}}+C_{\rho}a^{2-n\gamma_{ad}}||v||_{H^{s-1}})\mathcal{E}\\\nonumber 
&\lesssim&(\dot{a}a^{1-n\gamma_{ad}}+a^{1-n\gamma_{ad}})\mathcal{E}+\frac{1}{a}(||k^{tr}||_{H^{s-1}}+C_{\rho}a^{2-n\gamma_{ad}}||v||_{H^{s-1}})\mathcal{E}.
\end{eqnarray}
In the view of small data we may absorb $||k^{tr}||_{H^{s-1}}$ and $||v||_{H^{s-1}}$ in $\frac{1}{a}(||k^{tr}||_{H^{s-1}}+C_{\rho}a^{2-n\gamma_{ad}}||v||_{H^{s-1}})\mathcal{E}$ with a suitable constant. Therefore, the energy inequality becomes
\begin{eqnarray}
\frac{d\mathcal{E}}{dt}&\leq& -(n-1)\frac{\dot{a}}{a}\sum_{i=1}^{s}\langle k^{tr},\mathcal{L}^{i-1}_{\gamma,\gamma}k^{tr}\rangle_{L^{2}}\nonumber+\frac{n\dot{a}}{a}(\gamma_{ad}-\frac{n+1}{n})\sum_{i=1}^{s}\langle\phi,\Delta^{i}_{\gamma}\phi\rangle_{L^{2}}\\\nonumber 
 &&+\frac{n\dot{a}}{a}(\gamma_{ad}-\frac{n+1}{n})\sum_{i=1}^{s}\langle\xi,\Delta^{i-1}\xi\rangle_{L^{2}}+\frac{n\dot{a}}{a}(\gamma_{ad}-\frac{n+1}{n})\sum_{i=1}^{s}\langle\mathcal{H},\Delta^{i-1}_{\gamma}\mathcal{H}\rangle_{L^{2}}\\\nonumber
 &&+C(\dot{a}a^{1-n\gamma_{ad}}+a^{1-n\gamma_{ad}}+a^{-1})\mathcal{E},
\end{eqnarray}
which upon utilizing the first stability criterion
\begin{eqnarray}
\gamma_{ad}\leq \frac{n+1}{n}
\end{eqnarray}
 becomes 
 \begin{eqnarray}
 \frac{d\mathcal{E}}{dt}\leq C(\dot{a}a^{1-n\gamma_{ad}}+a^{1-n\gamma_{ad}}+a^{-1})\mathcal{E}.
 \end{eqnarray}
 Integration of the energy inequality yields 
 \begin{eqnarray}
 \mathcal{E}(t)\lesssim \mathcal{E}(t_{0})e^{C\left(-\frac{1}{n\gamma_{ad}-2}a^{2-n\gamma_{ad}}+\int_{t_{0}}^{t}a^{1-n\gamma_{ad}}dt^{'}+\int_{t_{0}}^{t}a^{-1}dt^{'}\right)}.
 \end{eqnarray}
 Noting that $n\gamma_{ad}-2>0$ ($n\geq 3$), a second stability criterion in the expanding universe is obtained as
\begin{eqnarray} 
\label{eq:integrability}
\int_{t_{0}}^{\infty}a^{1-n\gamma_{ad}}dt<\infty,~\int_{t_{0}}^{\infty}a^{-1}dt<\infty.
\end{eqnarray}
Analysis up to now holds for $\Lambda\geq 0$. However, in order to conclude that the FLRW models considered are stable, we need to consider two separate cases: 1. $\Lambda>0$, 2 $\Lambda=0$.  
\subsection{$\Lambda>0$ case}
We may immediately check whether the stability is satisfied by the spacetimes under consideration including a positive cosmological constant $\Lambda$. From lemma (1), we know that the scale factor exhibits the following asymptotic behaviour for $\Lambda>0$ 
\begin{eqnarray}
a\sim e^{\sqrt{\frac{2\Lambda}{n(n-1)}}t}
\end{eqnarray}
as $t\to\infty$. Let $\sqrt{\frac{2\Lambda}{n(n-1)}}=\alpha>0$. The integrability conditions $\int_{t_{0}}^{\infty}a^{1-n\gamma_{ad}}dt<\infty$ and $\int_{t_{0}}^{\infty}a^{-1}dt<\infty$ are trivially satisfied since, $1-n\gamma_{ad}<0$ for $\gamma_{ad}\in (1,\frac{n+1}{n}), n\geq 3$ and $\int_{t_{0}}^{\infty}e^{-(n\gamma_{ad}-1)\alpha t}dt<\infty, \int_{t_{0}}^{\infty}e^{-\alpha t}dt<\infty$. We may in fact establish the decay of the re-scaled perturbations (in a suitable norm) once we have established the uniform boundedness condition, that is, 
\begin{eqnarray}
\label{eq:uniformbounded}
||h(t)||_{H^{s}}+||k^{tr}(t)||_{H^{s-1}}+||\delta\rho(t)||_{H^{s-1}}+||v(t)||_{H^{s-1}}\\\nonumber
\lesssim ||h(t_{0})||_{H^{s}}+||k^{tr}(t_{0})||_{H^{s-1}}+||\delta\rho(t_{0})||_{H^{s-1}}+||v(t_{0})||_{H^{s-1}}<\delta
\end{eqnarray}
for $t>t_{0}$ and sufficiently small $\delta$. Following this uniform boundedness of the appropriate norm of the perturbations, we may immediately apply the local existence theorem to yield a global existence result. However, we may achieve better than the uniform boundedness in a rapidly expanding background since expansion is expected to `kill' off the perturbations. We may in fact obtain decay of the perturbations in an  appropriate norm.
Since, the rotational and harmonic parts of the $n-$velocity field exhibit decay, we will focus on the scalar field $\phi$, density perturbation $\delta\rho$, and the geometric entities $h$ and $k^{tr}$. First we obtain the estimates for the lapse and the shift. Utilizing the elliptic estimates (lemma 4 and 5) and boundedness of $||\delta\rho||_{H^{s-1}}$,$||v||_{H^{s-1}}$ (\ref{eq:uniformbounded}) one may obtain the following estimates in a straightforward way 
\begin{eqnarray}
\label{eq:lapsees}
||\delta N||_{H^{s+1}}\lesssim e^{-\alpha(n\gamma_{ad}-2)t},\\
\label{eq:shiftes}
||X||_{H^{s+1}}\lesssim e^{-\alpha(n\gamma_{ad}-2)t}.
\end{eqnarray}
Now let us evaluate the time derivative of the following entities 
\begin{eqnarray}
\mathcal{E}_{k}:=\frac{1}{2}\sum_{i=1}^{s-1}\langle k^{tr},\mathcal{L}^{i-1}_{\gamma,\gamma}k^{tr}\rangle_{L^{2}}
\end{eqnarray}
and 
\begin{eqnarray}
\mathcal{E}_{\phi}:=\frac{1}{2}\sum_{i=1}^{s-1}\langle\phi,\Delta^{i}_{\gamma}\phi\rangle_{L^{2}}.
\end{eqnarray}
An explicit calculation yields 
\begin{eqnarray}
\frac{\partial \mathcal{E}_{k}}{\partial t}\leq-2\frac{n\dot{a}}{a}\mathcal{E}_{k}+C(\frac{1}{a}||k^{tr}||_{H^{s-2}}||h||_{H^{s}}\nonumber+\frac{1}{a}||\delta N||_{H^{s}}||k^{tr}||_{H^{s-2}}\\\nonumber
+a^{1-n\gamma_{ad}}(t)||\delta\rho||_{H^{s-2}}||k^{tr}||_{H^{s-2}})+\frac{C}{a},\\
\frac{\partial \mathcal{E}_{\phi}}{\partial t}\leq \frac{2n\dot{a}}{a}(\gamma_{ad}-\frac{n+1}{n})\mathcal{E}_{\phi}+C(\frac{1}{a}||\delta N||_{H^{s-1}}||\phi||_{H^{s-1}}\nonumber+\frac{1}{a}||\delta\rho||_{H^{s-1}}||\phi||_{H^{s-1}})\\\nonumber
+\frac{C}{a}
\end{eqnarray}
Now notice the following extremely important fact. Since the dangerous term $\langle k^{tr},\mathcal{L}^{i}_{\gamma,\gamma}h\rangle_{L^{2}}$
does not get cancelled in $\frac{\partial \mathcal{E}_{k}}{\partial t}$, we lose one order of regularity i.e., we may only obtain a decay of $||k^{tr}||_{H^{s-2}}$ instead of $||k^{tr}||_{H^{s-1}}$ (similarly for $\phi$, we may only control $||\phi||_{H^{s-1}}$). However, since we can consider $s>\frac{n}{2}+2$, we will still obtain the desired point-wise decay. Utilizing the boundedness of the fields as $t\to\infty$, we may write the previous inequalities as follows 
\begin{eqnarray}
\label{eq:decayeqn}
\frac{\partial \mathcal{E}_{k}}{\partial t}\leq -2\frac{n\dot{a}}{a}\mathcal{E}_{k}+\frac{C}{a},\\
\frac{\partial \mathcal{E}_{\phi}}{\partial t}\leq \frac{2n\dot{a}}{a}(\gamma_{ad}-\frac{n+1}{n})\mathcal{E}_{\phi}+\frac{C}{a}.
\end{eqnarray}
Noticing $a\sim e^{\alpha t}$ and $\frac{\dot{a}}{a} \sim \alpha$ as $t\to\infty$, we have 
\begin{eqnarray}
\frac{\partial \mathcal{E}_{k}}{\partial t}\leq -2n\alpha \mathcal{E}_{k}+Ce^{-\alpha t},\\
\frac{\partial \mathcal{E}_{\phi}}{\partial t}\leq 2n\alpha (\gamma_{ad}-\frac{n+1}{n})\mathcal{E}_{\phi}+Ce^{-\alpha t}
\end{eqnarray}
yielding 
\begin{eqnarray}
\mathcal{E}_{k}\lesssim e^{-\alpha t},\\
\mathcal{E}_{\phi}\lesssim e^{-\zeta t},
\end{eqnarray}
where 
$0<\zeta:=\min \left(\alpha, 2n\alpha (\frac{n+1}{n}-\gamma_{ad})\right)$. Note here that we need $\gamma_{ad}<\frac{n+1}{n}$ for a decay. This is extremely important. These decay estimates yield 
\begin{eqnarray}
||k^{tr}||_{H^{s-2}}\lesssim e^{-\frac{\alpha}{2}t}, ||\phi||_{H^{s-1}}\lesssim e^{-\frac{\zeta}{2} t}.
\end{eqnarray}
One may actually obtain a better decay estimate for $||k^{tr}||_{H^{s-2}}$ by an iterative argument. Going back to the evolution equation of $\mathcal{E}_{k}$ 
\begin{eqnarray}
\frac{\partial \mathcal{E}_{k}}{\partial t}\leq-2\frac{n\dot{a}(t)}{a}\mathcal{E}_{k}+C(\frac{1}{a}||k^{tr}||_{H^{s-2}}||h||_{H^{s}}\nonumber+\frac{1}{a}||\delta N||_{H^{s}}||k^{tr}||_{H^{s-2}}\\\nonumber
+a^{1-n\gamma_{ad}}(t)||\delta N||_{H^{s}}||k^{tr}||_{H^{s-2}}+a^{1-n\gamma_{ad}}(t)||\delta\rho||_{H^{s-2}}||k^{tr}||_{H^{s-2}})
\end{eqnarray}
and substituting the estimate $||k^{tr}||_{H^{s-2}}\lesssim e^{-\alpha t/2}$ yields 
\begin{eqnarray}
\frac{\partial \mathcal{E}_{k}}{\partial t}\leq -2n\alpha \mathcal{E}_{k}+Ce^{-\alpha(1+\frac{1}{2}) t}
\end{eqnarray}
as $t\to\infty$. Integrating the previous inequality, one immediately obtains 
\begin{eqnarray}
\mathcal{E}_{k}\lesssim e^{-\alpha(1+\frac{1}{2})t}
\end{eqnarray}
and therefore, 
\begin{eqnarray}
||k^{tr}||_{H^{s-2}}\lesssim e^{-\frac{\alpha}{2}(1+\frac{1}{2})t}.
\end{eqnarray}
Notice that we have gained an extra factor $\alpha/4$ in the rate of decay. Continuing in a similar way, we obtain the final decay rate of $\mathcal{E}_{k}$ to be the following sum 
\begin{eqnarray}
\alpha\left\{1+\frac{1}{2}(1+\frac{1}{2}(1+\frac{1}{2}(1+.........\right\}=\alpha\sum_{k=0}^{\infty}\frac{1}{2^{k}}=2\alpha
\end{eqnarray}
and therefore we obtain 
\begin{eqnarray}
||k^{tr}||_{H^{s-2}}\lesssim e^{-\alpha t}~or~||k^{tr}||_{L^{\infty}}\lesssim e^{-\alpha t}.
\end{eqnarray}
Now using the equation of motion for the metric $g_{ij}$ (\ref{eq:fixed_new}), the estimates for lapse and shift (\ref{eq:lapsees},\ref{eq:shiftes}), and the uniform boundedness result (\ref{eq:uniformbounded}), one readily obtains 
\begin{eqnarray}
||\frac{\partial g}{\partial t}||_{H^{s-1}}\lesssim e^{-\alpha t}
\end{eqnarray}
which upon integration yields
\begin{eqnarray}
||g(t)-g^{\dag}||_{H^{s-1}}\lesssim e^{-\alpha t}~as~t\to\infty,
\end{eqnarray}
where $g^{\dag}\in H^{s-1}$. We need to identify the limit metric $g^{\dag}$. Since $M$ is compact, $H^{I}$ is compactly imbedded in $H^{I-1},~\forall I\geq 1$. From uniform boundedness (\ref{eq:uniformbounded}), the limit metric $g^{\dag}$ must therefore be in $H^{s}$. Now we need to estimate the density function. Using the evolution equation for $\rho=C_{\rho}+\delta\rho$, we obtain 
\begin{eqnarray}
||\rho(t)-\rho^{'}||_{H^{s-2}}\lesssim e^{-\beta t}~~as~~t\to\infty  
\end{eqnarray}
for $\rho^{'}\in H^{s-2}$ and $\beta=\min(\alpha(n\gamma_{ad}-2),\alpha+\zeta/2)$. From uniform boundedness (\ref{eq:uniformbounded}) and compact imbedding, $\rho^{'}\in H^{s-1}$.
Now utilizing the Hamiltonian constraint (\ref{eq:hamilton}), Sobolev embedding ($H^{I}(M)\hookrightarrow L^{\infty}(M),~I>\frac{n}{2}$), and the decay estimate of the relevant fields, one readily obtains 
\begin{eqnarray}
\lim_{t\to\infty}R(g)=-1
\end{eqnarray}
i.e., 
\begin{eqnarray}
||g-g^{\dag}||_{H^{s-1}}\lesssim e^{-\alpha t}
\end{eqnarray}
as $t\to\infty$. Here $g^{\dag}\in \mathcal{M}^{\epsilon}_{-1}$, where $\mathcal{M}^{\epsilon}_{-1}$ is a sufficiently small neighbourhood of the space of Einstein metrics $\mathcal{N}$ in the space of constant negative scalar curvature $-1$. Application of Sobolev embedding on compact domains (since we are now considering $s>\frac{n}{2}+2$), the final pointwise decay estimates of the fields are as follows
\begin{eqnarray}
\label{eq:decayprop}
||g(t)-g^{\dag}||_{L^{\infty}}\lesssim e^{-\alpha t},||k^{tr}(t)||_{L^{\infty}}\lesssim e^{-\alpha t}, ||\rho(t)-\rho^{'}||_{L^{\infty}}\lesssim e^{-\beta t},\\\nonumber 
||N(t)-1||_{L^{\infty}}\lesssim e^{-\alpha(n\gamma_{ad}-2)t},
||X(t)||_{L^{\infty}}\lesssim e^{-\alpha(n\gamma_{ad}-2)t}, ||l^{TT||}(t)||_{L^{\infty}}\lesssim e^{-2\alpha t}.
\end{eqnarray}
The individual constituents of $h$ and $k^{tr}$ may be estimated by using constraints. Using the momentum constraint (\ref{eq:MCNW}) and $k^{tr}_{ij}=k^{TT}_{ij}+(L_{Y}g-\frac{2}{n}\nabla_{m}Y^{m}g)_{ij}$, one obtains 
\begin{eqnarray}
\nabla[\gamma]^{i}\nabla[\gamma]_{i}Y^{j}+R[\gamma]^{j}_{i}Y^{i}+(1-\frac{2}{n})\nabla[\gamma]^{j}(\nabla[\gamma]_{i}Y^{i})=a^{2-n\gamma_{ad}}C_{\rho} v^{i}
\end{eqnarray}
Now of course the kernel of the elliptic operator on the left is trivial since the background metric is negative Einstein. Therefore, noting $||v||_{H^{s-1}}<C$, the following elliptic estimate holds 
\begin{eqnarray}
||Y||_{H^{s+1}}\lesssim e^{-\alpha(n\gamma_{ad}-2)t}
\end{eqnarray}
as $t\to\infty$ and therefore 
\begin{eqnarray}
||k^{TT}||_{H^{s-2}}\lesssim e^{-\alpha t}.
\end{eqnarray}
Notice that, by virtue of the transverse-traceless property of $k^{TT}$, we do not lose regularity of $Y$. Similarly, the Hamiltonian constraint (\ref{eq:HC}) together with the spatial harmonic gauge condition may be utilized to obtain decay of the individual parts of $h_{ij}=h^{TT}_{ij}+fg_{ij}+(L_{W}g)_{ij}$ \cite{koiso1979decomposition}. The first variation of the Hamiltonian constraint (\ref{eq:HC}) with respect to ($g,k^{tr},\rho,v$) at ($g=\gamma,~k^{tr}=0, \rho=C_{\rho},v=0$) yields 
\begin{eqnarray}
DR[\gamma]\cdot h&=&2a^{2-n\gamma_{ad}}\delta\rho,\\\nonumber
\Delta_{\gamma}tr_{g}h+\nabla^{i}\nabla^{j}h_{ij}-R[\gamma]_{ij}h^{ij}&=&2a^{2-n\gamma_{ad}}\delta\rho,
\end{eqnarray}
where $R[\gamma]_{ij}=\frac{R(\gamma)}{n}\gamma_{ij}=-\frac{1}{n}\gamma_{ij}$. Upon substituting the decomposition $h_{ij}=h^{TT}_{ij}+f\gamma_{ij}+(L_{W}\gamma)_{ij}$ into the variation of the Hamiltonian constraint (\ref{eq:HC}) and noticing that $\Delta_{\gamma}tr_{\gamma}(L_{W}g)+\nabla^{i}\nabla^{j}(L_{W}\gamma)_{ij}-R[\gamma]_{ij}(L_{W}\gamma)^{ij}\equiv0$ yields 
\begin{eqnarray}
n\Delta_{\gamma}f+\gamma^{ij}\nabla_{i}\nabla_{j}f-R[\gamma]f&=&2a^{2-n\gamma_{ad}}\delta\rho\\\nonumber
(n-1)\Delta_{\gamma}f-R[\gamma]f&=&2a^{2-n\gamma_{ad}}\delta\rho.
\end{eqnarray}
Following the injectivity of $2\Delta_{\gamma}-R[\gamma]$ for $R[\gamma]<0$ (which is the case here), we immediately obtain the following estimate 
\begin{eqnarray}
||f||_{H^{s+1}}\lesssim a^{2-n\gamma_{ad}}||\delta\rho||_{H^{s-1}}.
\end{eqnarray}
The remaining is the estimation of the vector field $W$. The vector field $W$ is estimated in terms of $f$ through the following elliptic equation, which follows from the spatial harmonic gauge, that is, `$id: (M,\gamma+h) \to (M,\gamma)$ is harmonic'. The linearized version of the spatial harmonic gauge yields
\begin{eqnarray}
\gamma^{ij}D\Gamma^{i}_{jk}[\gamma]\cdot h=0\\\nonumber 
2\nabla[\gamma]_{j}h^{ij}-\nabla^{i}tr_{\gamma}h=0\\\nonumber 
\Delta_{\gamma}W^{i}-R[\gamma]^{i}_{j}W^{j}=-\frac{n-2}{2}\nabla^{i}f.
\end{eqnarray}
Once again, $\delta^{i}_{j}\Delta_{\gamma}-R[\gamma]^{i}_{j}$ is injective since $\gamma$ is negative Einstein. The estimate of $W$ reads 
\begin{eqnarray}
||W||_{H^{s+2}}\lesssim ||f||_{H^{s+1}}. 
\end{eqnarray}
Since, $||\delta\rho||_{H^{s-1}}\lesssim C$ and $a\sim e^{-\alpha t}$, one immediately obtains the following decay 
\begin{eqnarray}
||f||_{H^{s+1}}\lesssim e^{-(n\gamma_{ad}-2)\alpha t},\\
||W||_{H^{s+2}}\lesssim e^{-(n\gamma_{ad}-2)\alpha t}
\end{eqnarray}
as $t\to\infty$. Using Sobolev embedding on compact domains, we may of course deduce the point-wise decay 
\begin{eqnarray}
\label{eq:decayprop2}
||k^{TT}(t)||_{L^{\infty}}\lesssim e^{-\alpha t}, ||Y||_{L^{\infty}}\lesssim e^{-\alpha(n\gamma_{ad}-2) t}, ||f(t)||_{L^{\infty}}\lesssim e^{-(n\gamma_{ad}-2)\alpha t},\\\nonumber ||W(t)||_{L^{\infty}}\lesssim e^{-(n\gamma_{ad}-2)\alpha t}, ||l^{TT||}(t)||_{L^{\infty}}\lesssim e^{-2\alpha t}
\end{eqnarray}
as $t\to\infty$. Notice that the decay property of the matter fields are only possible if the adiabatic index lies in the interval $(1,\frac{n+1}{n})$. In the border line case of $\gamma_{ad}=\frac{n+1}{n}$, one can only prove a uniform boundedness at the linear level. The uniform boundedness and decay property lead to the following main theorem.

\textbf{Main Theorem:} \textit{Let $(g_{0},k^{tr}_{0},\rho_{0},v_{0})\in B_{\delta}(\gamma,0,C_{\rho},0)\subset H^{s}\times H^{s-1}\times H^{s-1}\times H^{s-1}$, where $s>\frac{n}{2}+2$ and $\delta>0$ is sufficiently small. Also consider that $\mathcal{N}$ is the integrable deformation space of $\gamma$ and $\Lambda>0$ is the cosmological constant. Assume that the adiabatic index $\gamma_{ad}$ lie in the interval $(1,\frac{n+1}{n})$. Let $t \mapsto (g(t), k^{tr}(t), \rho(t), v(t))$ be the maximal development of the Cauchy problem for the linearized Einstein-Euler-$\Lambda$ system about (\ref{eq:backgroundsolution}) in constant mean extrinsic curvature spatial harmonic gauge (CMCSH) (\ref{eq:metric1}-\ref{eq:MCNW}) with initial data $(g_{0},k^{tr}_{0},\rho_{0},v_{0})$. Then there exists a $\gamma^{\dag}\in \mathcal{M}^{\epsilon}_{-1}\cap \mathcal{S}_{\gamma}$ and $\gamma^{*}\in \mathcal{N}$ such that $(g,\gamma, k^{tr}, \rho, v)$ flows toward $(\gamma^{\dag},\gamma^{*},0, \rho^{'}, 0)$ in the limit of infinite time, that is, 
\begin{eqnarray}
\lim_{t\to\infty}(g(t,x),\gamma(t,x),k^{tr}(t,x),\rho(t,x),v(t,x))\nonumber=(\gamma^{\dag}(x),\gamma^{*}(x),0,\rho^{'}(x),0)
\end{eqnarray}
and moreover either $\gamma^{\dag}=\gamma^{*}$ or  $\gamma^{*}$ is the shadow of $\gamma^{\dag}$. Here $\mathcal{M}^{\epsilon}_{-1}$ denotes the space of metrics of constant negative ($-1$) scalar curvature sufficiently close to and containing the deformation space $\mathcal{N}$. $\mathcal{S}_{\gamma}$ denotes a harmonic slice through the metric $\gamma$. $\rho^{'}$ is dependent on the initial density. The convergence is understood in the strong sense i.e., with respect to the available Sobolev norms. 
}

\subsection{\textbf{$\Lambda=0$ Case}}
In our linearized analysis, the scale factor $a$ associated with the background solution plays an important role. In the time coordinate $t$, we needed an integrability condition on $a(t)$ (\ref{eq:integrability}) in addition to the suitable range of the adiabatic index $\gamma_{ad}$ (\ref{eq:adiabat}). This integrability condition was satisfied by the scale factor associated with our background solutions since it exhibits asymptotically exponential behavior (\ref{eq:asymp})
\begin{eqnarray}
a\sim e^{\alpha t}~as~t\to\infty,
\end{eqnarray}
where $\alpha=\sqrt{\frac{2\Lambda}{n(n-1)}}$. Now an important question arises. Do we need a positive cosmological constant? In other words: do we need accelerated expansion to stabilize the self-gravitating relativistic fluid on a background that is already expanding? If we turn off the cosmological constant, then we have 
\begin{eqnarray}
a\sim t~as~t\to\infty,
\end{eqnarray}
and therefore $a$ satisfies the first integrability condition of (\ref{eq:integrability}) (i.e., $\int_{t_{0}}^{\infty}a^{1-n\gamma_{ad}}dt<\infty$) but $\int_{t_{0}}^{\infty}a^{-1}dt<\infty$ is no longer satisfied. Therefore, we do not even obtain boundedness. However, we cannot claim that turning off the cosmological constant leads to instability but are simply unable to establish stability with the current method. Note that recently, using a special technique, \cite{fajman2021slowly} showed asymptotic stability of Milne universe in the presence of dust, where the cosmological constant is turned off (the scale factor behaves $\sim t$). However, we want to point out that the FLRW models considered here are different from the Milne model since the latter has a vanishing background energy density.     

\begin{center}
\begin{figure}
\begin{center}
\includegraphics[width=13cm,height=60cm,keepaspectratio,keepaspectratio]{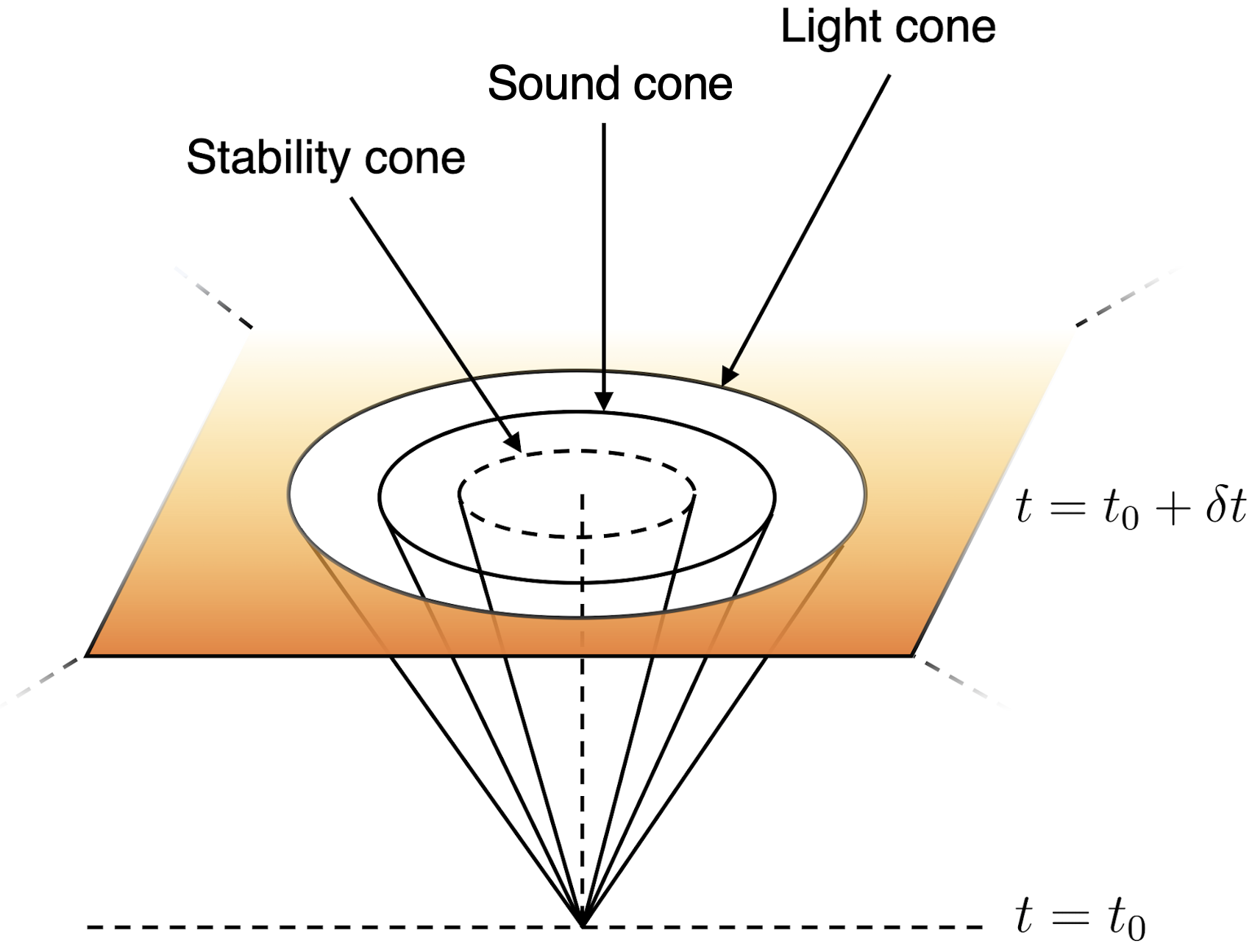}
\end{center}
\begin{center}
\caption{The tangent space picture of null-geometry associated with the Einstein-Euler equations. The stability criteria is essentially defined by the fact that the perturbations to the matter part do not travel at a speed more than $\sqrt{1/n}$ i.e., they are restricted within the innermost cone.}
\label{fig:pdf}
\end{center}
\end{figure}
\end{center}

\subsection{\textbf{Geodesic Completeness}}
In order to establish the future completeness of the perturbed spacetime, we need to show that the solutions of the geodesic equation must exist for an infinite interval of the associated affine parameter. Here we need to be more precise about what we mean by the perturbed spacetimes. If the first variation of the metric is denoted by $\delta \hat{g}$ which satisfies the linearized equations along with a certain smallness condition (a re-scaled version of the physical perturbations to be precise). We show the geodesic completeness for the perturbed spacetimes $\hat{g}=\hat{g}_{FLRW}+\delta\hat{g}$. We provide a rough sketch of the proof in this context following \cite{andersson2004future}. 
Let's designate a timelike geodesic in the homotopy class of a family of timelike curves by $\mathcal{C}$. The tangent vector $\beta=\frac{d\mathcal{C}}{d\lambda}=\beta^{\mu}\partial_{\mu}$ to $\mathcal{C}$ for the affine parameter $\lambda$ satisfies $\hat{g}(\beta,\beta)=-1$, where $\hat{g}$ is the spacetime metric. As $\mathcal{C}$ is causal, we may parametrize it as $(t,\mathcal{C}^{i})$, $i=1,2,3$. We must show that $\lim_{t\to\infty}\lambda=+\infty$, that is, 
\begin{eqnarray}
\lim_{t\to\infty}\int_{t_{0}}^{t}\frac{d\lambda}{dt^{'}}dt^{'}=+\infty.
\end{eqnarray}
Noting that $\beta^{0}=\frac{dt}{d\lambda}$, we must show 
\begin{eqnarray}
\lim_{t\to\infty}\int_{t_{0}}^{t}\frac{1}{\beta^{0}}dt^{'}=+\infty.
\end{eqnarray}
Showing that $|\tilde{N}\beta^{0}|$ is bounded and therefore $\lim_{t\to\infty}\int_{t_{0}}^{t}\tilde{N}dt^{'}=+\infty$ is enough to ensure the geodesic completeness. We first show that $|\tilde{N}\beta^{0}|$ is bounded. Proceeding the same way as that of \cite{mondal2019attractors} together with the estimates obtained in section 5.2, we obtain
\begin{eqnarray}
\tilde{N}^{2}(\beta^{0})^{2}\leq C
\end{eqnarray}
as $t\to\infty$ for some $C<\infty$. Therefore, we only need to show that the following holds 
\begin{eqnarray}
\lim_{t\to\infty}\int_{t_{0}}^{t}\tilde{N}dt=\infty
\end{eqnarray}
in order to finish the proof of timelike geodesic completeness. Notice that $\tilde{N}=N=1+\delta N$. Now $|\delta N(t)|\leq C^{'}e^{-\alpha(n\gamma_{ad}-2)t}, 0<C^{'}<\infty$ as $t\to\infty$ yielding 
\begin{eqnarray}
\lim_{t\to\infty}\left(\int_{t_{0}}^{t}dt-C^{'}\alpha(n\gamma-2)(e^{-\alpha(n\gamma_{ad}-2)t_{0}}\nonumber-e^{-\alpha(n\gamma_{ad}-2)t})\right)\leq \lim_{t\to\infty}\int_{t_{0}}^{t}\tilde{N}dt\\\nonumber 
\leq\lim_{t\to\infty}\left(\int_{t_{0}}^{t}dt+C^{'}\alpha(n\gamma_{ad}-2)(e^{-\alpha(n\gamma_{ad}-2)t_{0}}-e^{-\alpha(n\gamma_{ad}-2)t})\right)\\\nonumber 
=>\lim_{t\to\infty}\int_{t_{0}}^{t}\tilde{N}dt=\infty.
\end{eqnarray}
This completes the proof of the geodesic completeness of the spacetimes of interest under linearized perturbations. The following theorem summarizes the results obtained so far.\\ 
\textbf{Theorem:} \textit{Let $\mathcal{N}$ be the integrable deformation space of $\gamma^{*}$. Then $\exists \delta>0$ such that for any $(g(t_{0}),k^{tr}(t_{0}),\delta\rho(t_{0}),\phi(t_{0}),\xi(t_{0}),\chi(t_{0})\in B_{\delta}(\gamma^{*},0,C_{\rho},0,0,0)\subset H^{s}\times H^{s-1}\times H^{s-1}\times H^{s}\times H^{s-1}\times H^{s-1}\times H^{s-1}$,$s>\frac{n}{2}+1$, the Cauchy problem for the re-scaled linearized Einstein-Euler system in constant mean extrinsic curvature (CMC) and spatial harmonic (SH) gauge is globally well posed to the future and the space-time is future complete.}

\section{Concluding Remarks}
We have proved a global existence result for the linearized Einstein-Euler-$\Lambda$ system about a spatially compact negative spatial curvature FLRW solution (and its higher-dimensional generalization). In addition to the uniform boundedness of the evolving data in terms of its initial value, we proved a decay (in suitable norm) of the perturbations as well. This is an indication of the fact that the accelerated expansion-induced linear decay term may dominate the nonlinearities at the level of small data in a potential future proof of the fully non-linear stability. However, a serious difficulty - namely the possibility of shock formation - arises at the non-linear level in the perfect fluid matter model even without the presence of gravity. Coupling to gravity could make the scenario even more troublesome since pure gravity could itself blow up in finite time through curvature concentration. On the other hand, \cite{rodnianski2009stability, speck2012nonlinear} utilized the accelerated expansion (i.e., included a positive $\Lambda$) of the background solution on the spacetime of topological type $\mathbb{T}^{3}\times \mathbb{R}$ to avoid the shock formation at the small data limit. Our study is significantly different from that of Rodnianski and Speck \cite{rodnianski2009stability} or Speck \cite{speck2012nonlinear} in a sense that we are concerned with negatively curved background (for physically interesting three spatial dimensions, we consider compact hyperbolic manifolds) unlike their study where compact flat torus $\mathbb{T}^{3}$ is used. As mentioned earlier, our study is motivated by the recent observations that do tend to indicate a negatively curved spatial universe. Secondly, they study the fully nonlinear stability (of small data perturbations) while we restrict ourselves with linear theory only. In terms of choice of gauge, \cite{rodnianski2009stability} and \cite{speck2012nonlinear} used a space-time harmonic gauge or wave gauge first by Choquet-Bruhat to prove a local existence theorem for vacuum Einstein's equations. We, on the other hand, used a constant mean extrinsic curvature spatial harmonic gauge (CMCSH) introduced by \cite{andersson2003elliptic}. An advantage of choosing the CMCSH gauge is that one readily has nice elliptic estimates for the associated lapse function and the shift vector field.  
Despite these distinctions, we will observe a similar decay property in the current study with the presence of a positive cosmological constant. We also obtain a similar stability criterion that is similar to \cite{rodnianski2009stability, speck2012nonlinear}. A fully nonlinear analysis of the spacetimes of our interest is currently under intense investigation.

Even though there are numerous studies of cosmological perturbation theory \cite{lifshitz1992gravitational, hawking1966perturbations, weinberg1972gravitation, bardeen1980gauge, kodama1984cosmological, ma1995cosmological}, they are primarily `formal' mode stability results and in a sense (described in the introduction) do not address the `true' linear stability problem. Our study, on the other hand, casts the Einstein-Euler-$\Lambda$ system into a coupled initial value PDE problem (a `Cauchy' problem to be precise). This facilitates the study of the `true' linear stability of suitable background solutions via energy methods. In addition, even the `formal' mode stability analysis is absent in the case of the spatially compact negatively curved FLRW model. This model is particularly interesting since there are infinitely many topologically distinct manifolds mathematically constructible by taking quotients of $\mathbb
{H}^{3}$ by discrete and torsion-free subgroups of $SO^{+}(1,3)$ acting properly discontinuously. The existence of non-trivial harmonic forms precisely encodes such topological information, which as we have seen, inevitably appear in the field equations. In addition, we study the more general `$n+1$' model where the spatial part is described by a compact negative Einstein manifold (which, while restricted to the $n=3$ case becomes compact hyperbolic through the application of the Mostow rigidity theorem). Our study recovers the mode stability results in a straightforward way (briefly described in the appendix). One interesting aspect of our study is that it hints towards the following fact: the FLRW spacetimes with a compact spatial manifold of constant negative sectional do not exhibit an asymptotic stability property with $\Lambda>0$. Simply put, any perturbations to FLRW solutions may not return to solutions of the same type rather converge to nearby solutions characterized by a compact spatial manifold with constant negative scalar curvature. Simultaneously, the matter density perturbations may not approach zero as well. Since the local homogeneity and isotropy criteria require constancy of \textit{sectional} curvature, not just the \textit{scalar} curvature and a homogeneous matter distribution, a natural conclusion would be that the Einsteinian evolution asymptotically drives the homogeneous and isotropic universe (local) to an inhomogeneous and anisotropic state. This may indicate a possibility of `structure formation' \cite{peebles2020large} which tends to break the homogeneity and isotropy of the physical universe. We do not however claim that such a mechanism is yet so compelling based on only a linear stability analysis.

Lastly, it is worth mentioning that our results indicate the linear stability if the adiabatic index $\gamma_{ad}$ lies in the range $(1,\frac{n+1}{n})$. Such a result, while is a necessary condition for stability, is only valid in the linear regime. It would certainly be interesting to check if this criterion persists in the fully nonlinear level as well. As mentioned previously, in the presence of accelerated expansion, Rodnianski and Speck \cite{rodnianski2009stability} and Speck \cite{speck2012nonlinear} obtained the same condition for nonlinear stability of small data on $\mathbb{R}\times \mathbb{T}^{3}$. In addition to \cite{rodnianski2009stability} and \cite{speck2012nonlinear}, avoidance of shock formation by exponential expansion was also studied by \cite{lubbe2013conformal} for radiation fluid ($\mathbb{S}^{3}$ spatial topology), by \cite{oliynyk2021future} for $\gamma_{ad}\in (4/3,3/2)$ ($\mathbb{T}^{3}$ spatial topology),  \cite{hadvzic2015global} for dusts ($\mathbb{T}^{3}$ spatial topology). Recently by \cite{fajman2021slowly} removed the positive cosmological constant and proved the asymptotic stability of the Milne universe including dust. We will extend our study to the fully non-linear case in the future.

\section{Acknowledgement}
P.M would like to thank Prof. Vincent Moncrief for numerous useful discussions related to this project and for his help improving the manuscript. P.M would also like to thank the reviewer and the editorial board member for important comments. This work was supported by Yale university and CMSA at Harvard University.

\section{Appendix: Mode stability analysis}
Now, we briefly describe the mechanics of `mode' stability analysis of spatially compact and (spatially) negatively curved `FLRW' spacetimes. A subtlety is that while decomposing the sections of the vector bundles of the spatial manifold $M$, the topological invariants provided by the harmonic forms enter into the picture. We need to keep track of these topological invariants in the mode equations carefully. First, we start by briefly describing the relevant Hodge decomposition of forms on $M$.

 Our main strategy is to decompose the sections of vector bundles over $M$ in terms of eigenmodes of suitable self-adjoint operators. For this matter, we will briefly introduce Hodge theory and discuss how it is applied in the present context. Even though our background is a closed (compact without boundary) Einstein manifold, the method of this section applies to the case of any closed Riemannian manifold.
 The Hodge theorem (\cite{warner2013foundations}) states that the space of $k-$forms may be written as the following $L^{2}-$orthogonal decomposition (with respect to the given Riemannian metric $g$ on $M$)
\begin{eqnarray}
\Omega^{k}(M)&=&Im(d)\oplus Im(\delta)\oplus\ker(\Delta=d\delta+\delta d),
\end{eqnarray}
where $\delta: \Omega^{k+1}(M)\to\Omega^{k}(M)$ is the $L^{2}$ adjoint of $d: \Omega^{k-1}(M)\to \Omega^{k}(M)$. In this particular case, the Riemannian metric is our background metric $\gamma$ satisfying $R[\gamma]_{ij}=-\frac{1}{n}\gamma_{ij}$. Following the decomposition, a 1-form $A$ may be written as 
\begin{eqnarray}
A&=&d\phi+Z+Y,
\end{eqnarray}
where $\phi\in \Omega^{0}(M)$ and $Z\in \Omega^{1}(M)$ with $Z=\delta\beta, \beta\in \Omega^{2}(M)$ and $(d\delta+\delta d)Y=0$. Now notice an important fact. Clearly $Z$ is divergence-free (and through the metric $\gamma$, its vector field counterpart is also divergence-free). The 1-form $Z$ corresponds to the `curl' part of $A$ and is responsible for the local rotational degrees of freedom. This may be interpreted as follows. Noting the dimension of $\Omega^{2}(M)$ to be $^{n}C_{2}=\frac{n(n-1)}{2}$ and the skew symmetry of an element $\beta$ of $\Omega^{2}(M)$, one could identify the $n(n-1)/2$ basis elements with the Lie algebra basis of $so(T_{x}M), x\in M$. But the generators of the Lie group $SO(T_{x}M)$ are nothing but the generators of local rotation. However, not every divergence-free 1-form may be written as the co-differential of a 2-form. That deficit is precisely fulfilled by the harmonic form $Y$. This additional contribution is topological and is related to the non-vanishing of the first de-Rham cohomology group (or singular cohomology group $H^{1}(M;\mathbb{R})$) the dimension of which is finite for a compact manifold $M$ and corresponds to its first Betti number. Equivalently, this may be described as follows. Lets us consider the map $\delta:\Omega^{k}\to \Omega^{k-1}$. Now, of course $\delta\circ \delta\zeta=0$ for $\zeta_{1}\in\Omega^{k}(M)$. However, not all the elements $\kappa$ of $\Omega^{k-1}(M)$ which belong to the kernel of $\delta$ (such that $\delta\kappa=0$) may be written as $\delta \zeta_{2}$ for $\zeta_{2}\in \Omega^{k}$. This failure of exactness measured by $\ker(\delta:\Omega^{k-1}\to \Omega^{k-2})/image(\delta:\Omega^{k}\to \Omega^{k-1})$ is the space of harmonic forms $\mathcal{H}^{k-1}(M)$. Naturally, $\mathcal{H}^{k}(M)$ is isomorphic to $H^{k}_{deRham}(M)$).

Because of the Hodge decomposition, any non-harmonic eigenform may be decomposed as a sum of exact and co-exact eigenforms. In the case of a closed and connected manifold $M$ (which we consider throughout) harmonic $0-$forms are constant functions and therefore any non-trivial $\phi\in \Omega^{0}(M),$ admits the following formal eigenfunction decomposition, that is, 
\begin{eqnarray}
\phi(t,x)&=&\sum_{a}\alpha^{a}(t)\xi_{a}(x),
\end{eqnarray}
where 
\begin{eqnarray}
\Delta_{\gamma}\xi_{a}&=&\lambda_{a}\xi_{a},
\end{eqnarray}
that is, $\{\xi_{a}\}_{1}^{\infty}$ form a complete set of $L^{2}$ basis elements. The 1-form $A$ may be written formally as 
\begin{eqnarray}
A(t,x)&=&\sum_{a}\alpha^{a}(t)\Theta_{a}(x)+\sum_{c}\phi^{c}(t)\Phi_{c}(x)+\sum_{b=1}^{N}\beta^{b}(t)\mathcal{\zeta}_{b}(x),
\end{eqnarray}
where $N$ is the dimension of the first Cohomology group ($H^{1}(M;\mathbb{R})$), which also equals to first Betti number of $M$ which encodes topological information of $M$. Elementary calculations yield the following for  $\Theta_{a}=d\xi_{a}, \Phi_{c}, \zeta_{b}\in\Omega^{1}(M)$
\begin{eqnarray}
\Delta\Theta_{a}&=&(d\delta+\delta d)\Theta_{a}=\theta_{a}\Theta_{a},\\
\Delta\Phi_{a}&=&(d\delta+\delta d)\Phi_{a}=\phi_{c}\Phi_{a},\\
\Delta\zeta_{b}&=&(d\delta+\delta d)\zeta_{b}=0,
\end{eqnarray}
that is 
\begin{eqnarray}
\Delta_{\gamma}\Theta_{a}+Ric(.,\Theta_{a})&=&\theta_{a}\Theta_{a},\\
\Delta_{\gamma}\Phi_{c}+Ric(.,\Phi_{c})&=&\phi_{c}\Phi_{c},\\
\Delta_{\gamma}\zeta_{b}+Ric(.,\zeta_{b})&=&0.
\end{eqnarray}
($d\Theta_{a}=0, \delta\Phi_{c}=0, d\zeta_{b}=\delta\zeta_{b}=0$)
Note that the following orthonormality holds
\begin{eqnarray}
\int_{M}\Theta_{a}\wedge^{*}\Theta_{b}=\delta_{ab},
\int_{M}\Phi_{a}\wedge^{*}\Phi_{b}=\delta_{ab},
\int_{M}\zeta_{a}\wedge^{*}\zeta_{b}=\delta_{ab},\\
\int_{M}\Theta_{a}\wedge^{*}\Phi_{b}=0,
\int_{M}\Theta_{a}\wedge^{*}\zeta_{b}=0,
\int_{M}\Phi_{a}\wedge^{*}\zeta_{b}=0,
\end{eqnarray}
where the pairing by Hodge star $^{*}: \Omega^{k}(M)\to \Omega^{n-k}(M)$ is defined as follows 
\begin{eqnarray}
\zeta_{a}\wedge ^{*}\zeta_{b}:=\gamma(\zeta_{a},\zeta_{b})\mu_{\gamma}.
\end{eqnarray}
Assuming the background metric to be smooth, one may use a variational principle to establish existence of solutions to these eigenvalue equations. For a rigorous proof, the reader is referred to the excellent book \cite{warner2013foundations}. Note that through the isomorphism between the space of $1-$forms and vector fields induced by the Riemannian metric $\gamma$, we may as well use this decomposition for the vector fields by simply `raising' the index by $\gamma$. The associated scalar and vector fields are expressed formally in the following way
\begin{eqnarray}
\delta N(t,x)=\sum_{a}N^{a}(t)\xi_{a}(x), \delta\rho(t,x)=\sum_{a}D^{a}(t)\xi_{a}(x),\\\nonumber
X=\sum_{a}X^{IRa}(t)\Theta_{a}(x)+\sum_{a}X^{Ra}(t)\Phi_{a}(x)+\sum_{b=1}^{N}X^{Hb}(t)\Theta_{a}(x),\\\nonumber
v=\sum_{a}v^{IRa}(t)\Theta_{a}(x)+\sum_{a}v^{Ra}(t)\Phi_{a}(x)+\sum_{b=1}^{N}v^{Hb}(t)\Theta_{a}(x).
\end{eqnarray}
Here $IR$, $R$, and $H$ are used to denote the irrotational, rotational, and the harmonic parts of the associated vector fields.
The two symmetric $2-$tensors relevant to our analysis may be decomposed as follows (when expressed in local coordinates) 
\begin{eqnarray}
h_{ij}&=&h^{TT}_{ij}+f\gamma_{ij}+(L_{W}\gamma)_{ij},\\
k^{tr}_{ij}&=&k^{TT}_{ij}+(L_{Y}\gamma)_{ij}-\frac{2}{n}\nabla[\gamma]_{m}Y^{m}\gamma_{ij}
\end{eqnarray}
where $f\in \Omega^{0}(M), W,Y\in sections\{TM\}$, and $(h^{TT},k^{TT})$ are  symmetric and transverse-traceless with respect to $\gamma$. $h^{TT}$ may formally be written as a linear combination of the eigensolutions of the Lichnerowicz type Laplacian
\begin{eqnarray}
\mathcal{L}_{\gamma,\gamma}m^{TT}_{ij}&=&\Delta_{\gamma}m^{TT}_{ij}+(R[\gamma]^{k}~_{ij}~^{m}+R[\gamma]^{k}~_{ji}~^{m})m^{TT}_{km}=lm^{TT}_{ij},l\geq 0.
\end{eqnarray}
In general, the kernel of $\mathcal{L}_{\gamma,\gamma}$ would be finite dimensional on compact manifolds. Expressing $Y$ and $W$ in formal modes, 
\begin{eqnarray}
Y(t,x)=\sum_{a}\mathcal{Y}^{IRa}(t)\Theta_{a}(x)+\sum_{a}\mathcal{Y}^{Ra}(t)\Phi_{a}(x)+\sum_{b=1}^{N}\mathcal{Y}(t)^{H b}\zeta_{a}(x),\\
W(t,x)=\sum_{a}\mathcal{W}^{IRa}(t)\Theta_{a}(x)+\sum_{a}\mathcal{W}^{Ra}(t)\Phi_{a}(x)+\sum_{b=1}^{N}\mathcal{W}^{H b}(t)\zeta_{a}(x)
\end{eqnarray}
the formal harmonic decompositions of $h_{ij}$ and $k_{ij}$ are as follows 
\begin{eqnarray}
h(t,x)&=&\sum_{a}h^{TTa}(t)m^{TT}_{a}(x)+(\sum_{a}\alpha^{a}(t)\xi_{a}(x))\gamma\nonumber +\sum_{a}\mathcal{W}^{IRa}(t)L_{\Theta_{a}}\gamma\\\nonumber
&&+\sum_{a}\mathcal{W}^{Ra}(t)L_{\Phi_{a}}\gamma+\sum_{b=1}^{n}\mathcal{W}(t)^{H b}L_{\zeta_{a}}\gamma,\\\nonumber 
k^{tr}(t,x)&=&\sum_{a}k^{TTa}(t)m^{TT}_{a}(x) +\sum_{a}\mathcal{Y}^{IRa}(t)L_{\Theta_{a}}\gamma+\sum_{a}\mathcal{Y}^{Ra}(t)L_{\Phi_{a}}\gamma\\\nonumber 
&&+\sum_{b=1}^{n}\mathcal{Y}^{H b}(t)L_{\zeta_{a}}\gamma.
\end{eqnarray}
One might raise a natural question of why decompose the symmetric $2-$tensor fields in terms of the eigenfunctions of $\mathcal{L}_{\gamma,\gamma}$ instead of $\Delta_{\gamma}$ (as done in \cite{lifshitz1992gravitational, bardeen1980gauge} and elsewhere). This is a crucial point of which the natural answer may be stated as follows. Note the evolution equation (\ref{eq:SF}) of $k^{tr}$. The term `$(R[\gamma]^{k}~_{ij}~^{m}+R[\gamma]^{k}~_{ji}~^{m})h^{TT}_{km}$' appears on the right hand side contained in $\mathcal{L}_{\gamma,\gamma}h$ (it originates from the variation of the Ricci tensor (of $M$)). In the case of $n=3$, one may utilize the constant sectional curvature property of $\gamma$ (satisfying $R[\gamma]_{ij}=-\frac{1}{3}\gamma_{ij}$) to express $R[\gamma]_{ijkl}=-\frac{1}{6}(\gamma_{ik}\gamma_{jl}-\gamma_{il}\gamma_{jk})$ and therefore, `$(R[\gamma]^{k}~_{ij}~^{m}+R[\gamma]^{k}~_{ji}~^{m})h^{TT}_{km}$' simply becomes $-\frac{1}{6}h^{TT}_{ij}$. However, in higher dimensions ($n>3$), $R[\gamma]_{ij}=-\frac{1}{n}\gamma_{ij}$ does not imply that $\gamma$ has constant sectional curvature. Therefore, the reduction mentioned above fails. This problem does not arise if we perform the formal mode decomposition of $TT$ tensors by the eigentensors of $\mathcal{L}_{\gamma,\gamma}$ since in CMCSH gauge the right hand side of the evolution equation (\ref{eq:SF}) contains $\mathcal{L}_{\gamma,\gamma}h^{TT}$. The following ODEs are obtained from the evolution equations in a straightforward way
\begin{eqnarray}
\frac{dh^{TTa}}{dt}=-\frac{2}{a}k^{TTa}+l^{TTa},l^{TTa}=l^{TT||a}, l^{TT||a}=\frac{2}{a}k^{TT||a},\\
\frac{dk^{TTa}}{dt}=-(n-1)\frac{\dot{a}}{a}k^{TTa}+\frac{1}{2a}k^{TTa},\\
\frac{dv^{IRa}}{dt}=\frac{n\dot{a}}{a}(\gamma_{ad}-\frac{n+1}{n})v^{IRa}-\frac{1}{a}N^{a}-\frac{\gamma_{ad}-1}{C_{\rho}\gamma_{ad}a}D^{a},\\
\frac{dD^{a}}{at}=-\frac{C_{\rho}n\gamma\dot{a}}{a}N^{a}+\frac{C_{\rho}\gamma_{ad}}{a}\theta_{a}v^{IRa},\\
\frac{dv^{Ra}}{dt}=\frac{n\dot{a}}{a}(\gamma_{ad}-\frac{n+1}{n})v^{Ra}=> v^{Ra}(t)=v^{Ra}(t_{0})\left(\frac{a(t)}{a(t_{0})}\right)^{n(\gamma_{ad}-\frac{n+1}{n})},\\
\frac{dv^{Hb}}{dt}=\frac{n\dot{a}}{a}(\gamma_{ad}-\frac{n+1}{n})v^{Hb}=> v^{Hb}(t)=v^{Hb}(t_{0})\left(\frac{a(t)}{a(t_{0})}\right)^{n(\gamma_{ad}-\frac{n+1}{n})}.
\end{eqnarray}
Once the entities $(h^{TTa},k^{TTa},D^{a},v^{IRa},v^{Ra},v^{Hb})$ are obtained, the rest of the unknowns may be obtained in terms of these primary variables utilizing the constraint equations and gauge fixing conditions (\ref{eq:metric1}, \ref{eq:MCNW}-\ref{eq:shiftnew}) (which become simply algebraic equations upon implementing the mode decomposition). Therefore, we avoid writing out these equations. Note that the condition `$\gamma\in(1,\frac{n+1}{n})$' is readily obtained to be a necessary condition for stability.\\

\author{Puskar Mondal$^{*}$}
\address{$^{*}$ Center of Mathematical Sciences and Applications,\\
Department of Mathematics,\\
Harvard University, Cambridge, MA02138, USA}
\ead{puskar@cmsa.fas.harvard.edu}

\end{document}